%% file: thesis.tex
\documentclass[11pt, a4paper, oneside]{Thesis} 

\graphicspath{{./Pictures/}} 

\usepackage[square, numbers, comma, sort&compress]{natbib} 
\usepackage{amsmath}
\usepackage{amssymb}
\usepackage{amsthm}
\usepackage[utf8]{inputenc}
\usepackage{graphicx}
\usepackage{dsfont,amsthm,enumerate,amsfonts,caption}
\usepackage[vcentermath]{youngtab}
\hypersetup{urlcolor=black, colorlinks=true} 
\title{\ttitle} 
\newcommand{\bra}[1]{\ensuremath{\left\langle#1\right|}}
\newcommand{\ket}[1]{\ensuremath{\left|#1\right\rangle}}
\newcommand{\ketbra}[2]{\ensuremath{\left|#1\right\rangle\left\langle#1\right|}}
\newcommand{\braket}[2]{\ensuremath{\left\langle#1|#2\right\rangle}}

\DeclareMathOperator{\Tr}{Tr}
\newtheorem{problem}{Problem}
\begin{document}

\frontmatter 

\setstretch{1.3} 

\fancyhead{} 
\rhead{\thepage} 
\lhead{} 

\pagestyle{fancy} 

\newcommand{\HRule}{\rule{\linewidth}{0.5mm}} 

\hypersetup{pdftitle={\ttitle}}
\hypersetup{pdfsubject=\subjectname}
\hypersetup{pdfauthor=\authornames}
\hypersetup{pdfkeywords=\keywordnames}


\begin{titlepage}
\begin{center}

\textsc{\LARGE \univname}\\[1.5cm] 
\begin{figure}[h]
\centering
\includegraphics[scale=0.15]{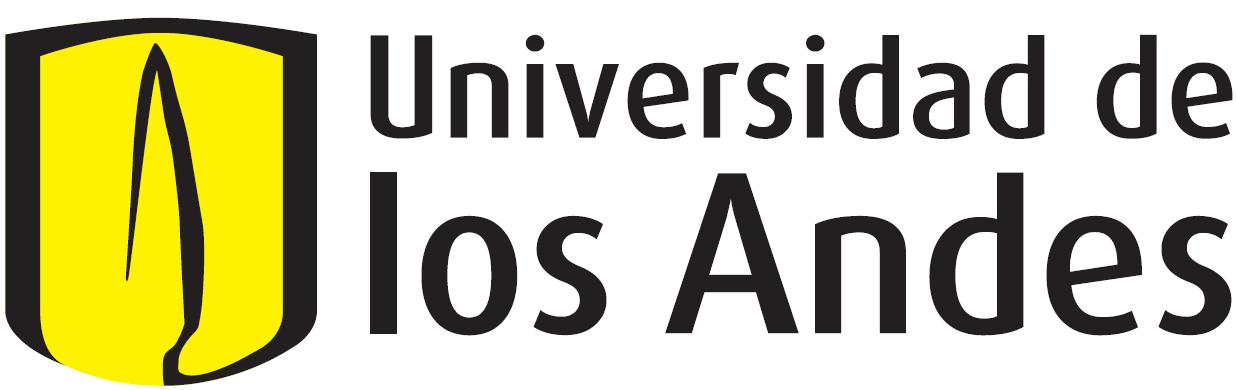}
\end{figure}

\HRule \\[0.4cm] 
{\huge \bfseries \ttitle}\\[0.4cm] 
\HRule \\[1.5cm] 
 
\begin{minipage}{0.4\textwidth}
\begin{flushleft} \large
\emph{Author:}\\
\authornames 
\end{flushleft}
\end{minipage}
\begin{minipage}{0.4\textwidth}
\begin{flushright} \large
\emph{Supervisor:} \\
\supname 
\end{flushright}
\end{minipage}\\[3cm] 
\large \textit{A document submitted in fulfilment of the requirements\\ for the degree of \degreename}\\[0.3cm] 
\textit{in the}\\[0.4cm]
 
{\large \today}\\[4cm] 

\vfill
\end{center}

\end{titlepage}


%
%
%
%
%


\addtotoc{Abstract} 

\abstract{\addtocontents{toc}{\vspace{1em}} 
The aim of this work is to examine the exponential rates at which entanglement distillation occur in three-qubits systems. The approach we will follow to elucidate the entanglement concentration is based on the Schur-Weyl decomposition and the Keyl-Werner theorem. In order to clearly state the main results of this work we will have to introduce the notion of SLOCC equivalence and the covariant algebra which will allow us to classify states in different \emph{entanglement classes}. Our main results comprehend the asymptotic rates for the probability of being in an invariant subspace in the Wedderburn decomposition of the state with effective Kronecker coefficient $g_{\alpha\beta\gamma}=1$. We will also present a combinatorial formula for the effective Kronecker coefficient for two-row Young diagrams. With this work we hope to shed some light in the way a multipartite system is entangled and at which rate can we convey entanglement for further applications.
}

\clearpage 


\addtotoc{Resumen} 

\abstract{\addtocontents{toc}{\vspace{1em}} 
	El propósito de este trabajo es examinar las razones asintóticas a las cuales se presenta distilación de enredamiento en sistemas de tres qubits. El enfoque que se seguirá para ver esta concentración está basado en la dualidad de Schur-Weyl y el teorema de Keyl-Werner. Para poder presentar los principales resultados de este trabajo primero debemos introducir la noción de equivalencia SLOCC y el álgebra de covariantes, la cual nos permitirá clasificar los estados en diferentes \emph{clases de enredamiento}. Uno de nuestros resultados principales son las razones para la probabilidad de estar en un subespacio invariante en la descomposición de Wedderburn en el caso en que el estado tenga un coeficiente efectivo de Kronecker $g_{\alpha,\beta,\gamma}=1$. También presentaremos una formula combinatoria para el coeficiente efectivo de Kronecker para diagramas de Young de dos filas. Con este trabajo esperamos ampliar nuestro conocimiento acerca del enredamiento en sistemas multipartitos, como también las razones en las que podemos transmitir de forma eficiente este recurso, el cual se pretende usar en posteriores aplicaciones.
}

\clearpage 


\setstretch{1.3} 

\acknowledgements{\addtocontents{toc}{\vspace{1em}} 
I would like to thank my advisor, \emph{Prof.} \emph{Alonso Botero}, for guidance and support throughout the years I had worked with him. Also for introducing me to this interesting and engaging topic that is quantum entanglement and its distillation. Besides learning a set of useful mathematical skills, this project leaves me with a more intuitive view about entanglement in multipartite qubit systems.
  
I would also like to thank the Physics department at Universidad de los Andes for giving me all the amenities to develop this research project. Also from the great classes I took during the course of my Master's degree, specially the outstanding classes from \emph{Prof. Pedro Bargueño}, from whom I learned gravitation in a marvellous way.

Moreover, I would like to thank all the post-graduate students and friends I met during the last two years: \emph{Jean Carlos}, \emph{J.Luís}, \emph{Fabián}, \emph{Diego}, for the discussions on many interesting issues and \emph{Andrés},\emph{Nestor},\emph{Luís},\emph{Ronald} for making my stay enjoyable.

I am also indebted to my family that gave me their unconditional support and helped me through the journey that makes this work possible.
}
\clearpage 


\pagestyle{fancy} 

\lhead{\emph{Contents}} 
\tableofcontents 

\lhead{\emph{List of Figures}} 
\listoffigures 



\clearpage 

\setstretch{1.5} 

\title{\emph{Abbreviations}} 
\listofsymbols{ll} 
{
\textbf{LOCC} & \textbf{L}ocal  \textbf{O}perations and  \textbf{C}lassical \textbf{C}ommunication \\
\textbf{SLOCC} & \textbf{S}tochastic \textbf{L}ocal  \textbf{O}perations and  \textbf{C}lassical \textbf{C}ommunication\\
\textbf{LU} & \textbf{L}ocal \textbf{U}nitaries \\
\textbf{GHZ} & \textbf{G}reenberger-\textbf{H}orne-\textbf{Z}eilinger\\
}

\listofnomenclature{lll} 
{
	$\mathcal{H}$ & Hilbert spaces \\
	$S(\rho)$ & von Neumann entropy \\
	$\ket{\psi}$ & Quantum state \\
	$U$ & Unitary matrix\\
	$\Tr$ & Trace\\
	$\phi(\cdot|\psi)$ & Rate function for a given state $\psi$
	%
}




\setstretch{1.3} 



\addtocontents{toc}{\vspace{2em}} 


\mainmatter 

\pagestyle{fancy} 

\input{Chapter1}

\input{Chapter2}

\input{Chapter3test3}
\input{Chapter4}



\addtocontents{toc}{\vspace{2em}} 

\appendix 


\input{AppendixA}

\addtocontents{toc}{\vspace{2em}} 

\backmatter


\label{Bibliography}

\lhead{\emph{Bibliography}} 

\bibliographystyle{alpha} 

\bibliography{bibli} 

\end{document}

%% file: Chapter1.tex

\chapter{Introduction} 

\label{Chapter1} 

\lhead{Chapter 1. \emph{Introduction}} 


Since the origins of the quantum theory in the 1930's, Einstein, Podolski, Rosen \cite{EPR} and Schrödinger\cite{Schro1,Schro2} noted a  striking feature in the behaviour of coupled systems: entanglement (or as Schrödinger called it \emph{Verschränkung}). This phenomenon has been crucial in the development of quantum physics throughout the years. Especially nowadays, entanglement is highly relevant in quantum information theory protocols, such as \emph{quantum state teleportation} or \emph{quantum cryptography}. The theory of \emph{quantum computation} also relies heavily on entangled states as a resource to function properly and outperform its classical counterpart \cite{horodeckis}.
\\
\\
Despite its many applications and extensive study over the years, many aspects of entanglement are still a mystery. In particular, our understanding of entanglement in multipartite systems is very limited \cite{Amico}. In the present work we will focus our attention to the entanglement concentration phenomenon first noticed by Bennet et al. \cite{Bennet_1996}, specifically the asymptotic rates at which such concentration occurs. By analysing this concentration rates we hope to understand the structure of the states on which entanglement is concentrated when we perform a distillation in a three qubit system. The analysis will be framed in Schur-Weyl/ Wedderburn decomposition of $n$ copies of a quantum state.  
\\
\\
The basics of entanglement distillation consist on taking $n$ copies of the same state and convert it to $m<n$ copies of maximally entangled states. For the bipartite case, the maximally entangled states will correspond to the well-known Bell states or EPR. However, for systems of more than two qubits, the mere definition of a maximally entangled state becomes ambivalent and a whole structure of entanglement arises. It is for this reason that classification of states under the SLOCC (stochastic local operations and classical communications) equivalence must be introduced to differentiate the amount and classes of entanglement present in a qubit system.
\\
\\
One approach to study this entanglement classification is by calculating the invariant and covariant algebra of the SLOCC associated group $G_{\text{SLOCC}}=SL_{d_1}\times\cdots\times SL_{d_n}$ acting over the Hilbert space $\mathcal{H}=\mathbb{C}^{d_1}\otimes\cdots\otimes\mathbb{C}^{d_n}$. In this case, entanglement classes correspond to orbits or equivalence classes of $\mathcal{H}$ under $G_{SLOCC}$ and a proposal is that such orbits can be classified with the vanishing of covariants and invariants\cite{Zimmerman}. This approach is highly non-trivial because of the exponential growth of the Hilbert space as we increase the number of particles $n$ or the dimension of each space $d_i$.
\\
\\
In the literature, entanglement classes for different qudit systems have been studied. One of the first results was for three  \cite{Dur1}, four qubits \cite{Verst,Luque} and later for five qubits \cite{Luque2}, where in this last paper the invariant quantitites are calculated. Other relevant results are the ones involving the invariants that can be used to classify entanglement classes in a system  of one qutrit and two qubits ($3\times 2\times 2$) \cite{Miyake}, three qutrits \cite{Briand}, systems $2\times2\times n$ y $2\times3\times3$ \cite{Holweck} and more general cases of $n$-odd qubits \cite{LiLi}. In the same line of work we can find polynomial invariants of order six for $n$-even qubits \cite{Li2} (see the works from \cite{Osterloh} for genuine entanglement in  4,5 and 6 qubits).
\\
\\
 So far, we have mentioned the works made for distinguishable particles, yet  entanglement classification can be studied for bosons and fermions as well, see for example \cite{Ghirardi,Eckert}. We have also some intriguing correspondences between fermion systems and qubit (distinguishable) systems \cite{Levay}. In this paper by Levay et al. it is shown  that  a fermion system with six possible states per fermion can be mapped to a three-qubit system resulting in two inequivalent ways of tripartite entanglement. Other references worth-mention studying the SLOCC classification of entanglement are \cite{Vrana,Chen,Chen2,Sarosi,Sarosi2} where the invariants are constructed in terms of Cayley hyperdeterminants and analogies with distinguishable particle system are made. 
\\
\\
In this document, we will present results for the three qubit case, based on previous results by Morales \cite{Morales} and Walter \emph{et al} \cite{walter2}, witnessing how in the asymptotic regime, the exponential rates for the probabilities for the Wedderburn decomposition subspaces can be written in terms of the covariants/invariants we have mentioned so far. We will also see evidence that this exponential rates can be written as a convex sum of exponential rates which are LU (locally unitary) invariant.
\\
\\
 Three qubit states have the intriguing property of having two inequivalent (under SLOCC) \emph{entanglement classes} with genuine tripartite entanglement. We will see how different entanglement classes result in different Wedderburn's factorizations and how is the effective Kronecker coefficient involved in entanglement classification. Additionally, using Keyl-Werner's theorem \cite{Keyl} we will present results on the characterization of entanglement in the asymptotic case, i.e., when we have a large number of copies of the same state. In the following lines we will introduce some basic concepts and will state the main problem we will address in a more specific way.
\section{Entanglement distillation in multipartite systems}
To describe entanglement within the mathematical framework of quantum mechanics, consider a state $\ket{\psi}\in\mathcal{H}_1\otimes\cdots\otimes\mathcal{H}_n$. Here $\mathcal{H}_i$ is the $i$-th Hilbert space we are considering for our description of a system in nature. We say a state is \emph{separable} or a product state if it can be written as $\ket{\psi}=\ket{\psi_1}\otimes\cdots\otimes\ket{\psi_n}$ with $\ket{\psi_i}\in\mathcal{H}_i$, otherwise we say the state is \emph{entangled}. In the case of a \emph{qudit} (a system with $d$ levels) the Hilbert space is $\mathcal{H}\simeq\mathbb{C}^d$. Particularly, for \emph{qubits} (two level system) we have that $\mathcal{H}\simeq \mathbb{C}^2$. The vectors spanning the Hilbert space for each qubit will be denoted as (a.k.a. computational basis) \begin{equation*}
\ket{0}=\left(\begin{array}{c}
1\\
0
\end{array}\right),\quad \text{and} \quad\ket{1}=\left(\begin{array}{c}
0\\
1
\end{array}\right).
\end{equation*} Then a normalized pure state can be written as
\begin{equation}
\ket{\psi_1}=\alpha\ket{0}+\beta\ket{1},
\end{equation}
where $\alpha,\beta\in\mathbb{C}$ and $|\alpha|^2+|\beta|^2=1$. In this basis, an example of an entangled two-qubit state  $\ket{\psi}\in\mathbb{C}^2\otimes\mathbb{C}^2$ is
\begin{equation}\label{eqBellstate}
\ket{\psi}=\dfrac{\ket{00}+\ket{11}}{\sqrt{2}},
\end{equation}
which is the so-called \emph{Bell state} or \emph{cat-state}. Note that we have used the short hand notation $\ket{00}=\ket{0}\ket{0}=\ket{0}\otimes\ket{0}$ and we will continue using it throughout the document. Now that we know what an entangled state is, the question is whether a state is more entangled than another. To answer this question we must appeal to entanglement measures which will be described in section \ref{secMeasures}. In fact, the state in equation \eqref{eqBellstate} has the property of being a \emph{maximally entangled} state which is the standard \emph{gold coin} for quantum information processes.

Aside from the striking non-locally features that come with entanglement \cite{bell}, it can also be used as a resource in quantum information \cite{horodeckis}. In particular, it is used in transmission of information where entangled states sent between sender and receiver are used to enhance the fidelity of the message. It is also useful to compress information by entanglement concentration - an idea that lies at the heart of the problem we would like to approach and that we will be explained in the next paragraphs using a two-qubit analogy.

Suppose two separated parties are supplied with $n$ copies of an entangled state as a resource for a quantum information protocol. Each state consisting of two entangled particles, each one given to each party. For simplicity we suppose the particles are qubits, e.g., a particle with spin $\frac{1}{2}$ . It is well known that through local operations the parties can concentrate entanglement in $m<n$ pairs of maximally entangled particles \cite{Bennet_1996,Matsumoto}. We will call the parties Alice ($A$) and Bob ($B$) and define their shared state as 
\begin{equation}\label{eqAB}
\ket{\psi}_{AB}=\alpha\ket{00}+\beta\ket{11},
\end{equation}
with $|\alpha|^2+|\beta|^2=1$. State $\ket{\psi}_{AB}$ may not seem to be a general state for a two-qubit system. However, any two-qubit entangled state can be transformed into the form \eqref{eqAB} via SLOCC ( stochastic local operations and classical communication), a set of operations we shall explain later on in section \ref{secSLOCC}. Now we take $n$ copies of state $\ket{\psi}_{AB}$
\begin{equation}
\ket{\psi}_{AB}^{\otimes n}=\left(\alpha\ket{00}+\beta\ket{11}\right)^{\otimes n}
\end{equation}
which can be expressed in terms of sequences $\ket{s_k}=\ket{01010\cdots}$ with $k$ zeroes and $n-k$ ones as
\begin{equation}
\ket{\psi}_{AB}^{\otimes n}=\sum_{k=0}^{n}\binom{n}{k}\alpha^{k}\beta^{n-k}\sum_{s_k}\ket{s_k}\ket{s_k}.
\end{equation}
If we normalize the states $\ket{s_k}\ket{s_k}$ we obtain a superposition of states
\begin{equation}
\ket{\psi}_{AB}^{\otimes n}=\sum_{k=0}^{n}\sqrt{\binom{n}{k}}\alpha^{k}\beta^{n-k}\sum_{s_k}\dfrac{\ket{s_k}\ket{s_k}}{\sqrt{\binom{n}{k}}},
\end{equation}
where the probability of being in a certain state with $k$ zeroes given that we start with vector $\ket{\psi}_{AB}$ is
\begin{equation}\label{eqAB1}
p(k|\psi_{AB})=\binom{n}{k}|\alpha|^{2k}|\beta|^{2(n-k)}.
\end{equation}
We may now ask how this probability behaves asymptotically; that is, in the limit $n\to\infty$. The answer will come in terms of the exponential rate $\phi$ defined as
\begin{equation}
\phi(\bar{k}|\psi_{AB})=-\lim\limits_{n\to\infty}\log \dfrac{p(k|\psi_{AB})}{n},
\end{equation}
where $\bar{k}=k/n$ and can be easily computed from \eqref{eqAB1} to give
\begin{equation}
\phi(\bar{k}|\psi_{AB})=D(\bar{k},1-\bar{k};|\alpha|^2,|\beta|^2)=\bar{k}\log\dfrac{\bar{k}}{|\alpha|^2}+(1-\bar{k})\log\dfrac{1-\bar{k}}{|\beta|^2},
\end{equation}
where $D$ is the well-known Kullback-Leibler divergence or \emph{relative entropy}. Knowing the rate function $\phi$ we will be able to tell asymptotically which state of the superposition will be found most likely. This most probable state will be the one such that minimizes the rate function, i.e., $\bar{k}\approx|\alpha|^2$. Note how this procedure enables us to know the spectrum of $\ketbra{\psi_{AB}}{\psi_{AB}}$ with one measurement and $n$ copies of the state. After the measurement, we will have the state in a subspace of dimension $\binom{n}{k}\sim2^{[n H(\bar{k},1-\bar{k})]}$, where $H$ is the Shannon entropy. Thus the concentration into a smaller subspace ($H\leq1$ here) is also evidenced.
\\
\\
What we will study in this work are the asymptotic rates $\phi$ for the probability $p(\alpha,\beta,\gamma|\psi)$ of tripartite qubits in a decomposition over ac certain triplet $(\alpha,\beta,\gamma)$ that will label the irreducible representation in the Wedderburn decomposition as we will see in section \ref{secWeylSchur} As in the case before, for each $(\alpha,\beta,\gamma)$ we will a have a different subspace, only this time it will not be of sequences with certain amount of zeroes but a rather more complex subspace.

In Chapter \ref{Chapter2} we will introduce all the mathematical machinery that will enable us to study the decomposition of states in the so-called Schur basis and find the asymptotic rates. In this chapter we will include also the description of entanglement via SLOCC and entanglement measures as well as the entanglement/compatibility polytope. 

In Chapter \ref{Chapter3} we will present our main results obtained in the grade project. The results include the relationship between the Kronecker coefficients and SLOCC covariants and also the rates for the GHZ and W entanglement classes calculated over a subregion of the entanglement/compatibility polytope.

In Appendix \ref{AppendixA} we will show detailed calculations that had to be made to obtain the results from chapter \ref{Chapter3}.



%% file: Chapter2.tex

\chapter{Mathematics and Physics Background} 

\label{Chapter2} 

\lhead{Chapter 2. \emph{Mathematical and Physical Background}} 

In our endeavour to study entanglement in multipartite systems, we must first become familiar with some mathematical tools which will prove useful to get an intuition on how entanglement is concentrated in this systems. In most cases each mathematical topic will be accompanied by a physical insight and its relation to the main problem. The physics required to describe and understand the results are also introduced in this chapter. So far we have introduced the notion and mathematical definition of entanglement. In the next section we will introduce a way to classify states upon their degree of entanglement. In order to accomplish such task we must be able to tell if a given state has at least the same amount of entanglement as another, namely we must introduce an equivalence class that deals with the entanglement contained in each state. For that purpose, we must introduce the notion of LOCC and SLOCC equivalence. 

\section{LOCC and SLOCC equivalence}\label{secSLOCC}
 Local operations and classical communication (LOCC) are a set of operations that can be performed on a many-party quantum state. These operations are implemented locally by the different parties that hold a part of the state. Local operations (LO) encompass local unitary operations, the addition of auxiliary systems (ancillas), performing local measurements and tracing out local systems \cite{horodeckis}. Classical communication (CC) refers to the possibility for the parties to communicate via classical channels, e.g. telephone, mail. It is important to remark that all the aforementioned operations can not increase quantum entanglement, although when performed they can decrease entanglement. For example, if we consider a two-party qubit entangled state $\ket{\psi}=\alpha\ket{00}+\beta\ket{11}$ and one of the parties performs a measurement in the ${\ket{0},\ket{1}}$ basis we will obtain either $\ket{00}$ or $\ket{11}$, which are both separable states.

An important property of LOCC is that one can differentiate classical correlations and quantum correlations (entanglement) (see \cite{Chitambar_2014} for an extensive review on LOCC). We can define classical correlations as the ones that we can generate via LOCC (because they can not generate quantum correlations). Thus, an equivalence class can be established between states whether they can be interrelated with certainty via LOCC. We call this equivalence \emph{LOCC equivalence}. Using the density matrix formalism we can see LOCC as a superoperator $\mathcal{E}$ that takes a quantum state into another
\begin{equation}\label{eqSLOCC}
\rho\mapsto\dfrac{\mathcal{E}(\rho)}{\Tr\mathcal{E}(\rho)},
\end{equation}
with the probability of occurrence equal to $\Tr\mathcal{E}(\rho)$. It is established that a LOCC equivalence class must transform a state to another with absolute certainty, thus we require $\Tr\mathcal{E}(\rho)=1$. However, there are protocols such as entanglement distillation \cite{Gisin_1996,Bennet_1996} where two parties can increase the entanglement of a shared state with a certain probability. We are then obliged to work with a more general type of operations, namely SLOCC where the \emph{S} stands for stochastic. Using the density matrix formalism exposed previously, the \emph{SLOCC equivalence} will transform a state into another just like in \eqref{eqSLOCC} with probability $0<\Tr\mathcal{E}(\rho)\leq 1$. Thus we say that a state is SLOCC equivalent to another if we can transform one into another with a non-zero probability using LOCC. In order to define an equivalence class we can also say that two states are SLOCC equivalent if they are related by invertible local operators \cite{Dur1}. The group associated to such operators will turn out to be (for a $n$ qudit system) $[SL(d,\mathbb{C})]^{\times n}$ and the equivalence classes or orbits of SLOCC equivalent states will be \cite{Dur1}
\begin{equation}
\dfrac{(\mathbb{C}^{d})^{\otimes n}}{[SL(d,\mathbb{C})]^{\times n}}.
\end{equation}
An approach to classify such orbits is to study the invariant and covariant quantities under de SLOCC group, their correspondence with some entanglement measures and see whether they vanish for a given orbit. This approach will be made clear in section \ref{secCovariant}, but first we must introduce a way to tell how much entanglement a state possesses.

\section{Entanglement measures}\label{secMeasures}
As its name suggests, entanglement measures quantify the amount of entanglement present in a quantum state. It is therefore a function of the state into the real numbers $\mathbb{R}$. Considering the discussion of the previous section, we must demand that an entanglement measure must be a non-increasing function under LOCC operations on the state, often referred as \emph{LOCC}-monotonicity \cite{Vedral_1997}. Additionally we require that for separable states such measure is zero; ideally it must be a convex function, additive and continuous \cite{Christandl_2004,Plenio_2007}. However, most of the entanglement measures proposed in the literature do not fulfil all of the requirements stated above. Particularly, most of the entanglement measures that we will present in this section are not manifestly SLOCC covariant, an important feature if we want to use this measures to classify different entanglement classes \cite{Zimmerman}.
\\
\\
 In the following we will present some entanglement measures relevant for this work. For a more complete compilation of entanglement measures see reference \cite{Plenio_2007}.
 \subsection{Distillable Entanglement}
 Suppose we have an arbitrary state $\rho$ and we want to transform $n$ copies of such state ($\rho^{\otimes n}$) into maximally entangled states (of two or three qubits in our particular case) only by means of SLOCC. This process is called entanglement concentration or entanglement distillation. An entanglement measure can be associated with this process by calculating  the efficiency at which entanglement concentration can be achieved. Mathematically, it can be expressed as \cite{Plenio_2007}
 \begin{equation}
 E_D(\rho)=\sup\left\lbrace r : \lim\limits_{n\to\infty}\left[\inf_{\Psi}\Tr|\Psi(\rho^{\otimes n})-(\ketbra{K}{K})^{\otimes rn}|\right]=0\right\rbrace,
 \end{equation} 
 where $\Psi$ is an SLOCC operation and $\ket{K}$ is the maximally entangled vector in the same space as $\rho$. For example, if $\rho$ is a two-qubit density matrix, $\ket{K}=\ket{EPR}=(\ket{00}+\ket{11})/\sqrt{2}$. Note that for more than 2 qudits, the vector $\ket{K}$ is not well defined. In this case we might choose a particular multipartite state which is not necessary the maximally entangled state (since there is not a clear hierarchy for more than three qubits). We will be highly interested in the rates $r$ for the case of three qubits and see how they change depending on the entanglement class the state $\rho$ belongs and which state $\ket{K}$ we choose.

 \subsection{Von Neumann Entropy} 
The von Neumann entropy is defined as
\begin{equation}
S(\rho)=-\Tr(\rho\log\rho),
\end{equation}
 we will use the notation
\begin{equation}
S_A=-\Tr(\rho_A\log\rho_A),
\end{equation}
to indicate the von Neumann entropy of a reduced density matrix $\rho_A=\Tr_{BC\cdots}(\rho)$. This measure is an indicator of purity of the state $\rho_A$. It acquires its maximum value for maximally mixed states, i.e. $\rho_A=\mathbb{I}/d_A$ where $d_A$ is the dimension of the Hilbert space associated with $\rho_A$. It is zero for pure states, i.e. $\rho_A=\ketbra{\psi_A}{\psi_A}$. 
\\
\\
For bipartite systems it is an entanglement measure in the following sense: when the state is separable, the local entropy (entropy of the reduced density matrices) is zero; otherwise the total state is entangled. 
\subsection{Entanglement of Formation and concurrence}
The entanglement of formation is an entanglement measure for mixed states that takes into account the decomposition 
\begin{equation}\label{eqmixed}
\rho=\sum_ip_i\ketbra{\psi_i}{\psi_i},
\end{equation}
is not unique. Thus we must work with an ensemble $\{p_i,\ket{\psi_i}\}$ to represent a mixed state $\rho$ in the form \eqref{eqmixed}. The entanglement of formation is defined as the minimum over the elements of the ensemble of the average entanglement contained in each vector $\ket{\psi_i}$. That is \cite{Plenio_2007}
\begin{equation}
E_F(\rho)=\min_{\{p_i,\ket{\psi_i}\}}\sum_i p_i S(\ketbra{\psi_i}{\psi_i}),
\end{equation} 
where $S(\ketbra{\psi_i}{\psi_i})$ is the von Neumann entropy of one of the subsystems.
\\
\\
The variational problem concerning the definition of $E_F$ is a generally a difficult task to perform. One has to consider either states with some sort of symmetry or consider low dimensional cases only. An exact formula to calculate $E_F$ in the case of two-qubit systems exists in terms of the \emph{concurrence}, which is defined as \cite{Plenio_2007}
\begin{equation}
C(\rho)=\max\{0,\lambda_1-\lambda_2-\lambda_3-\lambda_4\},
\end{equation}
with $\lambda_{i+1}\leq\lambda_i$ the square root of the eigenvalues of the matrix $R=\rho(\sigma_y\otimes\sigma_y \rho^* \sigma_y\otimes\sigma_y)$. Here $\rho^*$ is the complex conjugate of $\rho$ (by components) and $\sigma_y$ is a Pauli matrix. The formula for $E_F$ reads
\begin{equation}
E_F(\rho)=H\left(\dfrac{1+\sqrt{1-C^2(\rho)}}{2}\right),
\end{equation}
where $H(x)=-x\log_2(x)-(1-x)\log_2(1-x)$.

\subsection{Local Rank}
The local rank is defined for bipartite systems where the Schmidt decomposition of a state $\ket{\psi}\in\mathcal{H}_A\otimes\mathcal{H}_B\simeq\mathbb{C}^{n}\otimes\mathbb{C}^{m}$ is
\begin{equation}
\ket{\psi}=U_A\otimes U_B\sum_{i=1}^{r_\psi}\sqrt{\lambda_i}\ket{i}_A\ket{i}_B,
\end{equation}
where $U_A, U_B$ are local unitaries and $\lambda_i$ are the so-called Schmidt coefficients. The local rank will be $r_\psi\leq n,m$ and it is a measure of how entangled is the bipartite system. For $r_\psi=1$ the state is separable. Note that by local rank we also denote the rank of the reduced density matrices $\rho_A=\Tr_{BC\cdots}\rho$ for which we will use the notation $r(\rho_A)$. In our particular case of three qubits we will se that for tripartite entangled states the rank of the reduced density matrices will be maximal. However, as we mentioned before there are two inequivalent tripartite entanglement classes in three qubit systems: the W and the GHZ and in both cases the ranks of the reduced density matrices are maximal. This is why we introduce the next entanglement measure, which allows us to distinguish among these two classes.
\subsection{Tangles}
So far we have introduced some entanglement measures which are well defined for bipartite systems. For multipartite systems there is not a general entanglement measure and the main reason for this is that entanglement (maximally entangled states to be more precise) in multipartite systems are still a mystery. Nevertheless in \cite{Coffman} a way to measure multipartite entanglement between a qubit and he rest of the system is proposed in terms of an inequality involving an entanglement quantifier known as \emph{tangle}. The tangle is defined as
\begin{equation}
\tau(\rho)=\inf\sum_i p_i C^2(\ketbra{\psi_i}{\psi_i}),
\end{equation}
where the infimum is taken over the ensemble of $\{p_i,\ket{\psi_i}\}$ compatible with $\rho$. The aforementioned inequality is \cite{Coffman}
\begin{equation}
\tau(A:B)+\tau(A:C)+\cdots \leq\tau(A:BC\cdots),
\end{equation}
where the notation $\tau(A:X)$ means that the tangle is calculated in the bipartition of the total system $A-X$. This implies that the sum of the entanglement of system $A$ with several other systems $B,C\cdots$ is bounded by the entanglement of $A$ with the systems $BC\cdots$ as a whole. In the particular case of three qubit pure states we can define the \emph{residual tangle} or \emph{three tangle} as
\begin{equation}
\tau_3=\tau(A:BC)-\tau(A:B)-\tau(A:C),
\end{equation}
which is locally unitary invariant and it does not depend on the qubit we choose to be $A$. We will also see that this entanglement measure will be related with an invariant under SLOCC operations and will be crucial in the differentiation of the W and GHZ class. We can write explicitly the 3-tangle for a pure state $\ket{\psi}=\sum_{i,j,k=0}^{1}\psi_{ijk}\ket{ijk}$ as \cite{Zimmerman}
\begin{equation}
\tau_3=2|\psi_{ijk}\psi_{i'j'm}\psi_{npk'}\psi_{n'p'm'}\epsilon^{ii'}\epsilon^{jj'}\epsilon^{kk'}\epsilon^{nn'}\epsilon^{mm'}\epsilon^{pp'}|,
\end{equation}
where $\epsilon^{ii'}$ is the Levi-Civita tensor in two dimensions and the Einstein sum convention is used.
\section{SLOCC Entanglement classes of three qubits}\label{secEntanglementclasses}
\begin{figure}[h]
	\centering
	\includegraphics[scale=0.4]{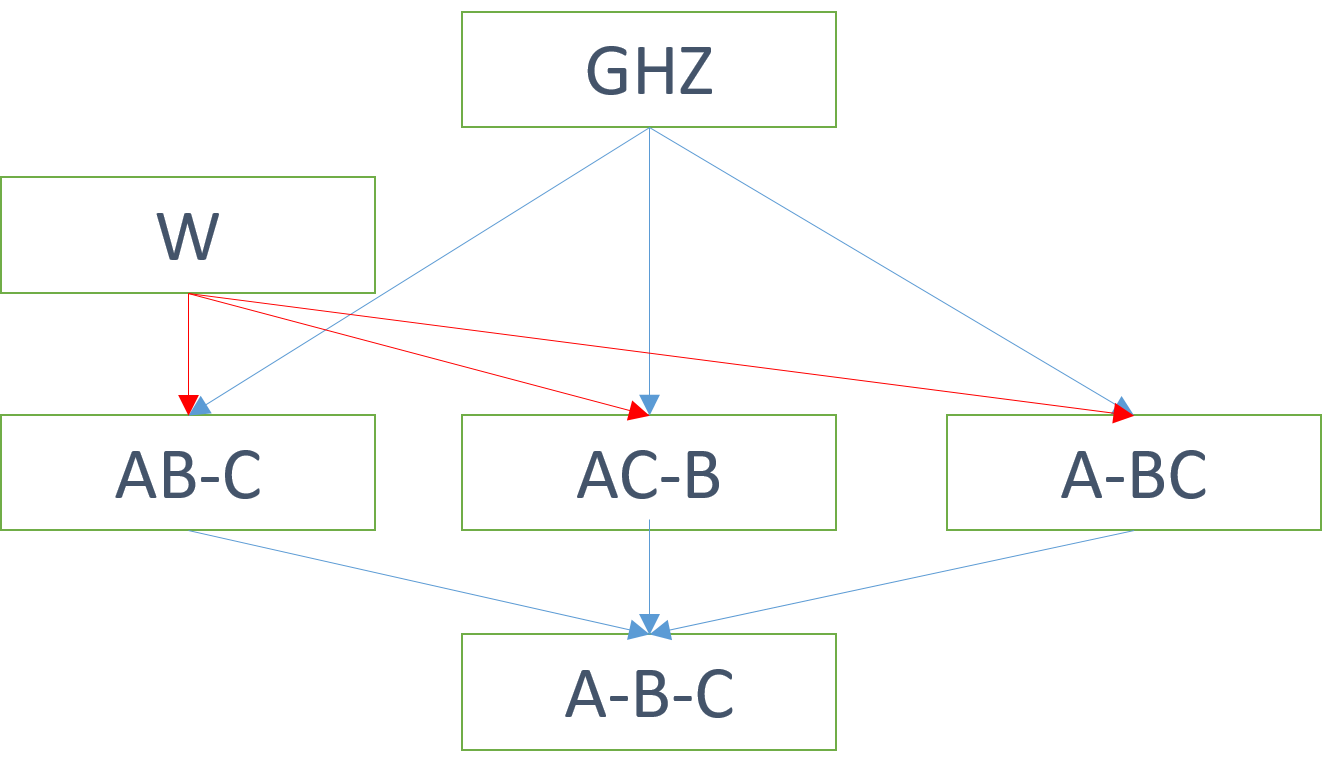}
	\caption{Hierarchy of entanglement classes for three qubits. The arrows indicate which transformations (non-invertible) are possible}
	\label{figHierarchy}
\end{figure}

Following \cite{Dur_2000}, it can be shown that under SLOCC there are six entanglement classes. These inequivalent classes have a hierarchy with the tripartite entanglement classes on top and the separable states at the bottom. Non-invertible operators are used to go downwards through the classes in the hierarchy as is schematically shown in Fig. \ref{figHierarchy}.
\\
The six entanglement classes found by Dür et al. are described in the following sections
\subsection{Separable states $\ket{A-B-C}$}
A typical state belonging to the separable class of entanglement can be written as
\begin{equation}
\ket{\psi_{A-B-C}}=\ket{000},
\end{equation} 
and all the others can be obtained via SLOCC operations (remember SLOCC can not increase entanglement). The rank of the reduced density matrices is equal to $r(\rho_A)=r(\rho_B)=r(\rho_C)=1$ and all the local entropies are zero $S_A=S_B=S_C=0$. The 3-tangle is also equal  to zero.
\subsection{Bipartite entangled states $\ket{A-BC} , \ket{AB-C} , \ket{AC-B}$}
There are three entanglement classes in which there is entanglement among two qubits and a third qubit that is not entangled to any of the former. The three entanglement classes can be differentiated by which position does this third qubit occupies. We denote this classes by the notation $\ket{AB-C}$ where in this case the qubit $C$ is the one that is not entangled to either $A$ or $B$. A typical state in class $\ket{AB-C}$ can be written as
\begin{equation}
\ket{\psi_{AB-C}}=(\alpha\ket{00}+\beta\ket{11})\otimes\ket{0},
\end{equation}
and in a similar fashion for the $\ket{AC-B},\ket{A-BC}$ entanglement classes. Using the entanglement measures introduced in the previous section it is easy to prove that (in the case of $\ket{AB-C}$) $r(\rho_A)=r(\rho_B)=2$ while $r(\rho_C)=1$. Regarding the local von Neumann entropy we have that $S_A\neq0,S_B\neq0$ and $S_C=0$. The 3-tangle for the three bipartite entangled states is identically zero.
\subsection{Tripartite entangled states: W and GHZ}\label{secTripent}
It is a remarkable fact that in a simple system such as three qubits, two inequivalent kinds of tripartite entanglement arise. According to Dür et al. in \cite{Dur_2000}, a three qubit state with tripartite entanglement can be converted via SLOCC to only one of the following states
\begin{equation}
\ket{W}=\dfrac{\ket{001}+\ket{010}+\ket{100}}{\sqrt{3}},
\end{equation}
or
\begin{equation}
\ket{GHZ}=\dfrac{\ket{000}+\ket{111}}{\sqrt{2}}.
\end{equation}
That is, if a state $\ket{\phi_1}$ can be converted to the state $\ket{GHZ}$ and a state $\ket{\phi_2}$ can be converted into the state $\ket{W}$, then the state $\ket{\phi_1}$ can not be converted to the state $\ket{\phi_2}$ nor the other way around. This result is justified in \cite{Dur_2000} with the relation among SLOCC transformations and the minimum number of terms in a linear combination of product states in which a tripartite entangled stated can be decomposed. In other words, not all states with tripartite entangled can be reduced to the sum of two product state (like the GHZ). However, a state in the W entanglement class can be approximated to arbitrary precision by a state in the GHZ class, the converse is not true \cite{Dur_2000}. Regarding the entanglement measures we use for the previous classes we have that $r(\rho_A)=r(\rho_B)=r(\rho_C)=2$ and $S_A\neq0,S_B\neq0,S_C\neq0$ for both W and GHZ classes. The 3-tangle is the entanglement measure that allows us to differentiate between the W and GHZ class, it vanishes for the W but not in the GHZ. In the next lines we will elaborate on each of these classes making emphasis in the differences.
\\
\\
On one hand we have that a typical state in the $\ket{W}$ class of entanglement depends on three parameters $a,b,c$ and can be written as
\begin{equation}\label{eqWgeneral}
\ket{\psi_W}=\sqrt{a}\ket{100}+\sqrt{b}\ket{010}+\sqrt{c}\ket{001}+\sqrt{d}\ket{000},
\end{equation}
with $a,b,c\leq0$ and $d=1-a-b-c\leq0$. A state in the form of \eqref{eqWgeneral} can be obtained from $\ket{W}$ by the following transformation
\begin{equation}
\ket{\psi_W}=\left(\begin{array}{cc}
\sqrt{c} & \sqrt{d}\\
0 & \sqrt{a}
\end{array}\right)\otimes \left(\begin{array}{cc}
\sqrt{3} & 0\\
0 & \sqrt{\dfrac{3b}{c}}
\end{array}\right) \otimes \left(\begin{array}{cc}
1 & 0\\
0 & 1
\end{array}\right) \ket{W}.
\end{equation}
Notice that when a system is traced out in $\ket{W}$, the remaining two qubits are in a maximally entangled state. On the other hand, a typical state in the GHZ entanglement class is described by five parameters, we will write such state as \cite{Dur_2000}
\begin{equation}\label{eqGHZgeneral}
\ket{\psi_{GHZ}}=\sqrt{K}(c_\delta\ket{000}+s_{\delta}e^{i\phi}\ket{\varphi_A}\ket{\varphi_B}\ket{\varphi_C}),
\end{equation}
with
\begin{eqnarray}
\ket{\varphi_A}=c_\epsilon\ket{0}+s_\epsilon\ket{1},\\
\ket{\varphi_B}=c_\theta\ket{0}+s_\theta\ket{1},\\
\ket{\varphi_C}=c_\varphi\ket{0}+s_\varphi\ket{1},
\end{eqnarray}
and $K=(1+2c_{\delta}s_\delta c_\epsilon c_\theta c_\varphi c_\phi)^{-1}$ a normalization factor. We have used the shorthand notation $c_\epsilon=\cos(\epsilon)$ and $s_\epsilon=\sin(\epsilon)$. The ranges for the angles are \cite{Dur_2000} $\delta\in(0,\pi/4]$, $\epsilon,\theta,\varphi\in(0,\pi/2]$ and $\phi\in[0,2\pi)$. A state of the form \eqref{eqGHZgeneral} can be obtained from $\ket{GHZ}$ via the invertible transformation
\begin{equation}
\ket{\psi_{GHZ}}=\sqrt{2K}\left(\begin{array}{cc}
c_\delta & s_\delta c_\epsilon e^{i\phi}\\
0 & s_\delta s_\epsilon e^{i\phi}
\end{array}\right)\otimes \left(\begin{array}{cc}
1 & c_\theta\\
0 & s_\theta
\end{array}\right) \otimes \left(\begin{array}{cc}
1 & c_\varphi\\
0 & s_\varphi
\end{array}\right) \ket{GHZ}.
\end{equation}
The main feature of the GHZ state is that it is the genuine maximally entangled state for three qubits. Nevertheless, when one of the qubits is traced out, a separable mixed state is obtained. 
\\
\\
In table \ref{tabla2}, we can see how to classify the entanglement classes according to the entanglement measures.
\begin{table}[h]
	\centering
	\begin{tabular}{|c|c|c|}
		Class & Entanglement Measure=0  & Entanglement Measure $\neq 0$ \\
		A-B-C & $S_A,S_B,S_C,\tau_3$& $-$\\
		AB-C & $S_A,S_B,\tau_3 $&$S_C $\\
		A-BC & $S_B,S_C,\tau_3$& $S_A$ \\
		AC-B  & $S_A,S_C,\tau_3$& $S_B$\\
		W & $\tau_3$ & $S_A,S_B,S_C$\\
		GHZ & - & $\tau_3,S_A,S_B,S_C$
	\end{tabular}
	\caption{Hierrarchic classification of three qubits based on entanglement measures}
	\label{tabla2}
\end{table}
The entanglement classes described above can be easily visualized using the \emph{entanglement polytope} which arises from the compatibility conditions of the eigenvalues of the reduced density matrices \cite{Han_2004}. A way to understand how this entanglement polytope is related to the entanglement classes we must first introduce de Schur-Weyl duality and the Keyl-Werner theorem, which are the cornerstones of this work.
\section{Representation theory and the Schur-Weyl duality}\label{secWeylSchur}
In this section we are going to present how the space $(\mathbb{C}^d)^{\otimes n}$ can be decomposed as a direct sum of irreducible representations of the general linear group $GL(d,\mathbb{C})$ and the symmetric group $S_n$. This result is known as the \emph{Wedderburn decomposition} and rely on the Schur-Weyl duality. Each irreducible representation in the Wedderburn decomposition is labelled by an integer partition of $n$, a concept we shall explain below.
\subsection{Partitions, Young Frames and Young Tableaux}
A partition $\lambda=(\lambda_1,\cdots,\lambda_{d})$ of an integer number $n$ is a non-decreasing sequence of integer numbers such that
\begin{equation}
\sum_{i=1}^{d}\lambda_i=n \quad;\quad \lambda_i\geq 0,
\end{equation}
where $d$ is the size of the partition (see \cite{Audanert_2006} and references within). Throughout this document we will use the notation $\lambda\vdash_{d}n$ to represent a partition of $n$ of size $d$. A partition $\lambda$ is often represented by a \emph{Young Frame} which have the form
\begin{equation*}
\yng(5,3,1,1),
\end{equation*}
where the $i$-th row has $\lambda_i$ boxes, it has a total of $n$ boxes and if $d$ denotes the numbers in the sequence different from zero, then the Young frame has $d$ rows. For example, the previous Young frame represents the partition $\lambda=(5,3,1,1)$, but also the partition $\lambda=(5,3,1,1,0,0)$ where the only difference is in the value of $d$. It is possible to define a partial ordering for the partitions known as \emph{dominance} order. We say that $\lambda$ \emph{dominates} $\mu$, expressed as $\lambda\succ\mu$, iff
\begin{equation}
\sum_{i=1}^{k}\lambda_k\geq\sum_{i=1}^{k}\mu_k \quad \forall k>0.
\end{equation}
A \emph{Young Tableaux} (YT) of $N$ objects and of shape $\lambda\vdash_d n$ is a Young frame $\lambda$ with the boxes labelled by the numbers $(1,\cdots,N)$, e.g., for $N=3,d=3,n=7,\lambda_1=3,\lambda_2=2,\lambda_3=2$
\begin{equation*}
\young(311,21,13),\quad \young(111,32,2).
\end{equation*}
A \emph{Standard Young Tableaux} (SYT) of $N$ objects and of shape $\lambda\vdash_d n$ is a Young frame with the boxes labelled by the numbers $(1,\cdots,N)$ with the condition that the numbers increase along each row (from left to right) and along each column (top to bottom). This implies that $N=n$, hence each number appears just once in the Young tableaux, e.g., for $\lambda_1=3,\lambda_2=2$ all the possible SYT are
\begin{equation*}
\young(135,24),\quad \young(123,45),\quad \young(124,35),\quad\young(125,34),\quad\young(134,25).
\end{equation*} 
A \emph{ Semistandard Young Tableaux}(SSYT) of $N$ objects and of shape $\lambda\vdash_d n$ is a Young frame with the boxes labelled by the numbers $(1,\cdots,N)$ such that $N\geq d$ and with the condition that the numbers along each row are non-decreasing (from left to right) and increase along each column(top to bottom),e.g., for $N=4,\lambda_1=2,\lambda_2=2,\lambda_3=1$ some SSYT are
\begin{equation*}
\young(111,23,3),\quad \young(112,33,4),\quad \young(222,34,4).
\end{equation*}
The number $f^{\lambda}$ of SYT for shape $\lambda\vdash n$ is given by the \emph{hook formula} \cite{Frame_1954}
\begin{equation}
f^{\lambda}=\dfrac{n!}{\prod_{(i,j)\in\lambda}h_{ij}},
\end{equation}
where the indices $(i,j)$ are coordinates in the Young Frame $\lambda$ indicating the row and column of a box. The number $h_{ij}$ is the \emph{hook lenght} and counts the number of boxes on the right of the box with coordinates $(i,j)$, the ones below it and the box itself. For example, if we have the shape $\lambda_1=3,\lambda_2=2$
\[
\yng(3,2)\quad,
\]
we have that $h_{11}=4,h_{12}=3,h_{13}=1,h_{21}=2,h_{22}=1$, then $f^{\lambda}=5$ as we verified before. The number of SSYT  of given shape $\lambda$ and $N$ objects, denoted by $r^{\lambda}$, is given by the formula \cite{Audanert_2006}
\begin{equation}
r^{\lambda}=\prod_{1\leq i<j\leq N}\dfrac{\lambda_j-\lambda_i+j-i}{j-i}.
\end{equation}
\subsection{Representation theory}
A representation of a group $G$ is a vector space $V$ together with an homomorphism from $G$ to $\text{End}(V)$(the set of linear maps from $V$ to itself). The homomorphism is a function $D:G\to\text{End}(V)$ with the property $D(g_1)D(g_2)=D(g_1g_2)$.
\\
\\
Consider the vector space of a physical system with $d$ levels $\mathcal{H}\simeq\mathbb{C}^{d}$. This vector space is the carrier of the defining representation of the general linear group $GL(d,\mathbb{C})$, the group of $d\times d$ invertible matrices with complex elements. The action of the group for $\ket{\psi},\ket{\psi'}\in\mathbb{C}^d$ and $g\in GL(d,\mathbb{C})$ will be $\ket{\psi'}=g\ket{\psi}$ with the usual matrix multiplication. Now if we consider the space $\mathcal{H}^{\otimes n}\simeq(\mathbb{C}^{d})^{\otimes n}$, which will account for $n$ copies of the original physical system, the group $GL(d,\mathbb{C})$ acts diagonally by tensor powers. Thus $(\mathbb{C}^{d})^{\otimes n}$ is the carrier space of a reducible representation of $GL$ under the action
\begin{equation}
D(g)\ket{\Psi}=g^{\otimes n}\ket{\Psi},
\end{equation}
where $\ket{\Psi}\in(\mathbb{C}^{d})^{\otimes n}$ and $D$ is the representation matrix of $g$. The element $g$ acts individually on each copy of $\mathcal{H}$. This particular representation is reducible \cite{Carter_1995}, then $(\mathbb{C}^{d})^{\otimes n}$ can be decomposed in irreducible representations of $GL(d,\mathbb{C})$. It is a well-known result in representation theory that there is one-to-one correspondence between irreducible representations of $GL(d,\mathbb{C})$ and the set of partitions of $n$ of at most $d$ rows, or equivalently, the set of Young frames with $n$ boxes and at most $d$ rows \cite{Carter_1995}. The carrier space of the irreducible representation $\lambda$ of $GL(d,\mathbb{C})$ will be denoted by $V^{d}_\lambda$ and the dimensions of such space will be the number of SSYT with shape $\lambda$ and $N=d$. Thus we can label a base for $V^{d}_\lambda$ by SSYT \cite{Audanert_2006}.
\\
\\
We can also consider the action of the symmetric group $S_n$ over $(\mathbb{C}^{d})^{\otimes n}$. In fact, the space $(\mathbb{C}^{d})^{\otimes n}$ is the carrier of a reducible representation of $S_n$ under the action
\begin{equation}
D(\pi)(\ket{\psi_1}\ket{\psi_2}\cdots\ket{\psi_n})=\ket{\psi_{\pi^{-1}(1)}}\ket{\psi_{\pi^{-1}(2)}}\cdots\ket{\psi_{\pi^{-1}(n)}},
\end{equation}
for $\ket{\psi_1}\ket{\psi_2}\cdots\ket{\psi_n}\in(\mathbb{C}^{d})^{\otimes n}$. In the same manner $(\mathbb{C}^{d})^{\otimes n}$ can be decomposed as a sum  of irreducible representations of $S_n$ \cite{Carter_1995}. Each irreducible representation can be labelled by a partition $\lambda$ of $n$. Will denote the carrier space of an irreducible representation of $S_n$ as $[\lambda]$. Note that the representation does not depend explicitly on $d$. The dimension of $[\lambda]$ will be the number of SYT $f^{\lambda}$ \cite{Audanert_2006}.
\\
\\
It is easy to see that actions of $GL(d,\mathbb{C})$ and $S_n$ commute. What is remarkable is that the multiplicity spaces, when we decompose $(\mathbb{C}^{d})^{\otimes n}$ as a sum of irreducible representations of $GL(d,\mathbb{C})$, are irreducible representations of $S_n$ and vice-versa. This result is known as the Schur-Weyl duality \cite{Carter_1995} and it enables us to write $(\mathbb{C}^{d})^{\otimes n}$ as
\begin{equation}\label{eqWedderburn}
(\mathbb{C}^{d})^{\otimes n}\simeq\bigoplus_{\lambda\vdash_d n}V_\lambda^{d}\otimes[\lambda].
\end{equation} 
This decomposition is known as the Wedderburn decomposition. We now introduce the Kronecker coefficients $g_{\alpha,\beta,\gamma}$ as the multiplicity of the irreducible representation $[\gamma]$ in the tensor product $[\alpha]\otimes[\beta]$ \cite{walter2}
\begin{equation}
[\alpha]\otimes[\beta]=\bigoplus_{\gamma\vdash n}g_{\alpha,\beta,\gamma}[\gamma].
\end{equation}
The Kronecker coefficients help us calculate the entanglement polytope for the three qubit systems as we will se in section \ref{secKronecker}.
\\
\\
In the space $\mathbb{C}^{d}$ it is usual to work with the computational basis $\{\ket{0},\ket{1},\cdots,\ket{d-1}\}$. Thus the tensor product space $(\mathbb{C}^{d})^{\otimes n}$ will inherit an extended version of  this basis. The elements in the extended computational basis will be sequences $\ket{s}$ of $n$ numbers in the set $\{0,1,\cdots,d-1\}$. Because of the Wedderburn decomposition, a natural basis arises in terms of $\lambda$ and the basis elements of $V^{d}_\lambda$ and $[\lambda]$. Since we would like to work with the new basis, we must found the appropriate transformation among the computational and the new basis, which is the topic of the next section. 
\section{The Schur transform for qubits}
As a consequence of the Wedderburn decomposition, we can find a basis for $(\mathbb{C}^{d})^{\otimes n}$ of the form $\ket{\lambda, m,\mu}$ where $\lambda$ labels the irreducible representation of $GL(d,\mathbb{C})$ and $S_n$. The symbols $m$ and $\mu$ label SSYT and SYT respectively.
In the case of qubits ($d=2$) the dimensions for the carrier spaces are $(\lambda=(\lambda_1,\lambda_2))$
\begin{eqnarray}
	\dim(V_{\lambda}^{2})&=& r^\lambda=\lambda_1-\lambda_2+1,\\
	\dim([\lambda])&=& f^\lambda=\dfrac{(\lambda_1-\lambda_2+1) n!}{\left(\dfrac{n+\lambda_1-\lambda_2}{2}+1\right)!\left(\dfrac{n-\lambda_1+\lambda_2}{2}\right)!}.
\end{eqnarray}

In $V^2_{\lambda}$ we have a basis $\ket{\lambda,m}$ where $m$ denotes a SSYT in the following sense. Define a new variable $j=(\lambda_1-\lambda_2)/2$, then the dimension $r^{\lambda}=2j+1$ in a similar fashion of a representation space of a particle with angular momentum $j$ \cite{Bacon_2006}. This angular momentum point of view will help us to gain some insight on the Schur transform. To see how $m$ relates to SSYT, we consider the set of all possible SSYT for $\lambda_1=7,\lambda_2=4$ (as an example) 
\begin{equation}
\young(1111222,2222),\young(1111122,2222),\young(1111112,2222),\young(111111,222).
\end{equation}
According to our definition $j=3/2$ which we will associate with the numbers $m=-j,-j+1,\cdots,j$ or $m=-3/2,m=-1/2,m=1/2$ and $m=3/2$ respectively. Note that the first $\lambda_2=4$ columns are always fixed, this will be case for a general $\lambda$. Then, the only degrees of freedom will come from the $\lambda_1-\lambda_2$ boxes that differentiate the two rows. If we make the following association $1\to1/2, 2\to-1/2$ and make the sum over the whole SSYT (actually is only necessary to make the sum over the $\lambda_1-\lambda_2$ boxes at the end) we obtain $m$. Thus we described a one-to-one equivalence between the label $m$ and SSYT.
\\
\\
 A basis for $[\lambda]$ is $\ket{\lambda,\mu}$ where $\mu$ is the Yamanouchi symbol. The Yamanouchi symbol is a sequence of $n$ numbers where the $i$-th element of the sequence corresponds to the number of the row $\{1,2\}$ in which the number $i$ is located in the SYT of shape $\lambda$. For example
\begin{equation*}
\young(124,35),
\end{equation*}
has a Yamanouchi symbol $\mu=(11212)$. The correspondence between the Yamanouchi symbol $\mu$ and a SYT is one-to-one \cite{Bacon_2006}.
\\
\\
The basis $\ket{\lambda,m,\mu}$ is known in literature as the Schur basis \cite{Bacon_2006}. Our objective now is to calculate the change of basis element, that is we want $\braket{\lambda,m,\mu}{s}$ where $\ket{s}$ is a sequence in the computational basis.
Note that the Yamanouchi symbol gives us a way(path) to build a Young frame in the following manner. The $i$-th element in the Yamanouchi symbol indicates in which row  we must add the box to construct the Young frame. We begin with one box in the first row (the Yamanouchi symbol always begins with a 1), the second box  must be added in the row that indicates the second Yamanouchi symbol such that the resultant frame is a valid Young frame (which is always the case for two rows). For example for $\lambda_1=2,\lambda_2=1$ we have two possible Yamanouchi numbers: $\mu_1=(112)$ and $\mu_2=(121)$, then the construction of the Young frame is

\begin{eqnarray*}
	\mu_1 &\rightarrow \quad &\varnothing\rightarrow \yng(1) \rightarrow \yng(2)\rightarrow \yng(2,1),\\
	\mu_2 & \rightarrow \quad &\varnothing\rightarrow \yng(1) \rightarrow \yng(1,1)\rightarrow \yng(2,1).
\end{eqnarray*}

 We can see that the Yamanouchi symbol contains implicitly information about $j$. If we think of the elements of the Yamanouchi symbol $\mu$ as $1/2$ spins with $1\rightarrow 1/2$ y $2\rightarrow -1/2$ we can define a relative $j$ to each step $k$ in the construction as \cite{Morales}
 \begin{equation}
 j_k=\sum\limits_{i=1}^{k}\mu_i \quad ; \quad j_n=j.
 \end{equation}
 Similarly we can define a relative $m_k$ to each step in terms of the sequence $\ket{s}$ (after identifying $1\rightarrow1/2$ y $0\rightarrow -1/2$) of the computational base that we would like to transform \cite{Morales}
  \begin{equation}
  m_k=\sum\limits_{i=1}^{k}s_i \quad ; \quad m_n=m.
  \end{equation}
  The Schur transform will look as a product of Clebsch-Gordan coefficients. Each of this coefficients calculated among the elements of the construction. The Clebsch-Gordan coefficient between the $k-1$ and $k$ step in the construction is
\begin{equation*}
CG_k=\braket{j_{k-1},m_{k-1}; 1/2,s_{k}}{j_{k-1}+\mu_{k},m_{k-1}+s_k}.
\end{equation*}
The Yamanouchi symbols tell us how to sum the angular momentums, and the $z$ component of the angular momentum are given by the sequence we are transforming and must be compatible with the total $m$ at the end of the sum.
Thus we obtain the result \cite{Bacon_2006,Morales} 
\begin{equation}
\braket{j,m,\mu}{s}=\prod_{k=2}^{n}CG_k.
\end{equation}
To make this procedure clearer, we will develop a couple of examples for small $n$
\subsection{Example n=2}
In this case there are only two partitions $\lambda=(2,0)$ and $\lambda=(1,1)$ that correspond to $j=1$ and $j=0$ respectively. The Schur transform will only consist of one Clebsch-Gordan coefficient
\begin{equation*}
\braket{j,m,\mu}{s}=CG_2=\braket{j_{1},m_{1}; 1/2,s_{2}}{j_{1}+\mu_{2},m_{1}+s_2}.
\end{equation*}
Let us first focus on the case $j=0$. Necessarily  $m=0$ and $\mu=(12)=\left(1/2,-1/2\right)$. The sequences of length two are $s^1=\{0,0\}=(-1/2,-1/2)$, $s^2=\{0,1\}=(-1/2,1/2)$, $s^3=\{1,0\}=(1/2,-1/2)$ and $s^4=\{1,1\}=(1/2,1/2)$. Thus, after the identification aforementioned we have
\begin{equation*}
\braket{j=0,m=0,\mu=(12)}{s^i}=CG_2=\braket{1/2,s_1^i; 1/2,s_2^i}{0,s_1^i+s_2^i},
\end{equation*}
where $i\in\{1,2,3,4\}$ and the sub-index indicates the element in the sequence . For this Clebsch-Gordan coefficient to be non-zero we must ask $s_1^i+s_2^i=0$. The only sequences that satisfy this condition are $s^2$ y $s^3$. Therefore, we obtain using the formula
\begin{equation*}
\braket{j=1,m=0,\mu=(11)}{s}=\dfrac{(-1)^{2 s_1+2 s_2}}{\sqrt{2}} \sqrt{\frac{\left(1-s_1-s_2\right)! \left(1+s_1+s_2\right)!}{\left(\frac{1}{2}-s_1\right)! \left(s_1+\frac{1}{2}\right)! \left(\frac{1}{2}-s_2\right)! \left(s_2+\frac{1}{2}\right)!}},
\end{equation*}
that
\begin{equation*}
\braket{j=0,m=0,\mu=(12)}{s^2}=-\dfrac{1}{\sqrt{2}},\quad \braket{j=0,m=0,\mu=(12)}{s^3}=\dfrac{1}{\sqrt{2}}.
\end{equation*}
We can then write (in this particular example)
\begin{equation}
\ket{j,m,\mu}=\sum_{i=1}^{4}\braket{j,m,\mu}{s^{i}}\ket{s^{i}}.
\end{equation}
Notice that the state we obtain coincides with the well-known singlet state
\begin{equation*}
\ket{j=0,m=0,\mu=(12)}=\dfrac{1}{\sqrt{2}}(\ket{10}-\ket{01}).
\end{equation*}
Now for the case $j=1$ we have three possible values for $m$, namely $\{-1,0,1\}$. Repeating the procedure we did for $j=0$ we obtain
\begin{equation*}
\braket{j=1,m=-1,\mu=(11)}{s^i}=CG_2=\braket{1/2,s_1^i; 1/2,s_{2}^i}{1,s_{1}^i+s_2^{i}}=\delta_{i,1},
\end{equation*}
since $s_1^i+s_2^i=m=-1$. Following a similar argument
\begin{equation*}
\braket{j=1,m=1,\mu=(11)}{s^i}=CG_2=\braket{1/2,s_1^i; 1/2,s_{2}^i}{1,s_{1}^i+s_2^{i}}=\delta_{i,4}.
\end{equation*}
In the case $m=0$ we have
\begin{equation*}
\braket{j=1,m=0,\mu=(11)}{s^i}=CG_2=\braket{1/2,s_1^i; 1/2,s_{2}^i}{1,s_{1}^i+s_2^{i}},
\end{equation*}
which is non-zero only if $s_1^i+s_2^i=0$. The sequences that satisfy this restriction are $s^2$ y $s^3$. Then
\begin{equation*}
\braket{j=1,m=0,\mu=(11)}{s^2}=\braket{j=1,m=0,\mu=(11)}{s^3}=\dfrac{1}{\sqrt{2}}.
\end{equation*}
These states correspond to the famous triplet in angular momentum theory
\begin{eqnarray}
\ket{j=1,m=-1,\mu=(11)}=\ket{00},\\
\ket{j=1,m=1,\mu=(11)}=\ket{11},\\
\ket{j=1,m=0,\mu=(11)}=\dfrac{\ket{01}+\ket{10}}{\sqrt{2}}.
\end{eqnarray}
\subsection{Example n=3}
We will now work a non-trivial example. For $n=3$ we have two possible partitions of two rows $\lambda^{(1)}=(3,0)\rightarrow j=3/2$ and $\lambda^{(2)}=(2,1)\rightarrow j=1/2$ with Yamanouchi symbols $\mu^{(1)}_1=(111)$ and $\mu^{(2)}_1=(112)$, $\mu^{(2)}_2=(121)$ respectively. Note that for $\lambda^{(2)}$, there are two Yamanouchi symbols, i.e., two paths to reach the Young frame \begin{tiny}
	\yng(2,1)
\end{tiny}. Explicitly, the paths are
\begin{eqnarray*}
	\mu^{(2)}_1 &\rightarrow \quad &\varnothing\rightarrow \yng(1) \rightarrow \yng(2)\rightarrow \yng(2,1),\\
	\mu^{(2)}_2 & \rightarrow \quad &\varnothing\rightarrow \yng(1) \rightarrow \yng(1,1)\rightarrow \yng(2,1).
\end{eqnarray*} 
In this example we only develop one of the paths, the other can be calculated in an analogous way.
We chose the path $\mu^{(2)}_1=(112)$, thus we would like to calculate
 \[\braket{j=1/2,m=1/2,\mu=(112)}{s}.\] 
 To keep the calculations short we will only analyse the sequence $s=\{1,1,0\}$. Now we can calculate
 $j_k\rightarrow j_1=1/2, j_2=1,j_3=1/2$ and $m_k\rightarrow m_1=1/2, m_2=1,m_3=1/2$. Then the CG coefficients are
\begin{equation*}
CG_2=\braket{j_{1},m_{1}; 1/2,s_{2}}{j_{1}+\mu_{2},m_{1}+s_2}=\braket{1/2,1/2; 1/2, 1/2}{1,1}=1,
\end{equation*}
\begin{equation*}
CG_3=\braket{j_{2},m_{2}; 1/2,s_{3}}{j_{2}+\mu_{3},m_{2}+s_3}=\braket{1,1; 1/2,-1/2}{1/2,1/2}=\sqrt{\dfrac{2}{3}}.
\end{equation*}
Hence
\begin{equation*}
\braket{j=1/2,m=1/2,\mu=(112)}{\{1,0,0\}}=\sqrt{\dfrac{2}{3}}.
\end{equation*}

\section{The Louck polynomials}
In the previous section we saw that working with the theory of angular momentum, Clebsch-Gordan coefficients and so on will be relevant to our results. Therefore, we introduce a  powerful combinatoric approach to the theory of angular momentum that can be found in the book of James D. Louck \cite{Louck}. The representation matrices in the angular momentum theory (or the Wigner $D$ functions) can be given in terms of a linear combination of polynomials known as Louck Polynomials. To study the Louck polynomials, we must first introduce some relevant notation regarding sequences and \emph{weight tensors}. We present this concepts by analysing $n$ copies of a state $\ket{\psi}$ and calculating its decomposition in terms of the Schur basis.  
\subsection{$n$-th tensor powers of a state}\label{secLouckstate}
As we see in section \ref{secWeylSchur}, we can expand the $n$th tensor product of $\underbrace{(\mathbb{C}^d\otimes\cdots\otimes\mathbb{C}^{d})}_{l \text{times}}{}^{\otimes n}$ using the Wedderburn decomposition \eqref{eqWedderburn}. In particular we are interested in the cases $l\leq 3$ and $d=2$. In the first part of this section we will work with $l=3$, although the same treatment applies for any integer $l\geq1$. If we take $n$ copies of the same state
\begin{equation}
\ket{\psi}=\sum_{i,j,k=0}^{d-1}\psi_{ijk}\ket{ijk},
\end{equation}
we would be able to decompose it as

\begin{equation}
\ket{\psi}^{\otimes n}=\sum_{\lambda,m,\mu}\psi_{m,\mu}^{\lambda}\ket{\lambda,m,\mu},
\end{equation}
or in the computational basis
\begin{equation}
\ket{\psi}^{\otimes n}=\sum_{\Omega}\sqrt{\binom{n}{\Omega}}[\psi]^{\Omega}\ket{\Omega} ,
\end{equation}
with the normalized vector $\ket{\Omega}$ defined as
\begin{equation}
\ket{\Omega}=\dfrac{1}{\sqrt{\binom{n}{\Omega}}}\sum_{s_1,s_2,s_3:W(s_1\circ s_2\circ s_3)=\Omega}\ket{s_1}\ket{s_2}\ket{s_3}.
\end{equation}
where $\Omega$ is a tensor with non-negative integer numbers as entries and satisfying the condition $\sum_{i,j,k}\Omega_{ijk}=n$. We will refer to this matrix as the \emph{weight matrix} $W$ and indicates the number of times a certain \emph{trisequence} is repeated in a \emph{concatenation} of three sequences $s_1\circ s_2 \circ s_3$. The concatenation can be understood in the following example ($d=2$): if $s_1=(10101),s_2=(00110),s_3=(11101)$ then the elements of $s_1\circ s_2\circ s_3$ are formed by taking the $i$-th element from each sequence, forming sequences of length three, explicitly
\begin{equation}
s_1\circ s_2 \circ s_3=\{101,001,111,010,101\}.
\end{equation}
For this example the tensor $\Omega$ has components $\Omega_{101}=2,\Omega_{001}=1,\Omega_{111}=1,\Omega_{010}=1$ and the rest fo the $\Omega_{ijk}$ are equal to zero. We also introduce the shorthand notation
\[
\Omega!=\prod_{i,j,k}\Omega_{ijk}! \qquad X^{\Omega}=\prod_{i,j,k}X_{ijk}^{\Omega_{ijk}},
\]
and
\[
\binom{n}{\Omega}=\dfrac{n!}{\Omega!}.
\]
The tensor $[\psi]$ is the one with entries $[\psi]_{ijk}=\psi_{ijk}$.
\subsection{Representation matrices $D^{\lambda}$}
The Louck polynomials appear when we want to calculate a matrix representation $D^{\lambda}(X)$ with $X\in GL(d,\mathbb{C})$. Note that since $X$ are $d\times d$ matrices, then $\Omega$ must be a $d\times d$ weight matrix. The explicit expression  for the representation matrix is \cite{Louck}
\begin{equation}
D^{\lambda}(X)=\sum_{\Omega\in T_n }\binom{n}{\Omega}C^{\lambda}(\Omega)X^{\Omega},
\end{equation}
where $T_n$ is the set of tensors with non-negative integer components such that $\sum_{i,j}\Omega_{i,j}=n$. The $C^{\lambda}(\Omega)$ are the matrix-valued coefficients of the Louck polynomial, it is a $r^{\lambda}\times r^{\lambda}$ matrix. In terms of the components we have
\begin{equation}\label{louckdef}
D^{\lambda}(X)_{m,m'}=\sum_{\Omega\in T_n}\binom{n}{\Omega}C^{\lambda}(\Omega)_{m,m'}X^{\Omega},
\end{equation}
and comparing to what it is known from representation theory (see \cite{Louck})
\begin{multline}
C^{\lambda}_{w,w'}(\Omega)=\dfrac{\sqrt{(\lambda_1-w)!(w-\lambda_2)!(\lambda_1-w')!(w'-\lambda_2)!}}{n!}\\
\times\sum_k \binom{\lambda_2}{k}\dfrac{\Omega!(-1)^k}{(\Omega_{12}-k)!(\Omega_{21}-k)!(\Omega_{11}-\lambda_2+k)!(\Omega_{22}-\lambda_2+k)!},
\end{multline}
where $w$ is the weight of a SSYT or in terms of the angular momentum variables
\begin{multline}
C^{j}_{m,m'}(\Omega)=\dfrac{\sqrt{(j+m)!(j-m)!(j+m')!(j-m')!}}{(2j)!}\\
\times\sum_k \binom{\frac{n}{2}-j}{k}\dfrac{\Omega!(-1)^k}{(\Omega_{12}-k)!(\Omega_{21}-k)!\left(\Omega_{11}-\frac{n}{2}+j+k\right)!\left(\Omega_{22}-\frac{n}{2}+j+k\right)!}.
\end{multline}
In both cases the sum over $k$ goes over all the integer values such that all factorials are nonzero.
To see some of the properties of the Louck polynomials let us consider the matrix $X^{\otimes n}$ in the Schur basis $\ket{j,m,\mu}$ 
\begin{equation}\label{eqDLouck}
\bra{\lambda,m,\mu}X^{\otimes n}\ket{\lambda,m',\mu'}=D^{\lambda}(X)_{m,m'}\delta_{\mu,\mu'},
\end{equation}
and also
\begin{equation}\label{eqXLouck}
\bra{\lambda,m,\mu}X^{\otimes n}\ket{\lambda,m',\mu'}=\sum_{s\sim m}\sum_{s'\sim m'}\braket{\lambda,m,\mu}{s}\braket{s'}{\lambda,m',\mu'}\bra{s}X^{\otimes n}\ket{s'},
\end{equation}
where the sum is taken over the sequences which are compatible to $m$, that is the ones that make the Schur transform to be non-zero. We denote this sequences by $s\sim m$.
Comparing equations \eqref{eqDLouck} and \eqref{eqXLouck} we see that $D^{\lambda}(X)$ is a linear combination of the entries of $X$. The term $\bra{s}X^{\otimes n}\ket{s'}$ can be expressed as $X^{\Omega}$ taking into account the  constrains $s\sim m$ and $s'\sim m'$ impose over $\Omega$. This conditions are over the partial sums of the tensor, namely
\[
\sum_{j=0}^{d-1}\Omega_{ij}=w'_i ,\quad \sum_{i=0}^{d-1}\Omega_{ij}=w_j,
\]
where $w_x$ is the number of elements (\emph{weight}) $x\in\{0,1,\ldots,d-1\}$ in the sequences $s'$ and $s$ respectively.
Thus we obtain that $\Omega$ is the weight matrix for the \emph{bisequence} $s\circ s'$, then
\[
D^{\lambda}(X)_{m,m'}\delta_{\mu,\mu'}=\sum_{s\sim m}\sum_{s'\sim m'}\braket{\lambda, m ,\mu}{s}\braket{s'}{\lambda,m',\mu'}X^{\Omega}.
\]
Comparing this expression to equation \eqref{louckdef} we obtain that
\begin{equation}\label{eqLouckyeah}
\sum_{s,s':W(s\circ s')=\Omega}\braket{\lambda, m ,\mu}{s}\braket{s'}{\lambda,m',\mu'}=\binom{n}{\Omega}C^{\lambda}_{m,m'}(\Omega)\delta_{\mu,\mu'}.
\end{equation}
Now consider two bisequences $s_1\circ s'_1$ and $s_2\circ s'_2$ with the same weight matrix $\Omega$. Then it should be possible to transform one into another by means of a permutation $\pi\in S_n$, $s_1\circ s'_1=\pi s_2 \circ \pi s'_2$. The transformation in the vector $\ket{s}$ reads $D(\pi)\ket{s}$, then for the internal product we have
\begin{equation}
\braket{\lambda, m ,\mu}{s_1}\braket{s'_1}{\lambda,m',\mu}=\bra{\lambda, m ,\mu}D(\pi)\ket{s_2}\bra{s'_2}D^{\dagger}(\pi)\ket{\lambda,m',\mu}.
\end{equation}
Because $D^{\dagger}(\pi)$ acts irreducibly over the Schur basis $\ket{\lambda,m,\mu}$ we have that
\begin{equation}
D^{\dagger}(\pi)\ket{\lambda,m,\mu}=\sum_{\mu'}S^{\lambda}_{\mu'\mu}(\pi^{-1})\ket{\lambda,m,\mu'}.
\end{equation}
Then we have
\begin{equation}
\braket{\lambda, m ,\mu}{s_1}\braket{s'_1}{\lambda,m',\mu}=\sum_{\mu',\mu''}S^{\lambda}_{\mu'\mu}(\pi^{-1})S^{\lambda}_{\mu\mu''}(\pi)\braket{\lambda, m ,\mu''}{s_2}\braket{s'_2}{\lambda,m',\mu'}.
\end{equation}
If we take the sum over $\mu$ we can use the orthogonality of the representations of $S_n$
\begin{equation}
\sum_{\mu}S^{\lambda}_{\mu'\mu}(\pi^{-1})S^{\lambda}_{\mu\mu''}(\pi)=\delta_{\mu'\mu''},
\end{equation}
to obtain
\begin{equation}
\sum_{\mu}\braket{\lambda, m ,\mu}{s_1}\braket{s'_1}{\lambda,m',\mu'}=\sum_{\mu}\braket{\lambda, m ,\mu}{s_2}\braket{s'_2}{\lambda,m',\mu}.
\end{equation}
We can assert that the quantity $\sum_{\mu}\braket{\lambda, m ,\mu}{s_1}\braket{s'_1}{\lambda,m',\mu'}$ is constant among all the sequences with the same matrix $\Omega$. Then we can go back to equation \eqref{eqLouckyeah} and sum over $\mu$ to obtain
\begin{equation}\label{eqLouckident1}
\sum_{\mu}\braket{\lambda, m ,\mu}{s}\braket{s'}{\lambda,m',\mu}=C^{\lambda}_{m,m'}(\Omega)f^{\lambda},
\end{equation}
where $\Omega=W(s\circ s')$. This is an important result we will use later on when we calculate internal products of vectors in the computational basis and Schur basis.
\section{The Keyl-Werner theorem}
The Keyl-Werner theorem shows us an asymptotic correspondence between the eigenvalues of the density matrix and the partitions labelling an irreducible representation in the Wedderburn decomposition.
\begin{theorem}\cite{Keyl}
	Let $\rho$ be a density matrix with spectrum $\vec{r}^\downarrow$ organized in decreasing order. For a partition $\lambda\vdash n$ such that $\frac{\lambda}{n}$ tends to $\vec{s}^\downarrow$ as $n\to\infty$ we have that
	\begin{equation*}
	\lim\limits_{n\rightarrow\infty}\Tr\left[P^{\lambda}\rho^{\otimes n}\right]=\exp\left[-n D(\vec{s}^\downarrow||\vec{r}^\downarrow)\right],
	\end{equation*}
	where $P^{\lambda}$ is the projector in the carrier space of the representation labelled by $\lambda$, explicitly $V^{d}_\lambda\otimes\left[\lambda\right]$. The function $D$ is the relative entropy defined as $D(\vec{s}^\downarrow||\vec{r}^\downarrow)=\sum_{i=1}^{d}s_i\log\frac{s_i}{r_i}$.

\end{theorem}
Function $D$ is also known as the Kullback–Leibler divergence in Information Theory. Note that the probability to be in a space $V^{d}_\lambda\otimes\left[\lambda\right]$ decreases exponentially with $n$. Thus, for large $n$ the probability will be concentrated around the partition $\lambda$ that minimizes $D$, i.e. the partition $\lambda$ such that $\lambda_i/n\approx r_i$. As a result, this theorem gives us a way to measure the spectrum of the density matrix if we posses a large amount of copies of a state. It will also prove useful to characterize the entanglement polytope as we will se in section \ref{secKronecker}.

\section{Covariant and Invariants of three qubits}\label{secCovariant}
The invariant theory deals with the action of a group on algebraic varieties. The theory is mainly concerned on the calculation of a set of fundamental polynomial quantities that generate an invariant ring under the action of the group. We will also be interested in the relations among these fundamental quantities and in writing  an arbitrary invariant quantity in terms of the fundamental set \cite{Bernd_2008}. To apply the theory to vector spaces we write
a state in the Hilbert space $\ket{\psi}\in\mathbb{C}^{2}\otimes\mathbb{C}^2\otimes\mathbb{C}^2$ as
\begin{equation}
\ket{\psi}=\sum_{i,j,k=0}^{1}\psi_{ijk}\ket{ijk},
\end{equation}
which can be interpreted as a multilinear form in $\mathbb{C}^2\times\mathbb{C}^2\times\mathbb{C}^2$
\begin{equation}
A_{111}=\sum_{i,j,k=0}^{1}\psi_{ijk}x_{i}y_{j}z_{k}.
\end{equation}
Now we may ask how does the multilinear form transforms under the action of a group, in particular the one associated with the SLOCC.
As we mentioned in section \ref{secSLOCC}, the orbits will be the equivalence classes 
\begin{equation}
\dfrac{(\mathbb{C}^{2})^{\otimes n}}{[SL(2,\mathbb{C})]^{\times n}},
\end{equation}
so the calculation of the invariants will be closely related to the SLOCC entanglement classes. The advantage this covariant formulation gives is that the calculations will be made with explicitly SLOCC-invariant quantities, whereas there are some entanglement measures (like the entropy) that are not manifestly SLOCC-invariant.

We can say two states are equivalent if they belong to the same orbit under the SLOCC associated group. To tell whether two states are equivalent we can use invariant theory to describe polynomials in the variables $\psi_{ijk}$ and the auxiliary variables $x_{i},y_{j},z_{k}$ that are invariant under the action of $[SL(2,\mathbb{C})]^{\times n}$ (with $n=3$ in our case) \cite{Luque}. The set of polynomials which only depend on the variables $\psi_{ijk}$ are often referred as \emph{invariants} while the ones that depend on $\psi_{ijk}$ and the auxiliary variables are the so-called \emph{covariants}. The latter are used because only with the invariants it is not possible to make a complete classification of entanglement classes.

 The idea is to classify the multilinear forms by comparing their evaluations on this polynomials. This classification problem is as old as one of the Hilbert problems \cite{Hilbert,Zariski} and is computationally out of reach in the general case. However for the case we are considering, the algebra of invariants and covariants is well known. To obtain the number of covariants and invariants and their degrees the multivariate Hilbert series defined as \cite{Bernd_2008}
\begin{equation}
H_{\text{Cov}}(t;\vec{u}):=\sum_{\vec{d}} \dim\text{Cov}(\vec{d}\,)t^{d_0}u_1^{d_1}\ldots u_k^{d_k},
\end{equation}
where the vector $\vec{d}$ is the multidegree of the polynomial and $\dim\text{Cov}(d)$ is the dimension of the algebra of covariants of multidegree equal to $\vec{d}$. In the case of three qubits it takes the form \cite{Holweck}
\begin{equation}
H_{\text{Cov}}(t;\vec{u})=\dfrac{1-t^6 u_1^2 u_2^2 u_3^2}{(1-tu_1u_2u_3)(1-t^2u_1^2)(1-t^2u_2^2)(1-t^2u_3^2)(1-t^3u_1u_2u_3)(1-t^4)},
\end{equation}
or equivalently
\begin{equation}
H_{\text{Cov}}(t;\vec{u})=\dfrac{1+t^3 u_1 u_2 u_3}{(1-tu_1u_2u_3)(1-t^2u_1^2)(1-t^2u_2^2)(1-t^2u_3^2)(1-t^4)},
\end{equation}
which enable us to identify a set of fundamental covariants.
By examining the terms in the denominator we conclude we will have 6 covariant polynomials. The term in the numerator indicates a relationship between the covariants, known in the literature as \emph{syzygy}. The degrees at which the variables $t,u_1,u_2,u_3$ appear in the denominator indicates the degrees of each covariant in the variables $\psi_{ijk},x,y,z$. Thus the degrees of the covariants will be: 
\begin{itemize}
	\item One covariant of degree one in $\psi_{ijk},x,y,z$.
	\item Three covariants of degree two in $\psi_{ijk}$, degree two in one fo the variables $x,y,z$ and degree zero in the other two.
	\item One covariant of degree three in $\psi_{ijk}$ and degree one in $x,y,z$.
	\item One invariant of degree four in  $\psi_{ijk}$.
\end{itemize}
 The notation we will use with covariants will the be same as in  \cite{Zimmerman}: $X^{m}_{pqr}$ where $X$ indicates the degree of the covariant in  $\psi_{ijk}$ (e.g. $A=1,B=2$) and ${p,q,r}$ indicate the degree in the auxiliary variables ${x,y,z}$ respectively. The factor $m$ is used to differentiate covariants that under the previous notation are written the same but are explicitly different. Using this notation we have that the set of covariants for three qubit systems is $\{A_{111},B_{200},B_{020},B_{002},C_{111},D_{000}\}$. According to the invariant theory for finite matrix groups $\Gamma\subset GL(\mathbb{C}^n)$(like the SLOCC), the ring of invariants $\mathbb{C}[\vec{\psi},\vec{x},\vec{y},\vec{z}]^{\Gamma}$ which is the set of all the invariant polynomials under the group $\Gamma$ on the variables $\psi_{ijk}, x,y,z$, has a very nice decomposition. This property is known as \emph{Cohen-Macaulay}, and it enables us to classify the fundamental invariants/covariants into \emph{primary} and \emph{secondary} invariants/covariants \cite{Bernd_2008}. In our analysis we found that there is only a secondary covariant, the one we called $C_{111}$. As a consequence of the \emph{Cohen-Macaulayness} we can write the ring as a free module over the ring generated by the primary covariants, explicitly
 \begin{equation}
 \mathbb{C}[\vec{\psi},\vec{x},\vec{y},\vec{z}]^{G_{SLOCC}}=\mathbb{C}[A_{111},B_{200},B_{020},B_{002},D_{000}]+C_{111}\mathbb{C}[A_{111},B_{200},B_{020},B_{002},D_{000}].
 \end{equation}
 Therefore, any covariant quantity can be written as
 \begin{equation}
 I(\vec{\psi},\vec{x},\vec{y},\vec{z})=p_0(A_{111},B_{200},B_{020},B_{002},D_{000})+C_{111}p_1(A_{111},B_{200},B_{020},B_{002},D_{000}),
 \end{equation}
 where $p_0$ and $p_1$ are proper polynomials. In later chapters we will see how to choose such polynomials and how can we simplify further the decomposition.

  As we will see below, we will also be interested in relating these covariants to different entanglement measures, thus we analyse the cases whether a covariant is equal to zero. This approach will allows us to classify the entanglement classes by the nullity of a subset of the covariants as it is done in \cite{Zimmerman}. In the next lines we will write the explicit form of the covariant for a general three qubit state following \cite{Borsten}.
 \\
 \\ 
 The first covariant is the linear form
\begin{equation}
A_{111}=\sum\limits_{i,j,k=0}^{1}\psi_{ijk}x_i y_j z_k,
\end{equation}
which as we saw above is in one-to-one correspondence with the state $\ket{\psi}$.  This covariant is only zero for the null vector, for any othe state we have  $A_{111}\neq0$. 
\\
\\
The covariants of degree two in $\psi_{ijk}$ can be written as
\begin{eqnarray}
B_{200}=\varepsilon^{j_1j_2}\varepsilon^{k_1k_2}\psi_{i_1j_1k_1}\psi_{i_2j_2k_2}x^{i_1}x^{i_2}=(\gamma^{A})_{i_1i_2}x^{i_1}x^{i_2},\\
B_{020}=\varepsilon^{i_1i_2}\varepsilon^{k_1k_2}\psi_{i_1j_1k_1}\psi_{i_2j_2k_2}y^{j_1}y^{j_2}=(\gamma^{B})_{j_1j_2}y^{j_1}y^{j_2},\\
B_{002}=\varepsilon^{i_1i_2}\varepsilon^{j_1j_2}\psi_{i_1j_1k_1}\psi_{i_2j_2k_2}z^{k_1}z^{k_2}=(\gamma^{C})_{k_1k_2}z^{k_1}z^{k_2},
\end{eqnarray}
where $\varepsilon$ in the Levi-Civita tensor and we have used the Einstein summation convention. The matrices $(\gamma^{A},\gamma^{B},\gamma^{C})$ are defined by the above equations. Remarkably the local entropies can be written in terms of the $\gamma$ matrices as \cite{Borsten}
\begin{equation*}
S_A=4\Tr\left[\gamma^{B\dagger}\gamma^{B}+\gamma^{C\dagger}\gamma^{C}\right],
\end{equation*}
and cyclic permutations of $(A,B,C)$. 
\\
\\
The next covariant is $C_{111}$ is cubic in the state variables $\psi_{ijk}$ and linear in the auxiliary variables. Explicitly
\begin{eqnarray}
C_{111}&=&\varepsilon^{i_1i_2}\psi_{i_1j_1k_1}\left(\gamma^{A}\right)_{i_2i_3}x^{i_3}y^{j_1}z^{k_1}=T_{i_3j_1k_1}x^{i_3}y^{j_1}z^{k_1},\nonumber\\
&=&\varepsilon^{j_1j_2}\psi_{i_1j_1k_1}\left(\gamma^{B}\right)_{j_2j_3}x^{i_1}y^{j_3}z^{k_1}=T_{i_1j_3k_1}x^{i_1}y^{j_3}z^{k_1},\\
&=&\varepsilon^{k_1k_2}\psi_{i_1j_1k_1}\left(\gamma^{C}\right)_{k_2k_3}x^{i_1}y^{j_1}z^{k_3}=T_{i_1j_1k_3}x^{i_1}y^{j_1}z^{k_3}\nonumber,
\end{eqnarray}
where $T$ is defined by the above equations and can thought as a vector (compare to $A_{111}$). It is related to the Kempe invariant \cite{Borsten} defined as
\begin{equation*}
K=\Tr(\rho_A\otimes\rho_B\rho_{AB})-\Tr(\rho_A^3)-\Tr(\rho_B^3),
\end{equation*}
via the relation
\begin{equation*}
\braket{T}{T}=\dfrac{2}{3}(K-|\braket{\psi}{\psi}|^3)+\dfrac{\braket{\psi}{\psi}}{16}(S_A+S_B+S_C).
\end{equation*} 
The last covariant $D_{000}$ which is actually an invariant
\begin{eqnarray}
D_{000} &=& 2\det\gamma^{A},\nonumber\\
& = & 2\det\gamma^{B},\\
& = & 2\det\gamma^{C},\nonumber.\\
\end{eqnarray}
 The relation with the 3-tangle is
\begin{equation*}
\tau_3(\rho)=4 |D_{000}|.
\end{equation*}
We can make a classification of the entanglement classes based on which covariants are equal to zero as it is seen in Table  \ref{tabla1}.
The hierarchy of the covariants  in ascendant order is $\{A_{111},B,C_{111},D_{000}\}$, where $B=\{B_{200},B_{020},B_{002}\}$ where this three covariants have the same hierarchy. This organization into hierarchies  tell us that if a covariant is zero, all the above in the hierarchy are zero too. The converse is true as well, if a covariant is not equal to zero, then the covariants below in the hierarchy are different from zero too. Thus we can classify the entanglement classes based on the nullity (or non-nullity) of a set of covariants as it is seen in Table \ref{tabla1}. 
\begin{table}[h]
	\centering
	\begin{tabular}{|c|c|c|}
		Class & Covariants=0  & Covariants$\neq0$ \\
		$\varnothing$& $A_{111}$ & -\\
		A-B-C & $B$& $A_{111}$\\
		AB-C & $C_{111},B_{200},B_{020}$&$B_{002} $\\
		A-BC & $C_{111},B_{020},B_{020}$& $B_{200}$ \\
		AC-B  & $C_{111},B_{200},B_{002}$& $B_{020}$\\
		W & $D_{000}$ & $C_{111}$\\
		GHZ & - & $D_{000}$
	\end{tabular}
	\caption{Hierarchic classification of the entanglement classes for three qubits using covariants}
	\label{tabla1}
\end{table}

Finally, the syzygy, i.e. the relation between the covariants takes the form
\[
C_{111}^2+\dfrac{1}{2}B_{200}B_{020}B_{002}+D_{000}A_{111}^2=0.
\]
In section \ref{secKronecker} we will see how to relate this covariants with Young frames an the calculation of the Kronecker coefficient. They will also prove useful when constructing a state in the Wedderburn decomposition space $V^{d}_\lambda$ as we will see in section \ref{secSLOOCcovariant}. 

\section{Constructing LU Invariants and states from Covariants}\label{secCovstaInv}
Let $X_{a,b,c}$ be a covariant for the 3 qubit SLOCC associated group. Then a tensor(vector) can be assigned with correspondence one-to-one via the next procedure. We write the expression for the covariant
\begin{equation}\label{eqXabc}
X_{abc}=\sum_{a_0=0}^{a}\sum_{b_0=0}^{b}\sum_{c_0=0}^{c}X_{a_0b_0c_0}x_0^{a_0}x_1^{a-a_0}y_0^{b_0}y_1^{b-b_0}z_0^{c_0}z_1^{c-c_0},
\end{equation}
and we want to translate this information into a vector in certain representation $(\alpha,\beta,\gamma)$. We are interested in the part of the representation which is not a determinant (boxes in the upper and lower rows $\begin{tiny}
\yng(2,2)
\end{tiny}$). Then we make the degree of the covariant in each variable equal to the number of free boxes in the upper row, i.e. $a=\alpha_1-\alpha_2=n-2\alpha_2$, $b=\beta_1-\beta_2=n-2\beta_2$ and $c=\gamma_1-\gamma_2=n-2\gamma_2$.\\
\\
 The number $a_0$ will indicate the number of zeros in a sequence that such state has in the first component and $b_0,c_0$ will do the same for the second and third components. Knowing the number of zeroes we follow a prescription to build a un-normalized state derived from \eqref{eqXabc}, explicitly
 \begin{equation}
 \ket{X_{abc}}=\sum_{a_0=0}^{a}\sum_{b_0=0}^{b}\sum_{c_0=0}^{c}X_{a_0b_0c_0}\dfrac{\ket{a_0}\ket{b_0}\ket{c_0}}{\sqrt{\binom{n-2\alpha_2}{a_0}\binom{n-2\beta_2}{b_0}\binom{n-2\gamma_2}{c_0}}},
 \end{equation}
 where the states of the form $\ket{a_0}$ are
 \begin{equation}
 \ket{a_0}=\ket{j=n/2-\alpha_2,m=j-a_0}\propto\sum_{s\sim n/2-\alpha_2-a_0}\ket{s},
 \end{equation}
 where the sum is over all the sequences with $a_0+\alpha_2$ zeroes and $n-\alpha_2-a_0$ ones; which in terms of the angular momentum basis is equivalent to $m=n/2-\alpha_2-a_0$.
 
  Now that we know how to assign a state to a covariant, we would like to calculate invariants under local unitaries from the SLOCC covariants. We will follow the procedure proposed in \cite{Toumazet_2006} where Local Unitary (LUT) invariants are obtained from \emph{internal products} of SLOCC covariants. We will denote by $\Phi^{a}_{d}$ an arbitrary product of SLOCC covariants of degree $d$ in the state variables and multidegree $m=(a,b,c)$ in the auxiliary variables. Then the scalar products $\braket{\Phi_{d}^{m}}{\Phi_{d}^{m}}$ with respect to the auxiliary variables forms a basis for the space of LUT invariants when we choose certain combinations of the SLOCC covariants (see\cite{Toumazet_2006}). Furthermore, if in the scalar product we consider degrees $d$ and $d'$ for the state variables, then the scalar product will form a basis for the Local Special Unitary LSUT invariants. The inner product for the auxiliary variables is defined as
\begin{equation}
({x_1\cdots x_n},{y_1\cdots y_n})=\sum_{\pi\in S_n}\prod_{i=1}^{n}\delta_{x_i,y_{\pi(i)}}.
\end{equation} 
For the case of qubits, this product is easy to calculate, for it to be non-zero the number of terms must be equal on both sides, resulting in
\begin{equation}
({\underbrace{x_0\cdots x_0}_{a_0 \text{times}}\underbrace{x_1\cdots x_1}_{a_1 \text{times}}},{\underbrace{x_0\cdots x_0}_{a'_0 \text{times}}\underbrace{x_1\cdots x_1}_{a'_1 \text{times}}})= a! \dfrac{a_0!a_1!}{a!}\delta_{a,a'}.
\end{equation}
Or, in terms of the associated state 
\begin{equation}
\braket{\Phi^{m}_{d}}{\Phi^{m}_{d}}=\sum_{a_0=0}^{a}\sum_{b_0=0}^{b}\sum_{c_0=0}^{c} |\Phi^{a_0,b_0,c_0}_{d}|^2 \dfrac{a_0!a_1!b_0!b_1!c_0!c_1!}{a!b!c!}.
\end{equation}
For our three qubit example we can construct the following LU invariants \cite{Toumazet_2006}
\begin{eqnarray}
\braket{A_{111}}{A_{111}},\\
\braket{B_{200}}{B_{200}},\\
\braket{B_{020}}{B_{020}},\\
\braket{B_{002}}{B_{002}},\\
\braket{C_{111}}{C_{111}},\\
\braket{D_{000}}{D_{000}},\\
\braket{A_{111}^2D_{000}}{C_{111}},
\end{eqnarray}
which will appear surprisingly when we are calculating the asymptotic rates for the probability to be in a triplet $(\alpha,\beta,\gamma)$ in the Wedderburn representation. 

\section{Quantum marginal problem and the entanglement polytope}\label{secpolytope}
In this section we will state the quantum marginal problem and its relation to this work. The quantum marginal problem, analogous to its classical counterpart, deals with local density matrices(marginal probability distributions) compatible with a global pure state density matrix (joint probability distribution). It can be stated as \cite{walter}:
\begin{problem}[Quantum Marginal Problem]
		Let $\mathcal{H}=\bigotimes_{k=1}^{n}\mathcal{H}_k$ be the Hilbert space describing a physical system. Given $\rho_k$ to be the density matrices on each $\mathcal{H}_k$ for $k=1,\ldots,n$. Does there exist a global pure state $\ket{\psi}$ such that the partial density matrices in each subspace $\mathcal{H}_k$ are $\rho_k$?
	\end{problem}
A general solution to this problem in the case of indistinguishable particles was found by Klyachko \cite{Klyachko_2004,Klyachko_2006} where a set of inequalities for the eigenvalues of the reduced density matrices are obtained in order to ensure the local compatibility. Such inequalities define a the facets of a convex region which we will call the \emph{compatibility polytope}.
\\
\\
A more recent approach to solve the quantum marginal problem involves the moment map of a symplectic manifold as  in \cite{walter,walter2}. For completeness, we will only mention the basic features of this approach, but an interested reader should consult references \cite{walter,walter2}.
In this approach the authors consider $(M,\omega)$ a symplectic manifold which in this case will be the projective space $\mathbb{P}(\mathcal{H})$ of the corresponding Hilbert space $\mathcal{H}$. The action of the group $SL(d)\times\cdots\times SL(d)$ will generate orbits or equivalence classes as we have mentioned before. The orbits will be taken as symplectic manifolds as well. The \emph{moment polytope} will be the image of the moment map $\mu: M \rightarrow \mathfrak{k}^*$ where $\mathfrak{k}^*$ is the dual Lie algebra of the maximal compact group of $SL(d)\times\cdots\times SL(d)$, which is $U(d)\times\cdots\times U(d)$. The image of this moment map (its intersection with the positive Weyl chamber to be more precise ) will be a convex region, hence the name entanglement polytope (for more details see \cite{walter}). In this approach the entanglement classes correspond to convex subsets of the moment polytope which are called \emph{entanglement polytopes}. 
\\
\\
The compatibility polytope and the the moment polytope are basically the same as shown in \cite{walter} and are defined in terms of inequalities of the reduced density eigenvalues. Specialising to $n$-qubit systems we have that in \cite{Higuchi_2003} the inequalities that are necessary and sufficient to be compatible with a global pure state are
\begin{equation}
\sum_{k\neq l}\lambda^1_{k}\leq(n-2)+\lambda^1_{l}\quad ;\quad\forall l=1,\ldots,n,
\end{equation}
where $\lambda^1_k$ is the largest eigenvalue of the reduced density matrix $\rho_k$. 
\begin{figure}[h!]
	\centering
	\includegraphics[scale=0.5]{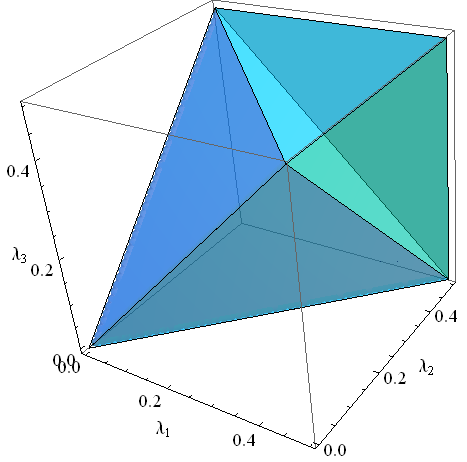}
	\caption{Compatibility polytope for three qubits}
	\label{figcompoly}
\end{figure}

In the case of three qubits $n=3$ the inequalities define a convex region as it is shown in Fig. \ref{figcompoly} where the axes represent the lowest eigenvalue of the three reduced density matrices. The entanglement polytopes for the six entanglement classes must be convex subsets of this compatibility polytope. For separable states the set is just the point at the origin $(0,0,0)$. For the bi-separable entanglement classes we will have the edge of the polytope that goes from the origin to the points $(1/2,1/2,0)$, $(1/2,0,1/2)$ and $(0,1/2,1/2)$. The W entanglement polytope corresponds to the lower region of the compatibility polytope, the one comprehended between the origin and the plane $\lambda_1+\lambda_2+\lambda_3=1$. The GHZ entanglement polytope is the whole compatibility polytope \cite{walter}.
\section{Asymptotic rates}\label{secAsymptotictheory}
Our main objective throughout this thesis is to evaluate the asymptotic rates for the probability of being in a certain region of the compatibility/entanglement polytope when we take $n$ copies of a certain state $\psi$. We will denote such probabilities as $p(\alpha,\beta,\gamma|\psi)$ and the asymptotic rates will be generally defined as
\begin{equation}
\phi(\bar{\alpha},\bar{\beta},\bar{\gamma}|\psi)=-\lim\limits_{n\to\infty}\dfrac{\log p(\alpha,\beta,\gamma|\psi)}{n},
\end{equation}
where $\bar{\lambda}=\lambda/n$ and the limit exists and is finite. In the next chapter, we will show how to calculate the rate for different regions in the polytope and different $\psi$ belonging to distinct entanglement classes.
\\
\\
As we observed in Chapter \ref{Chapter1}, our main motivation is to see how the distillation occurs in entangled many-party systems. Thus the asymptotic rates give us detailed information on such distillation for the asymptotic limit where we have $n$ copies of the same state.

%% file: Chapter3test3.tex
\chapter{Kronecker states and asymptotic rates} 

\label{Chapter3} 
\lhead{Chapter 3. \emph{Kronecker states and asymptotic rates}} 

In this chapter we will present  results concerning the entanglement among three qubits. We will first present the case of two qubits to motivate the discussion for three qubits. We will concentrate on what we will call \emph{Kronecker states}, where entanglement of the state is concentrated. In section \ref{secKronecker} we will analyse the different entanglement classes to see how the different decompositions and asymptotic rates of entanglement concentration are. We will also use the Keyl-Werner theorem to see the correspondence between the polytope for Young frames and the entanglement polytope. In section \ref{secSLOOCcovariant} we will use an approach to calculate the state in the Wedderburn decomposition in terms of covariants. In section \ref{secasymplouck} we will use the Schur-Weyl duality to express $n$ copies of a single state in terms of the Schur basis with the help of the Schur transform and the Louck polynomials. In the following sections \ref{secTwoqubit} and \ref{secThreequbits} we will elaborate the problem we wish to attack throughout the chapter.
\section{The state $\ket{\psi}^{\otimes n}$ for two qubits}\label{secTwoqubit}
Let us first consider a state $\ket{\psi}_{AB}$ in a bipartite qubit system. We will call the parties Alice and Bob. The Hilbert space for the $n$-copy state $\ket{\psi}_{AB}^{\otimes n}$ shared between Alice and Bob will be $\mathcal{H}\simeq(\mathbb{C}^2\otimes \mathbb{C}^2)^{\otimes n}$. Using the Schur-Weyl duality \cite{Weyl_1997} we can write the one-qubit Hilbert space as

\begin{equation}
(\mathbb{C}^2)^{\otimes n}\cong\bigoplus_{\lambda\underset{2}{\vdash} n}V_\lambda^2\otimes[\lambda],
\end{equation}
where $[\lambda]$ and $V_\lambda^2$ are irreducible representations of the symmetric group $S_n$ and the general linear group $GL(2,\mathbb{C})$ respectively (see section \ref{secWeylSchur}). The term $\lambda$ labels integer partitions of $n$ with length two, or equivalently, Young frames with two rows and $n$ boxes. For the qubit bipartite case we obtain
\begin{equation*}
(\mathbb{C}^2\otimes\mathbb{C}^2)^{\otimes n}=(\mathbb{C}^2)^{\otimes n}\otimes(\mathbb{C}^2)^{\otimes n}\cong\bigoplus_{\alpha\underset{2}{\vdash} n, \beta\underset{2}{\vdash} n}V_\alpha^2 \otimes U_{\beta}^2\otimes[\alpha]\otimes[\beta].
\end{equation*}
Since we are taking $n$ copies of the same state, the tensor product must be invariant under permutations , i.e. for any $\pi\in S_n$, $\pi\ket{\psi}_{AB}^{\otimes n}=\ket{\psi}_{AB}^{\otimes n}$. Thus we are interested in the symmetric subspace
\begin{equation*}
\text{Sym}^n(\mathbb{C}^2\otimes\mathbb{C}^2)^{\otimes n}\cong \bigoplus_{\alpha\underset{2}{\vdash} n, \beta\underset{2}{\vdash} n}V_\alpha^2 \otimes U_{\beta}^2\otimes[[\alpha]\otimes[\beta]]^{\text{Sym}_n}.
\end{equation*}
We now introduce the Kronecker coefficients $g_{\alpha,\beta,\gamma}$ as the multiplicity of the irreducible representation $[\gamma]$ in the tensor product $[\alpha]\otimes[\beta]$
\begin{equation*}
[\alpha]\otimes[\beta]=\bigoplus_{\gamma\vdash n}g_{\alpha,\beta,\gamma}[\gamma].
\end{equation*}
From the symmetric condition we must ensure that [$\gamma$] is the trivial irreducible representation, which we will denote as $[(n)]$. It can be shown that $g_{\alpha,\beta,(n)}=\delta_{\alpha,\beta}$ (the Kronecker delta), then the symmetric subspace is simplified to
\begin{equation*}
\text{Sym}^n(\mathbb{C}^2\otimes\mathbb{C}^2)^{\otimes n}\cong\bigoplus_{\alpha\underset{2}{\vdash} n} V_\alpha^2\otimes U_\alpha^2\otimes [[\alpha]\otimes[\alpha]]^{\text{Sym}_n}.
\end{equation*}
Therefore, we can write $\ket{\psi}_{AB}^{\otimes n}$ in the generic form
\begin{equation}
\ket{\psi}_{AB}^{\otimes n}=\sum_{\lambda\underset{2}{\vdash} n}c_\lambda\ket{\Phi(\psi)}_{\lambda}\otimes\ket{\mathcal{K}}_\lambda,
\end{equation}
with $c_\lambda\in \mathbb{C}$ and $\sum_{\lambda}|c_\lambda|^2=1$. We have introduced the normalized vectors $\ket{\Phi(\psi)}_{\lambda}\in V_\lambda^2\otimes U_\lambda^2$ and $\ket{\mathcal{K}}_\lambda\in[[\lambda]\otimes[\lambda]]^{\text{Sym}_n}$. The former vector carries the information about the original state $\ket{\psi}_{AB}$ while the latter, which we will call the Kronecker vector, carries the information about the entanglement (and permutation symmetry) and only depends on the entanglement class of the original state. If we make a measurement of the Young frame ($\lambda$) and trace out the part from the $GL$ representations it can be shown that the resulting vector $\ket{\mathcal{K}}_\lambda$ is a maximally entangled state in the subspace $[\lambda]$of dimension $f^\lambda$. Because the dimension of the symmetric group representation grows exponentially while the one from the general linear group $V_{\lambda}$ grows polynomially, in the asymptotic regime we will be concerned with the symmetric part of the state which will contain most of the entanglement information from the original state. Notice that an important feature is that the effective Kronecker coefficient is always in this bipartite case equal to one. In the next lines we will show what is stated above for a particular two qubit state
\\
\\
Consider the state $\ket{\psi_{AB}}=\alpha\ket{00}+\beta\ket{11}$, with $|\alpha|^2+|\beta|^2=1$. Then we can express the $n$th tensor power as
\begin{equation}\label{louck}
\ket{\psi_{AB}}^{\otimes n}=\sum_{\Omega}\sqrt{\binom{n}{\Omega}}[\psi]^{\Omega}\ket{\Omega},
\end{equation}
where $\Omega$ is defined as we saw in section \ref{secLouckstate}. The normalized vector $\ket{\Omega}$ is defined as
\begin{equation}
\ket{\Omega}:=\dfrac{1}{\sqrt{\binom{n}{\Omega}}}\sum_{s,s':W(s\circ s')=\Omega}\ket{s}\ket{s'}.
\end{equation}
The binomial term is defined as
\begin{equation}
\binom{n}{\Omega}=\dfrac{n!}{\Omega!}.
\end{equation}
The matrix $[\psi]$ is the one with entries $[\psi]_{i,j}=\psi_{ij}$. In our case  $[\psi]=\text{diag}(\alpha,\beta)$. Thus we can write

\begin{equation}
\ket{\psi_{AB}}^{\otimes n}=\sum_{\Omega}\sqrt{\binom{n}{\Omega}}\alpha^{\Omega_{00}}\beta^{\Omega_{11}}\ket{\Omega}.
\end{equation}
From the symmetry of the state $\ket{\psi_{AB}}$ we will have that $s=s'$, therefore the vector $\ket{\Omega}$ is
\begin{equation}
\ket{\Omega}=\dfrac{1}{\sqrt{\binom{n}{\Omega}}}\sum_{s,s:W(s\circ s)=\Omega}\ket{s}\ket{s}.
\end{equation}
Applying the Schur transform to he term $\ket{s}\ket{s}$ we obtain
\begin{equation}
\ket{s}\ket{s}=\sum_{j,j'}\sum_{m,m'}\sum_{\mu,\mu'}\braket{jm\mu}{s}\braket{j'm'\mu'}{s}\ket{jm\mu}\ket{j'm'\mu'},
\end{equation}
which after summing over all sequences we obtain
\begin{equation}
\ket{\Omega}=\dfrac{1}{\sqrt{\binom{n}{\Omega}}}\sum_{s,s:W(s\circ s)=\Omega}\sum_{j,j'}\sum_{m,m'}\sum_{\mu,\mu'}\braket{jm\mu}{s}\braket{j'm'\mu'}{s}\ket{jm\mu}\ket{j'm'\mu'}=\dfrac{1}{\sqrt{\binom{n}{\Omega}}}\sum_{j,\mu}\ket{jm\mu}\ket{jm\mu},
\end{equation}
where it is clear that the $m$ is determined by the matrix $\Omega$ such that $s\sim m$. The variables $\mu$ and $j$ do not depend on the weight matrix. Then,
\begin{equation}
\ket{\psi_{AB}}^{\otimes n}=\sum_{\Omega}\alpha^{\Omega_{00}}\beta^{\Omega_{11}}\sum_{j,\mu}\ket{jm(\Omega)\mu}\ket{jm(\Omega)\mu}.
\end{equation}
Now let us take a look at matrix $\Omega$, we know it must be diagonal. Now denote the number of bi-sequences $(0,0)$ by $n_0$, then analogously $n_1=n-n_0$ indicates the number of bi-sequences $(1,1)$. Thus $\Omega$ is explicitly
\begin{equation}
\Omega=\text{diag}(n_0,n-n_0).
\end{equation}
The sum over $\Omega$ is changed by a sum in $n_0$:
\begin{equation}
\ket{\psi_{AB}}^{\otimes n}=\sum_{n_0=0}^{n}\alpha^{n_0}\beta^{n-n_0}\sum_{j,\mu}\ket{jm(n_0)\mu}\ket{jm(n_0)\mu},
\end{equation}
or equivalently a sum over $m$ via the relation among the number of zeroes and the component of the \emph{angular momentum} in the $z$ axis (when $n_0=0$ all spins are down, therefore $m=-j$ and when $n_0=n$ all the spins are up, $m=j$). Thus, we have the relation $n_0=j+m$
\begin{equation}
\ket{\psi_{AB}}^{\otimes n}=\sum_{j}\sum_{m=-j}^{j}\alpha^{j+m}\beta^{n-j-m}\ket{jm}\ket{jm}\sum_{\mu}\ket{j\mu}\ket{j\mu},
\end{equation}
which if we normalize the states we obtain the desired form
\begin{equation}\label{eqTwoqubits}
\ket{\psi_{AB}}^{\otimes n}=\sum_{j}c_j\underbrace{\sum_{m=-j}^{j}\dfrac{1}{\sqrt{A_j}}\alpha^{j+m}\beta^{j-m}\ket{jm}\ket{jm}}_{\ket{\Phi(\psi_{AB})}_j}\otimes\underbrace{\dfrac{1}{\sqrt{f^{j}}}\sum_{\mu}\ket{j\mu}\ket{j\mu}}_{\ket{\mathcal{K}}_j},
\end{equation}
where 
\begin{equation}\label{eqbipartite}
A_j=\frac{\left(|\beta| ^{4 j+2}-|\alpha| ^{4 j+2}\right)}{|\beta| ^2-|\alpha| ^2},
\end{equation}
and $|c_j|^2=f^{j}A_j$ is the probability to be in a representation $j$. There are several things to notice in equation \eqref{eqTwoqubits}. The most important is that as we say the state $\ket{\mathcal{K}}_j$ is a maximally bipartite entangled state in a Hilbert space of dimension $f^{j}$. It is also noteworthy that only the state $\ket{\Phi(\psi_{AB})}$ depends on the variables of the original state $\alpha,\beta$. As we mentioned before, the dimension $f^{j}$ grows exponentially with $n$ while the dimension of the space of the state  $\ket{\Phi(\psi_{AB})}$ is $(2j+1)$ which we see that grows polynomially. We can see this procedure as a kind of distillation where asymptotically we will keep only the state $\ket{\mathcal{K}}_j$ which is maximally entangled to describe the entanglement among the $n$ copies of a state $\ket{\psi_{AB}}$. 
\\
\\
Now the main problem of this work will be to see how is this distillation in the case of three qubits, where the entanglement structure is richer as we saw in section \ref{secEntanglementclasses}.

\section{The state $\ket{\psi}^{\otimes n}$ for three qubits}\label{secThreequbits}
We will follow the same approach as in the previous section to analyse the space of Kronecker vectors $\ket{\mathcal{K}}_\lambda$ and the rates at which the distillation previously addressed occurs in this states. We begin with the Schur-Weyl decomposition for a $n$-copy space of three qubits 
\begin{equation}
\text{Sym}^n\left(\mathbb{C}^{2}\otimes\mathbb{C}^{2}\otimes\mathbb{C}^{2} \right)^{\otimes n}=\bigoplus_{\alpha,\beta,\gamma \underset{2}{\vdash} n}V_\alpha\otimes V_\beta \otimes V_ \gamma \otimes( [\alpha]\otimes [\beta]\otimes [ \gamma])^{\text{Sym}_n}. 
\end{equation}
Following the same line of thought as in section \ref{secTwoqubit} we can reduce the tensor product of the symmetric group representation in terms of the Kronecker coefficient as
\begin{equation}
[\alpha]\otimes [\beta]\otimes [ \gamma]=\bigoplus_{\sigma\vdash n}g_{\alpha\beta\sigma}[\sigma]\otimes[\gamma]=\bigoplus_{\sigma,\zeta \vdash n}g_{\alpha\beta\gamma}g_{\sigma\gamma\zeta}[\zeta],
\end{equation}
by the symmetry conditions we know $[\zeta]=[(n)]$ then
\begin{equation}
[\alpha]\otimes [\beta]\otimes [ \gamma]=g_{\alpha\beta\gamma}[(n)].
\end{equation}
Hence, the physical cases will be the ones with $g_{\alpha\beta\gamma}\neq0$. An explicit formula for the Kronecker coefficients is an open problem in the general case (i.e. qudits); nevertheless, there are many efficient algorithms to calculate such coefficients for special cases (see \cite{Rosas,walter}). Therefore, we can write a general state of three qubits as
\begin{equation}\label{eqThreequbits}
\ket{\psi}^{\otimes n}=\sum_{\alpha,\beta,\gamma}\sqrt{p({\alpha,\beta,\gamma}|\psi)}\sum\limits_{i=1}^{g_{\alpha\beta\gamma}}\ket{\Phi_{\alpha,\beta,\gamma}^i(\psi)}\ket{\mathcal{K}_{\alpha,\beta,\gamma}^i},
\end{equation}
where we have included the index $i$ to account for the degeneracy that comes with the Kronecker coefficient when $g_{\alpha\beta\gamma}>1$. The term $p({\alpha,\beta,\gamma}|\psi)$ is the probability of being in the representation $(\alpha,\beta,\gamma)$ given that we take $n$ copies of state $\ket{\psi}$. Our main objective in this work is to calculate the asymptotic rates for such probabilities in the particular case where $g_{\alpha\beta\gamma}=1$, where the vectors $\ket{\Phi(\psi)}$ and $\ket{\mathcal{K}}$ are not entangled. In such cases we have that
\begin{equation}
\ket{\psi}^{\otimes n}=\sum_{\alpha,\beta,\gamma}\sqrt{p_{\alpha\beta\gamma}(\psi)}\ket{\Phi_{\alpha,\beta,\gamma}(\psi)}\ket{\mathcal{K}_{\alpha,\beta,\gamma}},
\end{equation}
or if we choose an un-normalized state $\ket{\Phi(\psi)}$ we can write
\begin{equation}
\ket{\psi}^{\otimes n}=\sum_{\alpha,\beta,\gamma}c_{\alpha,\beta,\gamma}\ket{\Phi_{\alpha,\beta,\gamma}(\psi)}\ket{\mathcal{K}_{\alpha,\beta,\gamma}},
\end{equation}
where $c_{\alpha,\beta,\gamma}$ is a coefficient independent of the state $\psi$. Here the probability will be given by
\begin{equation}
p(\alpha,\beta,\gamma|\psi)=|c_{\alpha,\beta,\gamma}|^2\braket{\Phi_{\alpha,\beta,\gamma}(\psi)}{\Phi_{\alpha,\beta,\gamma}(\psi)}.
\end{equation}
Thus, it is easy to see that for two states $\psi_1$ and $\psi_2$
\begin{equation}\label{ecprincipal}
p(\alpha,\beta,\gamma|\psi_1)=p(\alpha,\beta,\gamma|\psi_2)\dfrac{\braket{\Phi_{\alpha,\beta,\gamma}(\psi_1)}{\Phi_{\alpha,\beta,\gamma}(\psi_1)}}{\braket{\Phi_{\alpha,\beta,\gamma}(\psi_2)}{\Phi_{\alpha,\beta,\gamma}(\psi_2)}}.
\end{equation}
Now, we will focus our attention on the elements of equation \eqref{ecprincipal}. On the left lies the probability distribution we would like to calculate for state $\psi_1$. On the right side we have the probability for state $\psi_2$, which we may choose as a representative state for a entanglement class, e.g. $\ket{W}$ or $\ket{GHZ}$, whose probabilities are calculated in a more tractable way as we will see in section \ref{secasymplouck}. The inner products can be calculated with the covariants of the state as we will explain in section \ref{secSLOOCcovariant}.
\\
\\
 Before we calculate the rates for the probability $p(\alpha,\beta,\gamma|\psi)$, we calculate the effective Kronecker coefficient using the covariants. Its relation with the entanglement polytope will be evidenced using the Keyl-Werner theorem.
\section{Covariants and the Kronecker coefficient for SLOCC entanglement classes}\label{secKronecker}
From sections \ref{secSLOCC} and \ref{secCovariant} we know that for three qubits there are six entanglement classes as well as six covariants that can be used to differentiate each entanglement class. We also saw that the components of state $\ket{\Phi(\psi)}$ are polynomials in the state variables, then they can be written as an algebraic combination of the six covariants. Furthermore, we can associate to each covariant a set of three Young diagrams that carry all the information about the multidegree of each covariant. The number of boxes in the diagram will indicate the degree in the variable $\psi_{ijk}$ and the difference of boxes between the first and second row will indicate the degree in the auxiliary variables $x,y,z$. Explicitly
\begin{equation}\label{eqtrip1}
A_{111}\to	\left( \yng(1) ,\yng(1),\yng(1) \right),\quad
\end{equation}
\begin{equation}\label{eqtrip2}
B_{200}\to \left(\yng(2),\yng(1,1),\yng(1,1)\right) \quad B_{020}\to\left(\yng(1,1),\yng(2),\yng(1,1)\right) \quad B_{002}\to \left(\yng(1,1),\yng(1,1),\yng(2)\right),
\end{equation}
\begin{equation}\label{eqtrip3}
	C_{111}\to \quad \left(\yng(2,1),\yng(2,1),\yng(2,1)\right)\quad \text{and} \quad D_{000}\to \left(\yng(2,2),\yng(2,2),\yng(2,2)\right).
\end{equation}
We will refer to these Young frame triplets as \emph{fundamental triplets} in the following sense: every possible Young frame triplet $\vec{\lambda}=(\alpha,\beta,\gamma)$ can be expressed as a linear combination of the fundamental triplets with non-negative integers as coefficients
\begin{equation}\label{eqlambda}
\vec{\lambda}=\sum\limits_{i=1}^{6}n_i v_i\quad; \quad \vec{n}\in \mathbb{Z}^6\quad  \wedge \quad n_i\geq 0\quad \wedge \quad n_5\leq 1,
\end{equation}
where $v_i$ are the fundamental triplets in the order of equations \eqref{eqtrip1},\eqref{eqtrip2} and \eqref{eqtrip3}. The sum is understood as the concatenation of the boxes without mixing rows, e.g.,
\begin{equation}
v_2+v_3=\left(\yng(2),\yng(1,1),\yng(1,1)\right)+\left(\yng(1,1),\yng(2),\yng(1,1)\right)=\left(\yng(3,1),\yng(3,1),\yng(2,2)\right).
\end{equation}
The additional restriction over $n_5$, which is the coefficient of the fundamental triplet associated with the covariant $C_{111}$, comes from the syzygy mentioned in section \ref{secCovariant}. This syzygy tells us that we can write $C_{111}^2$ as an algebraic combination of the rest of the covariants. The decomposition in $\vec{n}$ is not unique in general. However, the number of solutions gives us information about the Kronecker coefficient $g_{\alpha\beta\gamma}$. Recall that the Kronecker coefficient is defined as the multiplicity of a representation of the symmetric group $[\gamma]$ in the tensor product $[\alpha]\otimes [\beta]$
\begin{equation}
[\alpha]\otimes[\beta]=\bigoplus_\gamma g_{\alpha\beta\gamma}[\gamma],
\end{equation}
an in our particular Schur-Weyl decomposition we have
\begin{equation*}
[[\alpha]\otimes[\beta]\otimes[\gamma]]^{\text{Sym}_n}=\bigoplus_{\lambda,\sigma}g_{\alpha\beta\lambda}g_{\lambda\gamma\sigma}[\sigma]=\bigoplus_{\lambda}g_{\alpha\beta\lambda}\delta_{\lambda\gamma}[(n)]=g_{\alpha\beta\gamma}[(n)].
\end{equation*}
Thus if $g_{\alpha\beta\gamma}\neq 0$ we will have a possible state for the triplet $\vec{\lambda}=(\alpha,\beta,\gamma)$, the same is true if equation \eqref{eqlambda} has a solution. This lead us to relate both quantities in the following proposition
\begin{proposition}\label{prop1}
The number of solutions $\vec{n}$ to the equation 
\begin{equation}
\vec{\lambda}=(\alpha,\beta,\gamma)=\sum\limits_{i=1}^{6}n_i v_i\quad; \quad \vec{n}\in \mathbb{Z}^6\quad  \wedge \quad n_i\geq 0\quad \wedge \quad n_5\leq 1,
\end{equation}
	coincides with the Kronecker coefficient $g_{\alpha\beta\gamma}$ in the case of three qubits.
\end{proposition}
We will show how can this assertion be verified when we analyse the GHZ entanglement class. First we will use Proposition \ref{prop1} to gain some knowledge on the Kronecker coefficient in the remaining entanglement classes.

\subsection{Separable states: A-B-C}
For this class, all the covariants are equal to zero except for  $A_{111}$. Then, the only non-zero coefficient of $\vec{n}$ is $n_1$. If we label the Young frames by its second row then we have $\alpha=(\alpha_1,\alpha_2)=(n-\alpha,\alpha)$, a notation we will continue to use throughout the document. Then the equation to solve is $\vec{\lambda}=n_1v_1$, which by components is
\begin{eqnarray*}
	n-\alpha=n_1\\
	\alpha=0,
\end{eqnarray*}
\begin{eqnarray*}
	n-\beta=n_1\\
	\beta=0,
\end{eqnarray*}
\begin{eqnarray*}
	n-\gamma=n_1\\
	\gamma=0.
\end{eqnarray*}
Thus we obtain a unique solution $n_1=n$. Then the Kronecker coefficient is $g_{\alpha\beta\gamma}=\delta_{\alpha,0}\delta_{\beta,0}\delta_{\gamma,0}$. The separable states are thus the ones with $\alpha=\beta=\gamma=0$. If we think of the region in a $(\alpha,\beta,\gamma)$ space, the set of separable states will be a point at the origin.

\subsection{Bipartite entangled states $\ket{A-BC} , \ket{AB-C} , \ket{AC-B}$}\label{secA-BC}
Since the only difference between the three states is a permutation of which qubit is not entangled we will only analyse one case.
In the case $\ket{AB-C}$ we have that $D_{000}$, $C_{111}$, $B_{200}$ and $B_{020}$ are equal to zero.  Then
$n_2=n_3=n_5=n_6=0$. The equations to solve are
\begin{eqnarray*}
	n-\alpha=n_1+n_4\\
	\alpha=n_4,
\end{eqnarray*}
\begin{eqnarray*}
	n-\beta=n_1+n_4\\
	\beta=n_4,
\end{eqnarray*}
\begin{eqnarray*}
	n-\gamma=n_1+2n_4\\
	\gamma=0.
\end{eqnarray*}
In this case the solution is also unique
\begin{eqnarray*}
	n_4=\alpha=\beta\\
	n_1=n-2\alpha,
\end{eqnarray*}
with the additional restriction $\alpha=\beta$ and $\gamma=0$ for the solution to exist.
The Kronecker coefficient is $g_{\alpha\beta\gamma}=\delta_{\alpha,\beta}\delta_{\gamma ,0}$. To obtain the cases $\ket{AC-B}$ and $\ket{A-BC}$ the terms $(\alpha,\beta,\gamma)$ must be permuted accordingly. The region obtained in the space $(\alpha,\beta,\gamma)$ is a line in the plane $\gamma=0$. The three cases are represented as red lines in Fig.  \ref{regw}.

\subsection{W entanglement Class}\label{secWentclass}
\begin{figure}[h!]
	\centering
	\includegraphics[scale=0.6]{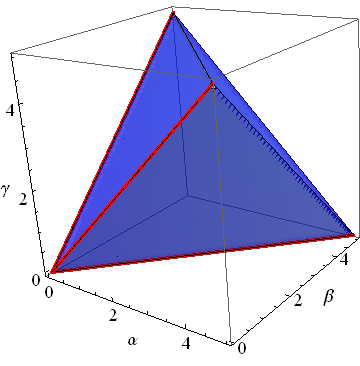}
	\caption{Region in the space $(\alpha,\beta,\gamma)$ of the permited states in the W entanglement class ($n=10$)}
	\label{regw}
\end{figure}

For the W entanglement class the only covariant which is zero is $D_{000}$, then $n_6=0$. The equation to solve for $\vec{n}$, (we will first consider the case $n_5=0$ and then $n_5=1$) are
\begin{eqnarray}
	n-\alpha=n_1+2n_2+n_3+n_4\\
	\alpha=n_3+n_4,
\end{eqnarray}
\begin{eqnarray}
	n-\beta=n_1+n_2+2n_3+n_4\\
	\beta=n_2+n_4,
\end{eqnarray}
\begin{eqnarray}
	n-\gamma=n_1+n_2+n_3+2n_4\\
	\gamma=n_2+n_3.
\end{eqnarray}
Note that $\alpha+\beta+\gamma=2(n_2+n_3+n_4)$. And making the sum over the equations of the form $n-\lambda$ we obtain
\begin{equation*}
3n-(\alpha+\beta+\gamma)=3n_1+4(n_2+n_3+n_4),
\end{equation*}
then
\begin{equation}
n_1=n-(\alpha+\beta+\gamma).
\end{equation}
Thus, the solution is unique $g_{\alpha\beta\gamma}=1$. The solution for $n_2,n_3$ and $n_4$ is:
\begin{eqnarray}
	n_2=\dfrac{\beta+\gamma-\alpha}{2},\\
	n_3=\dfrac{\alpha+\gamma-\beta}{2},\\
	n_4=\dfrac{\alpha+\beta-\gamma}{2}.
\end{eqnarray}
For the case $n_5=1$ we obtain, following the same procedure
\begin{eqnarray*}
	n-\alpha=n_1+2n_2+n_3+n_4+2\\
	\alpha=n_3+n_4+1,
\end{eqnarray*}
\begin{eqnarray*}
	n-\beta=n_1+n_2+2n_3+n_4+2\\
	\beta=n_2+n_4+1,
\end{eqnarray*}
\begin{eqnarray*}
	n-\gamma=n_1+n_2+n_3+2n_4+2\\
	\gamma=n_2+n_3+1.
\end{eqnarray*}
With $\alpha+\beta+\gamma-3=2(n_2+n_3+n_4)$. Taking the sum over the equation of the form $n-\lambda$, we have
\begin{equation*}
3n-(\alpha+\beta+\gamma)=3n_1+4(n_2+n_3+n_4)+6.
\end{equation*}
Then,
\begin{equation}
n_1={n-(\alpha+\beta+\gamma)},
\end{equation}
and we see that it remains the same as in the case $n_5=0$; however
\begin{eqnarray}
	n_2=\dfrac{\beta+\gamma-\alpha-1}{2},\\
	n_3=\dfrac{\alpha+\gamma-\beta-1}{2},\\
	n_4=\dfrac{\alpha+\beta-\gamma-1}{2}.
\end{eqnarray}
The solution is still unique and can be expressed in general as
\begin{eqnarray}
	n_2=\left\lfloor\dfrac{\beta+\gamma-\alpha}{2}\right\rfloor,\\
	n_3=\left\lfloor\dfrac{\alpha+\gamma-\beta}{2}\right\rfloor,\\
	n_4=\left\lfloor\dfrac{\alpha+\beta-\gamma}{2}\right\rfloor,\\
	n_5=(\alpha+\beta-\gamma)\mod{2},
\end{eqnarray}
where $\left\lfloor x\right\rfloor$ is the floor function which approximates $x$ to the nearest integer number below. To find the region in the $(\alpha,\beta,\gamma)$ space where the states satisfy $g_{\alpha\beta\gamma}\neq0$ we must appeal to the conditions over $\vec{n}$, specially $n_i\geq 0$. The region defined is plotted in Fig. \ref{regw} where we have considered $(\alpha,\beta,\gamma)$ to be continuous (which is true when we consider the normalized variables $(\bar{\alpha},\bar{\beta},\bar{\gamma})$ and we take the limit $n\to\infty$).
With the Keyl-Werner theorem in mind, the regions that are displayed in Fig. \ref{regw} (when $(\alpha,\beta,\gamma)$ is normalized by $n$) corresponds to the the entanglement polytope of the W class. We must expect that with the GHZ entanglement class we will be able to generate the entire entanglement polytope which is the result we will see in the next section.
\subsection{GHZ entanglement class and the entanglement polytope}\label{General}
\begin{figure}[h!]
\centering
\begin{tabular}{cc}
\includegraphics[scale=0.3]{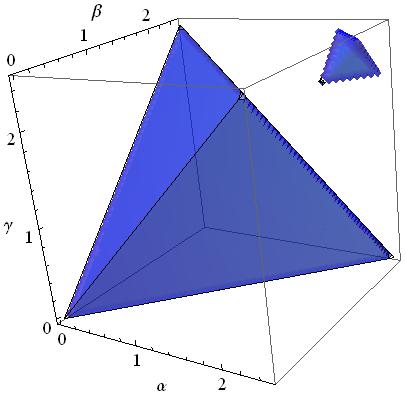} & \includegraphics[scale=0.3]{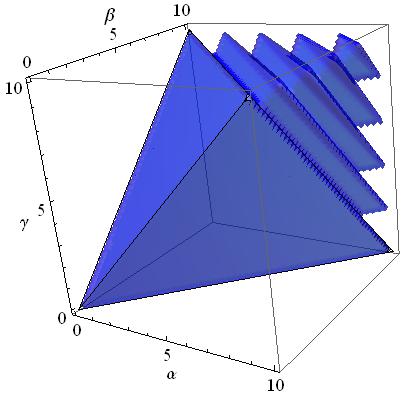}\\
(a) $n=5$& (b) $n=20$\\
\includegraphics[scale=0.3]{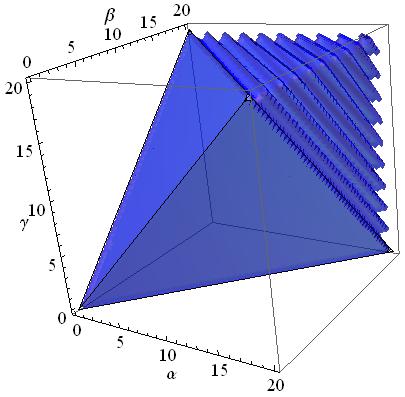} & \includegraphics[scale=0.3]{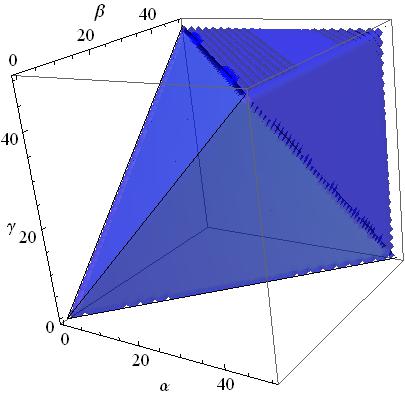}\\
(c) $n=40$&(d)$n=80$
\end{tabular}
\caption{Region in the $(\alpha,\beta\gamma)$ space where $g_{\alpha\beta\gamma}\neq0$ for increasing values of $n$.}
\label{figGHZ}
\end{figure}

\begin{figure}[h!]
	\centering
	\includegraphics[scale=0.4]{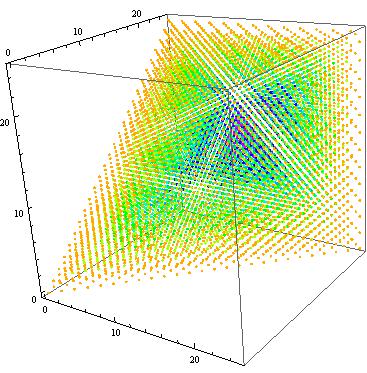}
	\caption{Kronecker coefficient inside the entanglement polytope $n=50$, the cold color regions indicate higher values for $g_{\alpha\beta\gamma}$  wheter hot colors indicate lower $g_{\alpha\beta\gamma}$.}
	\label{figkron}
\end{figure}

For this entanglement class none of the covariants is zero, then we do not have \emph{a priori} any component of $\vec{n}$ equal to zero. Nevertheless, we can analyse this case in a similar way to that of the W class. In the previous section we had $n_6=0$ and a unique solution for the rest of $\vec{n}$. The fundamental triplet $v_6$ is composed of three equal Young frames of two boxes per row. Therefore we can define a \emph{reduced frame} by removing the triplet $v_6$ certain amount of times. That is, a component of the triplet $\vec{\lambda}$ will transform as $(n-\alpha,\alpha)\rightarrow(n-\alpha-2k,\alpha-2k)$, where $k$ is the number of times we remove $v_6$ from $\vec{\lambda}$. This procedure can only be performed a maximum of $k_{\max}$ times. It is easy to see that
\begin{equation*}
k_\text{max}=\left\lfloor\dfrac{\min(\alpha,\beta,\gamma)}{2}\right\rfloor.
\end{equation*}
For each possible value of $k$, we have the same case as the W entanglement case, so we have a unique solution per $k$, explicitly
\begin{eqnarray}
	n_1 &=& n-(\alpha+\beta+\gamma)+2k,\label{ecghz1}\\
	n_2 &=& \left\lfloor\dfrac{\beta+\gamma-\alpha}{2}\right\rfloor-k,\\
	n_3 &=& \left\lfloor\dfrac{\alpha+\gamma-\beta}{2}\right\rfloor-k,\\
	n_4&=& \left\lfloor\dfrac{\alpha+\beta-\gamma}{2}\right\rfloor-k.\label{ecghz2}
\end{eqnarray}
Each time we remove $n_6$ we must verify that the solution obtained is feasible, that is we must verify that
\begin{equation}
n-(\alpha+\beta+\gamma)+2k\geq 0
\end{equation}
and
\begin{eqnarray}
	\left\lfloor\dfrac{\beta+\gamma-\alpha}{2}\right\rfloor-k\geq 0,\\
	\left\lfloor\dfrac{\alpha+\gamma-\beta}{2}\right\rfloor-k\geq 0,\\
	\left\lfloor\dfrac{\alpha+\beta-\gamma}{2}\right\rfloor-k\geq 0,
\end{eqnarray}
for all $k\in\{0,\ldots,k_{\max}\}$. The three inequalities above can be reduced to the single expression
\begin{equation*}
\left\lfloor\dfrac{\alpha+\beta+\gamma-2\max(\alpha,\beta,\gamma)}{2}\right\rfloor\geq k,\quad \forall k\in(0,\ldots,k_{\text{max}}).
\end{equation*}
Thus we can state our main result for the Kronecker coefficient in the next theorem
\begin{theorem}\label{teo2}
	Let $(n-\alpha,\alpha),(n-\beta,\beta)$ and $(n-\gamma,\gamma)$ be Young frames of two rows (integer partitions of $n$ of size two 2), then the Kronecker coefficient
	\begin{eqnarray}
		g_{\alpha\beta\gamma}&=&\sum\limits_{k=0}^{k_{\max}}\delta_{\alpha\beta\gamma}(k),\label{eqKronecker}\\
		\delta_{\alpha\beta\gamma}(k)&=&\left\{ \begin{array}{ccc}
			1 & & (y\geq k) \wedge (n-(\alpha+\beta+\gamma)+2k)\geq 0)\\
			0 & & \mathrm{otherwise}
		\end{array}\right.
	\end{eqnarray}
	where $y=\left\lfloor\dfrac{\alpha+\beta+\gamma-2\max(\alpha,\beta,\gamma)}{2}\right\rfloor$ and $k_{\max}=\left\lfloor\dfrac{\min(\alpha,\beta,\gamma)}{2}\right\rfloor.$ 
\end{theorem}
The region defined by the inequalities in which $g_{\alpha\beta\gamma}\neq0$ using equation \eqref{eqKronecker} is plotted in Fig.\ref{figGHZ}. We can see that as $n$ grows the region plotted converges to the well-known entanglement polytope mentioned in section \ref{secpolytope}. This is a consequence of the Keyl-Werner theorem that basically tells us that as $n\to\infty$, the Young frames triplet that we will almost surely obtain for any state $\ket{\psi}$ will be the ones that are compatible with the local spectrum of each susbsystem. Hence, the second row of the Young frames (the smallest eigenvalue of the local spectrum) must be compatible with the marginal conditions which are represented by the polytope. 
From all this analysis we have thus obtained a formula for the Kronecker coefficient of two row Young frames, a result analogous to the one reported in \cite{Rosas}. We have also learned how this coefficient varies among different entanglement classes, being $g_{\alpha\beta\gamma}=1$ in the separable, biseparable and W classes and in general greater than one in the GHZ class.
To see how the Kronecker coefficient changes inside the entanglement polytope, we compute it and make the plot shown in Fig. \ref{figkron}.
It can be seen that in the facet of the polytope $g_{\alpha\beta\gamma}=1$ or $g_{\alpha\beta\gamma}=0$ while it increases its value as we approach the plane $\alpha+\beta+\gamma=n$ where it attains its maximal value.
Now if we take a look at equation \eqref{eqThreequbits}
for a state $\ket{\psi_W}$ in the W entanglement class (or an inferior class in the hierarchy) we will obtain
\begin{equation}\label{eqmain}
\ket{\psi_W}^{\otimes n}=\sum_{\alpha,\beta,\gamma}\sqrt{p_{\alpha\beta\gamma}(\psi_W)}\ket{\Phi_{\alpha,\beta,\gamma}(\psi_W)}\ket{\mathcal{K}_{\alpha,\beta,\gamma}}.
\end{equation}

It is noteworthy that for this particular case the state for a given $(\alpha,\beta,\gamma)$ is separable, which is not the case for general $\ket{\psi}$. In the following sections we will focus on finding an expression for the inner product $\braket{\Phi_{\alpha,\beta,\gamma}(\psi)}{\Phi_{\alpha,\beta,\gamma}(\psi)}$ by calculating a proportional vector using the covariants.
\section{A SLOCC-Covariant approach to asymptotic rates}\label{secSLOOCcovariant}
We showed in section \ref{secCovstaInv} how we can construct a state out of the product of covariants. In this section we will use this result to build the state $\ket{\Phi_{\alpha,\beta,\gamma}}$ of equation \eqref{eqmain} and more important the inner products in equation \eqref{ecprincipal}. The state will be constructed up to a multiplicative constant $\eta_{\alpha\beta\gamma}$ that depends only on the triplet $(\alpha,\beta,\gamma)$ and not on the state $\psi$. The covariant which we will transform into a state will have the general form
\begin{equation}
\Phi(\vec{n})=A_{111}^{n_1}B_{200}^{n_2}B_{020}^{n_3}B_{002}^{n_4}C_{111}^{n_5}D_{000}^{n_6},
\end{equation}
with the constrain $n_1+2(n_2+n_3+n_4)+n_5+4n_6=n$ where $n$ is the number of copies we are considering. The vector $\vec{n}$ of integer values takes on the values assigned in the previous section for a given representation $(\alpha,\beta,\gamma)$. To see how with this approach we can reproduce the state $\ket{\Phi_{\alpha,\beta,\gamma}}\in V^{2}_{\alpha}\otimes V^{2}_{\beta}\otimes V^{2}_{\gamma}$, let us analyse a few simple cases and then turn to calculate such state for the W and GHZ entanglement classes.
\subsection{Separable and biseparable classes}
For the separable case we have that the only nonzero component of $\vec{n}$ is $n_1$, thus the covariant associated will be
\begin{equation}
\Phi(n_1)=A_{111}^{n_1}=A_{111}^n,
\end{equation}
and let us consider the more general separable state
\begin{equation}
\ket{\psi_{A-B-C}}=(c_1\ket{0}+d_1\ket{1})\otimes(c_2\ket{0}+d_2\ket{1})\otimes(c_3\ket{0}+d_3\ket{1}),
\end{equation}
then 
\begin{equation}
A_{111}=(c_1x_0+d_1x_1)(c_2y_0+d_2y_1)(c_3z_0+d_3z_1),
\end{equation}
and
\begin{equation}
A_{111}^{n}=\sum_{k_1,k_2k,_3}\binom{n}{k_1}\binom{n}{k_2}\binom{n}{k_3}c_1^{k_1}d_{1}^{n-k_1}x_0^{k_1}x_{1}^{n-k_1}c_2^{k_2}d_{2}^{n-k_2}y_0^{k_2}y_{1}^{n-k_2}c_3^{k_3}d_{3}^{n-k_3}z_0^{k_3}z_{1}^{n-k_3},
\end{equation}
and making the conversion to a state we have
\begin{equation}\label{eqcovsep}
\dfrac{\ket{\Phi(n_1)}}{\eta_{000}}=\left(\sum_{k_1}\binom{n}{k_1}\dfrac{c_1^{k_1}d_{1}^{n-k_1}\ket{k_1}}{\sqrt{\binom{n}{k_1}}}\right)\left(\sum_{k_2}\binom{n}{k_2}\dfrac{c_2^{k_2}d_{2}^{n-k_2}\ket{k_2}}{\sqrt{\binom{n}{k_2}}}\right)\left(\sum_{k_3}\binom{n}{k_3}\dfrac{c_3^{k_3}d_{3}^{n-k_3}\ket{k_3}}{\sqrt{\binom{n}{k_3}}}\right),
\end{equation}
with $\ket{k_i}=\ket{j=n/2-\alpha_2,m=j-k_i}=\ket{j=n/2,m=j-k_i}$, because for separable states $\alpha_2=\beta_2=\gamma_2=0$. Note that the state in  \eqref{eqcovsep} is also separable as we expected and it is normalized. Hence, in this case, we can write
\begin{equation}
\ket{\psi_{A-B-C}}^{\otimes n}=\eta_{000}\ket{\Phi(n_1)}\ket{\mathcal{K}},
\end{equation}
and the probability of having $p(\alpha,\beta,\gamma|\psi_{A-B-C})=\delta_{\alpha,0}\delta_{\beta,0}\delta_{\gamma,0}$ or as in equation \eqref{ecprincipal}
\begin{equation}
p(\alpha,\beta,\gamma|\psi)=p(\alpha,\beta,\gamma|\ket{000})\dfrac{\braket{\Phi(\psi)}{\Phi(\psi)}}{\braket{\Phi(\ket{000})}{\Phi(\ket{000})}},
\end{equation}
where we have chosen a representative state of the separable class: $\ket{000}$. It is then easy to verify that
\begin{equation}
p(\alpha,\beta,\gamma|\psi_{A-B-C})=\delta_{\alpha,0}\delta_{\beta,0}\delta_{\gamma,0}\dfrac{\braket{\Phi(\psi_{A-B-C})}{\Phi(\psi_{A-B-C})}}{\eta_{000}^2}=\delta_{\alpha,0}\delta_{\beta,0}\delta_{\gamma,0}.
\end{equation}
Although this case may seem trivial, we used it to show the approach we will use to attack more interesting instances. Now we consider a biseparable state in the entanglement class $A-BC$. A representative state of this class is ($c_\phi=\cos\phi,s_\phi=\sin\phi$)
\begin{equation}
\ket{\psi_{AB-C}}=(c_{\theta}\ket{0}+s_{\theta}\ket{1})\otimes(c_\delta\ket{00}+s_\delta\ket{11}),
\end{equation}
with non-vanishing components of $\vec{n}$: $n_1,n_2$. Then the covariant $\Phi(n_1,n_2)$ will be
\begin{eqnarray}
\Phi(n_1,n_4)&=&A_{111}^{n_1}B_{200}^{n_2}\nonumber\\
&=&\left(c_\theta x_0+s_\theta x_1\right)^{n_1}\left(c_\delta y_0z_0+s_\delta y_1z_1\right)^{n_1}(c_\delta s_\delta x_0^2)^{n_2},\\
&=&\sum_{j=0}^{n_1}\binom{n_1}{j}c_\theta^j x_0^j(s_\theta x_1)^{n_1-j}\sum_{k=0}^{n_1}\binom{n_1}{k}c_\delta^{n_2+k}s_\delta^{n_2+n_1-k}x_0^{n_1+2n_2}y_0^{k}y_1^{n_1-k}z_0^{k}z_1^{n_1-k}\nonumber,
\end{eqnarray}
which is converted to the vector
\begin{eqnarray}
\ket{\Phi_{A-BC}}&=&\eta_{0\beta\beta}\sum_{j=0}^{n_{_1}}\sqrt{\binom{n_{_1}}{j}}c_\theta^js_{\theta}^{n_{_1}-j}\ket{j}\sum_{k=0}^{n_{_1}}\binom{n_{_1}}{k}\dfrac{c^{n_2+k}d^{n_{_2}+n_{_1}-k}\ket{n}\ket{k}\ket{k}}{\binom{n_{_1}}{k}},\\
 &=&\eta_{0\beta\beta}\ket{\psi_A}\sum_{k=0}^{n_{_1}}c^{n_2+k}d^{n_2+n_1-k}\ket{k}\ket{k},\\
 &=&\eta_{0\beta\beta}\ket{\psi_A}\sum_{k=0}^{n-2\beta}c^{\beta+k}d^{n-\beta-k}\ket{k}\ket{k},
\end{eqnarray}
where we have used the results from section \ref{secA-BC} ($\alpha=0$ and $\beta=\gamma$ for biseparable states)  $n_1=n-2\beta$ and $n_2=\beta$ and
\begin{equation}
\ket{\psi_A}=\sum_{j=0}^{n_1}\binom{n_1}{j}c_\theta^js_{\theta}^{n_1-j}\dfrac{\ket{j}}{\sqrt{\binom{n_1}{j}}}.
\end{equation}
Note that we obtain a biseparable state as expected (the conversion cannot change the entanglement class). We can also calculate the inner product
\begin{equation}
\dfrac{\braket{\Phi_{A-BC}}{\Phi_{A-BC}}}{\eta_{0\beta\beta}^2}=\sum_{k=0}^{n-2\beta}|c|^{2\beta+2k}|d|^{2n-2\beta-2k}=\frac{(|d|^2)^{\beta } (|c|^2)^{-\beta +n+1}-(|c|^2)^{\beta } (|d|^2)^{-\beta +n+1}}{|c|^2-|d|^2},
\end{equation}
which is basically the same as equation \eqref{eqbipartite} cast in different variables. In the limit $c\to d$ we obtain the intuitive result $2^{-n}(1+n-2\beta)=2^{-n}\dim V^{2}_{\beta}$. The inner product in this case is related to the probability of being in a triplet $p(\alpha,\beta,\gamma)=p(0,\beta,\beta)$ by
\begin{equation}
p(\alpha,\beta,\gamma|\psi_{A-BC})=\delta_{\alpha,0}\delta_{\beta,\gamma}f^{\beta}\dfrac{\braket{\Phi_{A-BC}}{\Phi_{A-BC}}}{\eta_{0\beta\beta}^2},
\end{equation}
and the asymptotic rate 
\begin{equation}
\phi(\bullet,\bar{\beta},\bar{\beta}|\psi_{A-BC})=-H(\bar{\beta})-\lim\limits_{n\to\infty}\dfrac{\log \braket{\Phi_{A-BC}}{\Phi_{A-BC}}}{n},
\end{equation}
where we have used the shorthand notation $\bar{\alpha}=0\to\bullet$ and we will continue using it throughout the document together with $\bar{\alpha}=1/2\to\boxplus$. We may now analyse the following limit (with $\eta_{0\beta\beta}^2=1$)
\begin{equation}
\lim\limits_{n\to\infty}\dfrac{\log \braket{\Phi_{A-BC}}{\Phi_{A-BC}}}{n}=\lim\limits_{n\to\infty}\log\sum_{k=0}^{n-2\beta}|c|^{2\beta+2k}|d|^{2n-2\beta-2k}.
\end{equation}
First note that $|c|^2$ and $|d|^2$ are the spectrum of the reduced density matrix (which in this case coincides with the so-called Schmidt coefficients) of the system $BC$, that is $\text{spec}(\rho_{BC})=(|c|^2,|d|^2)$. Additionally, we can recast the sum as
\begin{equation}
\sum_{k=0}^{n-2\beta}|c|^{2\beta+2k}|d|^{2n-2\beta-2k}=\sum_{k=\beta}^{n-\beta}\left(|c|^2\right)^{k}\left(|d|^2\right)^{n-k},
\end{equation}
and in terms of the Schur polynomial for two row diagrams \cite{Audanert_2006}
\begin{equation}
s_{p,q}(x,y)=\sum_{k=q}^{p}x^{k}y^{p+q-k},
\end{equation}
as
\begin{equation}
\braket{\Phi_{A-BC}}{\Phi_{A-BC}}=s_{n-\beta,\beta}(|c|^2,|d|^2)=s_{\beta}(\text{spec}(\rho_{BC})).
\end{equation}
Now from the asymptotics of the Schur polynomials \cite{Audanert_2006} we have that the rate will be
\begin{equation}
\phi(\bullet,\bar{\beta},\bar{\beta}|\psi_{A-BC})=-H(\bar{\beta})-\sum_{i=1}^{2}\bar{\beta}_i\log\text{spec}(\rho_{BC})_i.
\end{equation}
In the general case, we will obtain a similar result and we will be able to write the above rate in terms of entanglement measures such as the concurrence and the geometric entanglement.
\\
\\
Motivated by this \emph{covariant-to-state} approach, we engage in calculating the ratio of the probabilities $p(\alpha,\beta,\gamma|\psi_1)/p(\alpha,\beta,\gamma|\psi_2)$ by calculating the ratio 
\begin{equation}
\braket{\Phi(\psi_1)}{\Phi(\psi_1)}/\braket{\Phi(\psi_2)}{\Phi(\psi_2)},
\end{equation}
 in the spirit of equation \eqref{ecprincipal}. We will focus our calculation on the asymptotics for the W and GHZ entanglement class. Notice that with this approach we have calculated the vector $\ket{\Phi(\psi)}$ directly from the covariant-to-state approach. However, in the more general case we are only able to calculate it up to a proportionality factor $\eta_{\alpha\beta\gamma}$ which depends only on $(\alpha,\beta,\gamma)$. This proportionality factor will not be a problem for further calculations because we are finally interested in the ratio of the inner products as seen in equation \eqref{ecprincipal}.
\subsection{W class}
We will begin with the states in the W entanglement class. A general state in this class can be written as
\begin{equation}
\ket{\psi_W}=\sqrt{a}\ket{100}+\sqrt{b}\ket{010}+\sqrt{c}\ket{001}+\sqrt{d}\ket{000},
\end{equation}
with $a+b+c+d=1$ and $a,b,c>0$ and $d\geq 0$. The only zero element from the vector $\vec{n}$ is $n_6$, thus we write
\begin{equation}
\Phi(\vec{n})_{W}=A_{111}^{n_1}B_{200}^{n_2}B_{020}^{n_3}B_{002}^{n_4}C_{111}^{n_5}.
\end{equation}
The decomposition in covariants is given by
\begin{multline}
\Phi_{W}=\left(\sqrt{a} x_1 y_0 z_0+\sqrt{b} x_0 y_1 z_0+\sqrt{c} x_0 y_0 z_1+\sqrt{d} x_0 y_0 z_0\right)^{n_1}\\
\times\left(-\sqrt{bc} x_0^2\right)^{n_2}\left(-\sqrt{ac} y_0^2\right)^{n_3}\left(-\sqrt{ab} z_0^2\right)^{n_4}(\sqrt{abc}x_0y_0z_0)^{n_5},
\end{multline}
expanding
\begin{multline}\label{eqsimplif}
\Phi_{W}=\sum_{\vec{k}=0}^{n_1}\dfrac{n_1!(-1)^{n_2+n_3+n_4}}{k_1!k_2!k_3!k_4!}\sqrt{a}^{k_1+n_3+n_4+n_5}\sqrt{b}^{k_2+n_2+n_4+n_5}\sqrt{c}^{k_3+n_2+n_3+n_5}\sqrt{d}^{k_4}\\
\times x_0^{k_2+k_3+k_4+2n_2+n_5}x_1^{k_1}y_0^{k_1+k_3+k_4+2n_3+n_5}y_1^{k_2}z_0^{k_1+k_2+k_4+2n_4+n_ 5}z_1^{k_3}.
\end{multline}
It is easy to see that whether $n_5=0$ or $n_5=1$ we obtain the same results since it is always accompanied in sums of two $n_{i}$ , $i=2,3,4$. Simplifying expression \eqref{eqsimplif} further using the condition $k_1+k_2+k_3+k_4=n_1$ we obtain
\begin{multline}
\Phi_{W}=\sum_{\vec{k}=0}^{n_1}\dfrac{n_1!(-1)^{n_2+n_3+n_4}}{k_1!k_2!k_3!k_4!}\sqrt{a}^{k_1+n_3+n_4}\sqrt{b}^{k_2+n_2+n_4}\sqrt{c}^{k_3+n_2+n_3}\sqrt{d}^{k_4}\\
\times x_0^{n_1+2n_2-k_1}x_1^{k_1}y_0^{n_1+2n_3-k_2}y_1^{k_2}z_0^{n_1+2n_4-k_3}z_1^{k_3}.
\end{multline}
Making the sum over $k_4=n_1-k_1-k_2-k_3$ we obtain
\begin{multline}
\Phi_{W}=\sqrt{a}^{n_3+n_4}\sqrt{b}^{n_2+n_4}\sqrt{c}^{n_2+n_3}\sqrt{d}^{n_1}\sum_{k_1,k_2,k_3}^{n_1}\dfrac{n_1!(-1)^{n_2+n_3+n_4}}{k_1!k_2!k_3!(n_1-k_1-k_2-k_3)!}\left(\dfrac{a}{d}\right)^{k_1/2}\left(\dfrac{b}{d}\right)^{k_2/2}\left(\dfrac{c}{d}\right)^{k_3/2}\\
\times x_0^{n_1+2n_2-k_1}x_1^{k_1}y_0^{n_1+2n_3-k_2}y_1^{k_2}z_0^{n_1+2n_4-k_3}z_1^{k_3},
\end{multline}
where the sum over the remaining $k_1,k_2,k_3$ s such that $n_1-k_1+k_2+k_3\geq0$. Now we translate the covariant to a state $\ket{\Phi_{\psi_W}}$, obtaining
\begin{multline}
\ket{\Phi_{\psi_W}}=\eta_{\alpha\beta\gamma}a^{(n_3+n_4)/2}b^{(n_2+n_4)/2}c^{(n_2+n_3)/2}d^{n_1/2}\sum_{k_1,k_2,k_3}^{n_1}\dfrac{n_1!(-1)^{n_2+n_3+n_4}}{k_1!k_2!k_3!(n_1-k_1-k_2-k_3)!}\\
\times\left(\dfrac{a}{d}\right)^{k_1/2}\left(\dfrac{b}{d}\right)^{k_2/2}\left(\dfrac{c}{d}\right)^{k_3/2}
\dfrac{\ket{n_1+2n_2-k_1}\ket{n_1+2n_3-k_2}\ket{n_1+2n_4-k_3}}{\sqrt{\binom{n_1+2n_2}{k_1}\binom{n_1+2n_3}{k_2}\binom{n_1+2n_4}{k_3}}},
\end{multline}
and replacing the values for $\vec{n}$ in terms of $(\alpha,\beta,\gamma)$ as in section \ref{secWentclass} we have
\begin{multline}
\ket{\Phi_{\psi_{W}}}=\eta_{\alpha\beta\gamma}a^{\alpha/2}b^{\beta/2}c^{\gamma/2}d^{(n-\alpha-\beta-\gamma)/2}\sum_{k_1,k_2,k_3}^{n-\alpha-\beta-\gamma}\dfrac{(n-\alpha-\beta-\gamma)!(-1)^{(\alpha+\beta+\gamma)/2}}{k_1!k_2!k_3!(n-\alpha-\beta-\gamma-k_1-k_2-k_3)!}\\
\times\left(\dfrac{a}{d}\right)^{k_1/2}\left(\dfrac{b}{d}\right)^{k_2/2}\left(\dfrac{c}{d}\right)^{k_3/2}
\dfrac{\ket{k_1}\ket{k_2}\ket{k_3}}{\sqrt{\binom{n-2\alpha}{k_1}\binom{n-2\beta}{k_2}\binom{n-2\gamma}{k_3}}},
\end{multline}
where we have used the simplified notation for the second row of the Young frames $\lambda_2=\lambda$. Now we calculate the inner product $\braket{\Phi_{\psi_{W}}}{\Phi_{\psi_{W}}}$ and relate it to the probability $p(\alpha,\beta,\gamma|\psi_{W})$. Explicitly,
\begin{multline}\label{eqbeast}
\braket{\Phi_{\psi_{W}}}{\Phi_{\psi_{W}}}=\eta_{\alpha\beta\gamma}^2a^{\alpha}b^{\beta}c^{\gamma}d^{(n-\alpha-\beta-\gamma)}\sum_{k_1,k_2,k_3}^{n-\alpha-\beta-\gamma}\dfrac{(n-\alpha-\beta-\gamma)!^2}{k_1!^2k_2!^2k_3!^2(n-\alpha-\beta-\gamma-k_1-k_2-k_3)!^2}\\
\times\left(\dfrac{a}{d}\right)^{k_1}\left(\dfrac{b}{d}\right)^{k_2}\left(\dfrac{c}{d}\right)^{k_3}
\dfrac{1}{\binom{n-2\alpha}{k_1}\binom{n-2\beta}{k_2}\binom{n-2\gamma}{k_3}},
\end{multline}
and in the case of the $\ket{W}$ state ($\lim a,b,c\to 3^{-1}$) we obtain 
\begin{multline}\label{eqbeastW}
\braket{\Phi_{W}}{\Phi_{W}}=\eta_{\alpha\beta\gamma}^23^{-n}\sum_{k_1+k_2+k_3=n-\alpha-\beta-\gamma}^{n-\alpha-\beta-\gamma}\dfrac{(n-\alpha-\beta-\gamma)!^2}{k_1!^2k_2!^2k_3!^2}
\dfrac{1}{\binom{n-2\alpha}{k_1}\binom{n-2\beta}{k_2}\binom{n-2\gamma}{k_3}},
\end{multline}
thus 
\begin{equation}\label{ecprobW}
p(\alpha,\beta,\gamma|\psi_{W})=p(\alpha,\beta,\gamma|W)\dfrac{\braket{\Phi_{\psi_{W}}}{\Phi_{\psi_{W}}}}{\braket{\Phi_{W}}{\Phi_{W}}}.
\end{equation}
So if we want to calculate the rate function for $p(\alpha,\beta,\gamma|\psi_{W})$ we must analyse the asymptotics of equation \eqref{ecprobW} and thus equation \eqref{eqbeast}. At a first glance it may seem quite complicated expression. However, it simplifies when the triplet $(\alpha,\beta,\gamma)$ lie in the plane $\alpha+\beta+\gamma=n$. We will discuss this particular case separately in the following section.
\subsubsection{The $\alpha+\beta+\gamma=n$ plane}
When we are in the plane $\alpha+\beta+\gamma=n$, equation \eqref{eqbeast} takes the simple form
\begin{equation}
\braket{\Phi_{\psi_{W}}}{\Phi_{\psi_{W}}}=a^{\alpha}b^{\beta}c^{\gamma},
\end{equation}
which has a rate function
\begin{equation}
\phi_1(\bar{\alpha},\bar{\beta},\bar{\gamma})=-\bar{\alpha}\log a-\bar{\beta}\log b-\bar{\gamma}\log c,
\end{equation}
with $\bar{\lambda}=\lambda/n$. Together with the asymptotics of the term
\begin{equation}
\braket{\Phi_{W}}{\Phi_{W}}=3^{-n},
\end{equation}
we obtain the asymptotic rate for the probability
\begin{equation}
p(\alpha,\beta,\gamma|\psi_{W})\sim\exp[-n\phi(\bar{\alpha},\bar{\beta},\bar{\gamma}|\psi_W)],
\end{equation}
namely
\begin{equation}\label{ecWabc}
\phi(\bar{\alpha},\bar{\beta},\bar{\gamma}|\psi_W)=\phi(\bar{\alpha},\bar{\beta},\bar{\gamma}|W)-\bar{\alpha}\log a-\bar{\beta}\log b-\bar{\gamma}\log c-\log 3.
\end{equation}
What is interesting of this expression, is that it can expressed as a convex combination of the rates at the vertex in the polytope $\bar{\alpha},\bar{\beta},\bar{\gamma}$ that lie in the plane $\bar{\alpha}+\bar{\beta}+\bar{\gamma}=1$. To prove this assertion explicitly, let us calculate the rate at the vertex. We will use the notation $\bar{\lambda}=0\to\bullet$, $\bar{\lambda}=1/2\to\boxplus$ and $\delta\phi(\bar{\alpha},\bar{\beta},\bar{\gamma}|\psi_{W})=\phi(\bar{\alpha},\bar{\beta},\bar{\gamma}|\psi_{W})-\phi(\bar{\alpha},\bar{\beta},\bar{\gamma}|W)$ to obtain
\begin{eqnarray}
\delta\phi(\boxplus,\boxplus,\bullet|\psi_{W})=-\log 3-\dfrac{1}{2}\log a-\dfrac{1}{2}\log b=-\log\left(3{(\braket{B_{002}}{B_{002}})^{1/4}}\right)\label{eqrateW1},\\
\delta\phi(\boxplus,\bullet,\boxplus|\psi_{W})=-\log 3-\dfrac{1}{2}\log a-\dfrac{1}{2}\log c=-\log\left(3{(\braket{B_{020}}{B_{020}})^{1/4}}\right),\\
\delta\phi(\bullet,\boxplus,\boxplus|\psi_{W})=-\log 3-\dfrac{1}{2}\log b-\dfrac{1}{2}\log c=-\log\left(3{(\braket{B_{200}}{B_{200}})^{1/4}}\right),\label{eqrateW3}
\end{eqnarray}
where we have see the explicit correspondence with the LU invariants (see section \ref{secCovstaInv}) constructed from the covariants of the state $\psi_W$, thus, this rates are unitarily invariant. We can use equations \eqref{eqrateW1}-\eqref{eqrateW3} to solve for $\log a, \log b$ and $\log c$
\begin{eqnarray}
\log a=\delta\phi(\bullet,\boxplus,\boxplus|\psi_{W})-\delta\phi(\boxplus,\bullet,\boxplus|\psi_{W})-\delta\phi(\boxplus,\boxplus,\bullet|\psi_{W})-\log 3,\\
\log b=\delta\phi(\boxplus,\bullet,\boxplus|\psi_{W})-\delta\phi(\bullet,\boxplus,\boxplus|\psi_{W})-\delta\phi(\boxplus,\boxplus,\bullet|\psi_{W})-\log 3,\\
\log c=\delta\phi(\boxplus,\boxplus,\bullet|\psi_{W})-\delta\phi(\boxplus,\bullet,\boxplus|\psi_{W})-\delta\phi(\bullet,\boxplus,\boxplus|\psi_{W})-\log 3,
\end{eqnarray}
and replace in equation \eqref{ecWabc} to obtain
\begin{equation}
\delta\phi(\bar{\alpha},\bar{\beta},\bar{\gamma}|\psi_{W})=\Delta\bar{\alpha}\delta\phi(\bullet,\boxplus,\boxplus|\psi_{W})+\Delta\bar{\beta}\delta\phi(\boxplus,\bullet,\boxplus|\psi_{W})+\Delta\bar{\gamma}\delta\phi(\boxplus,\boxplus,\bullet|\psi_{W}).
\end{equation}

A remarkable feature of this result is that the (relative) rate is expressed as a convex combination of the rates at the vertex of the plane $\alpha_2+\beta_2+\gamma_2=n$, which are LU invariant. In the following section we will analyse the vertex corresponding to the origin in the $(\alpha,\beta,\gamma)$ space and then we will focus on the more general case and find the asymptotics of \eqref{eqbeast}.
\subsubsection{The origin and the facets $\alpha=\beta+\gamma$, $\beta=\alpha+\gamma$, $\gamma=\alpha+\beta$}
In the case where $\alpha=\beta=\gamma=0$ we have that the inner product is
\begin{equation}
\braket{\Phi_{\psi_W}}{\Phi_{\psi_W}}=\eta_{000}^2d^{n}\sum_{k_1,k_2,k_3}^{n}\dfrac{n!^2}{k_1!^2k_2!^2k_3!^2(n-k_1-k_2-k_3)!^2}\dfrac{a^{k_1}b^{k_2}c^{k_3}}{d^{k_1+k_2+k_3}\binom{n}{k_1}\binom{n}{k_2}\binom{n}{k_3}},
\end{equation}
whose asymptotic limit can be expressed in terms of Shannon entropy and relative entropy as
\begin{multline}
\lim\limits_{n\to\infty}\dfrac{\log \braket{\Phi_{\psi_W}}{\Phi_{\psi_W}}|_{(\bullet,\bullet,\bullet)} }{n}=\sup_{k_1,k_2,k_3}\left(2H(\bar{k}_i,1-\bar{k})-H(\bar{k}_1,1-\bar{k}_1)-H(\bar{k}_2,1-\bar{k}_2)\right.\\
-H(\bar{k}_3,1-\bar{k}_3)\left.+\bar{k}_1\log a+\bar{k}_2\log b+\bar{k}_3\log c+(1-\bar{k})\log d\right),
\end{multline}
where $k=k_1+k_2+k_3$, which can be simplified as
\begin{multline}
\lim\limits_{n\to\infty}\dfrac{\log \braket{\Phi_{\psi_W}}{\Phi_{\psi_W}}|_{(\bullet,\bullet,\bullet)} }{n}=\sup_{k_1,k_2,k_3}
\left(H(\bar{k}_i,1-\bar{k})-H(\bar{k}_1,1-\bar{k}_1)-H(\bar{k}_2,1-\bar{k}_2)\right.\\
-H(\bar{k}_3,1-\bar{k}_3)\left.-D(\bar{k}_i,1-\bar{k}||a,b,c,d)\right).
\end{multline}
Whose asymptotics will be discussed in general in the next section. Notice that the inner product in this case is equal to
\begin{equation}
\braket{\Phi_{\psi_W}}{\Phi_{\psi_W}}|_{(\bullet,\bullet,\bullet)}=\braket{A_{111}^n}{A_{111}^n},
\end{equation}
which can be expanded as an algebraic combination of LU invariants, for example \cite{Toumazet_2006}
\begin{equation}
\braket{A_{111}^2}{A_{111}^2}=\braket{A_{111}}{A_{111}}^2-(\braket{B_{200}}{B_{200}}+\braket{B_{020}}{B_{020}}+\braket{B_{002}}{B_{002}}),
\end{equation}
and for higher powers the expressions will be more intricate.
Analogously we obtain previously that
\begin{equation}
\braket{\Phi_{\psi_W}}{\Phi_{\psi_W}}|_{(\bullet,\boxplus,\boxplus)}=\braket{B_{200}^{n/2}}{B_{200}^{n/2}},
\end{equation}
\begin{equation}
\braket{\Phi_{\psi_W}}{\Phi_{\psi_W}}|_{(\boxplus,\bullet,\boxplus)}=\braket{B_{020}^{n/2}}{B_{020}^{n/2}},
\end{equation}
\begin{equation}
\braket{\Phi_{\psi_W}}{\Phi_{\psi_W}}|_{(\boxplus,\boxplus,\bullet)}=\braket{B_{002}^{n/2}}{B_{002}^{n/2}},
\end{equation}
which are again, LU invariant and can be expressed in terms of algebraic combinations of the invariants defined in \ref{secCovstaInv}. In the case of the facets ($\alpha=\beta+\gamma$, $\beta=\alpha+\gamma$ and $\gamma=\alpha+\beta$) we will have for the covariant 
\begin{equation}
\Phi_{\psi_W}=A_{111}^{n-2(\beta+\gamma)}B_{020}^{\gamma}B_{002}^{\beta},
\end{equation}
thus we have that
\begin{eqnarray}
\braket{\Phi_{\psi_W}}{\Phi_{\psi_W}}|_{(\beta+\gamma,\beta,\gamma)}=\braket{A_{111}^{n-2(\beta+\gamma)}B_{020}^{\gamma}B_{002}^{\beta}}{A_{111}^{n-2(\beta+\gamma)}B_{020}^{\gamma}B_{002}^{\beta}},\\
\braket{\Phi_{\psi_W}}{\Phi_{\psi_W}}|_{(\alpha,\alpha+\gamma,\gamma)}=\braket{A_{111}^{n-2(\alpha+\gamma)}B_{200}^{\gamma}B_{002}^{\alpha}}{A_{111}^{n-2(\alpha+\gamma)}B_{200}^{\gamma}B_{002}^{\alpha}},\\
\braket{\Phi_{\psi_W}}{\Phi_{\psi_W}}|_{(\alpha,\beta,\alpha+\beta)}=\braket{A_{111}^{n-2(\alpha+\beta)}B_{200}^{\beta}B_{020}^{\alpha}}{A_{111}^{n-2(\alpha+\beta)}B_{200}^{\beta}B_{020}^{\alpha}}.
\end{eqnarray}
where it reads explicitly
\begin{multline}
\dfrac{\braket{\Phi_{\psi_W}}{\Phi_{\psi_W}}|_{(\beta+\gamma,\beta,\gamma)}}{\eta_{(\beta+\gamma)\beta\gamma}^2}=a^{\beta+\gamma}b^{\beta}c^{\gamma}d^{n-2(\beta+\gamma)}\\
\times\sum_{k_1,k_2,k_3}^{n-2(\beta+\gamma)}\dfrac{(n-2(\beta+\gamma))!^2}{k_1!^2k_2!^2k_3!^2(n-2(\beta+\gamma)-k_1-k_2-k_3)!^2}\\
\times\left(\dfrac{a}{d}\right)^{k_1}\left(\dfrac{b}{d}\right)^{k_2}\left(\dfrac{c}{d}\right)^{k_3}
\dfrac{1}{\binom{n-2(\beta+\gamma)}{k_1}\binom{n-2\beta}{k_2}\binom{n-2\gamma}{k_3}},
\end{multline}
whose asymptotics will be the concern of the next section.
\subsubsection{The asymptotics of $\braket{\Phi_{\psi_{W}}}{\Phi_{\psi_{W}}}$}
\begin{figure}[h]
	\centering
	\begin{tabular}{c}
		\includegraphics[scale=0.3]{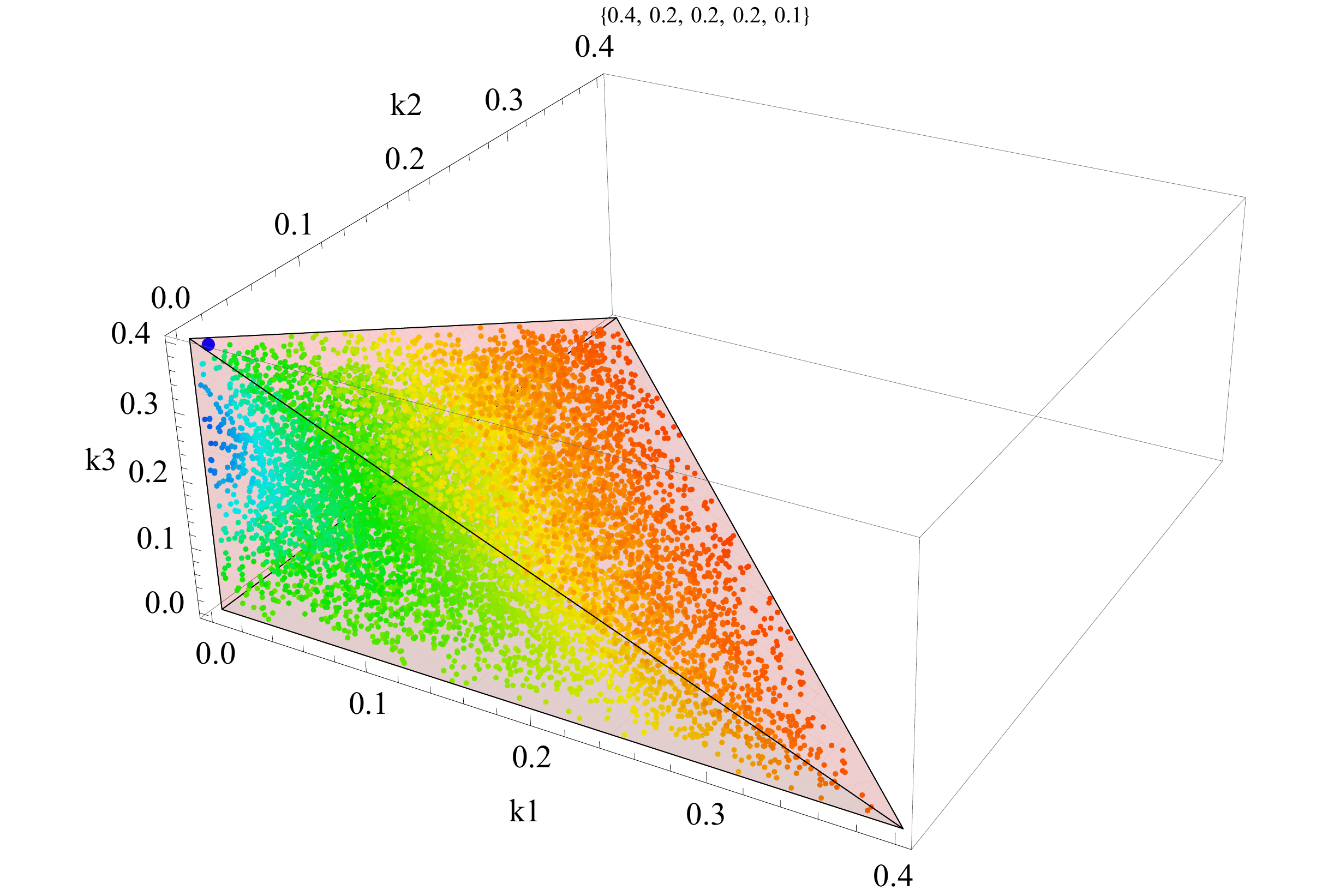}\\
		 (a) $\alpha=\beta=\gamma=0.2$,$a=b=0.1,c=0.7$
	\end{tabular}
	\caption{Numerical simulations for $\braket{\Phi_{\psi_{W}}}{\Phi_{\psi_{W}}}$, the cold colors indicate values close to the maximum in the region and hot colors values far from the maximum.}
	\label{fignumerasymp}
\end{figure}

In this section we will see the behaviour of $\braket{\Phi_{\psi_{W}}}{\Phi_{\psi_{W}}}$ in the asymptotic regime. The approach we will follow is to maximize the sum over  $k_1,k_2,k_3$ using the Laplace method ( see \cite{MacKay_2002}). We will begin by calculating the 
supremum of the logarithm of the term in the sum
\begin{equation}
s_{k_1,k_2,k_3}=\log\left(\dfrac{(n-\alpha-\beta-\gamma)!^{2}}{k_1!^2k_2!^2k_3!^2(n-\alpha-\beta-\gamma-k_1-k_2-k_3)!^2}\dfrac{\tilde{a}^{k_3}\tilde{b}^{k_2}\tilde{c}^{k_1}}{\binom{n-2\alpha}{k_1}\binom{n-2\beta}{k_2}\binom{n-2\gamma}{k_3}}\right),
\end{equation}
where we have used the shorthand notation $\tilde{x}=x/d$ for $x=a,b,c$. Using Stirling's approximation we get ($\lambda=\alpha+\beta+\gamma$ and $k=k_1+k_2+k_3$)
\begin{multline}
s_{k_1,k_2,k_3}\approx 2(n-\lambda)\log(n-\lambda)-2(n-\lambda)-2k_1\log(k_1)+2k_1-2k_2\log(k_2)+2k_2-2k_3\log(k_3)+2k_3\\
-2(n-\lambda-k)\log(n-\lambda-k)+2(n-\lambda-k)+k_1\log(\tilde{a})+k_2\log(\tilde{b})+k_3\log(\tilde{c})\\
-(n-2\alpha)\log(n-2\alpha)+(n-2\alpha)-(n-2\beta)\log(n-2\beta)+(n-2\beta)-(n-2\gamma)\log(n-2\gamma)+(n-2\gamma)\\
+k_1\log(k_1)-k_1+k_2\log(k_2)-k_2+k_3\log(k_3)-k_3\\
+(n-2\alpha-k_1)\log(n-2\alpha-k_1)-(n-2\alpha-k_1)+(n-2\beta-k_2)\log(n-2\beta-k_2)-(n-2\beta-k_2)\\
+(n-2\gamma-k_3)\log(n-2\gamma-k_3)-(n-2\gamma-k_3),
\end{multline}
which simplifies to
\begin{multline}\label{eqstriling}
s_{k_1,k_2,k_3}\approx 2(n-\lambda)\log(n-\lambda)-k_1\log(k_1)-k_2\log(k_2)-k_3\log(k_3)\\
-2(n-\lambda-k)\log(n-\lambda-k)+k_1\log(\tilde{a})+k_2\log(\tilde{b})+k_3\log(\tilde{c})\\
-(n-2\alpha)\log(n-2\alpha)-(n-2\beta)\log(n-2\beta)-(n-2\gamma)\log(n-2\gamma)\\
+(n-2\alpha-k_1)\log(n-2\alpha-k_1)+(n-2\beta-k_2)\log(n-2\beta-k_2)+(n-2\gamma-k_3)\log(n-2\gamma-k_3)\\
+\mu(n-\lambda-k),
\end{multline}
where we have included the Lagrange multiplier $\mu$ to account for the restriction $k_1+k_2+k_3\leq n-\alpha-\beta-\gamma$. Taking the gradient equal to zero we obtain three equations for $k_1,k_2,k_3$ 
\begin{eqnarray}
-\log(k_1)+2\log(n-\lambda-k)+\log(\tilde{a})-\log(n-2\alpha-k_1)-\mu=0,\\
-\log(k_2)+2\log(n-\lambda-k)+\log(\tilde{b})-\log(n-2\beta-k_2)-\mu=0,\\
-\log(k_3)+2\log(n-\lambda-k)+\log(\tilde{c})-\log(n-2\gamma-k_3)-\mu=0,
\end{eqnarray}
or in a more suitable form $(\mu\to e^{-\mu})$
\begin{eqnarray}
(n-\lambda-k)\tilde{a}\mu=k_1(n-2\alpha-k_1)\label{eqmonster1},\\
(n-\lambda-k)\tilde{b}\mu=k_2(n-2\beta-k_2),\\
(n-\lambda-k)\tilde{c}\mu=k_3(n-2\gamma-k_3)\label{eqmonster3},
\end{eqnarray}
along with the compatibility conditions
\begin{eqnarray}
k_1+k_2+k_3\leq n-\alpha-\beta-\gamma,\\
\mu(n-\lambda-k)=0,\\
\mu\geq 0.
\end{eqnarray}
We then have to consider two cases: $\mu=0$ or $(n-\lambda-k)=0$. If we consider the latter the solution to the equations will be
$k_1=\{0,n-2\alpha\},k_2=\{0,n-2\beta\},k_3=\{0,n-2\gamma\}$, where we have to choose the first root in all cases, i.e. $k_i=0$ which implies $n-\lambda=0$, a case we have already considered. Then we take $\mu=0$ and solve equations \eqref{eqmonster1}-\eqref{eqmonster3} with the condition $k_1+k_2+k_3\leq n-\alpha-\beta-\gamma$. However, the variables in this set of equations are highly correlated, which makes the analytical solution challenging and impractical for asymptotic purposes . This is why we will leave the expression as the maximization over rates such as entropies as we will see next.
\\
\\
If we take equation \eqref{eqstriling} and perform some algebraic manipulations we can obtain (see Appendix \ref{AppendixA} for details)
\begin{multline}\label{eqbeastasymp}
\lim\limits_{n\to\infty}\dfrac{s_{k_1,k_2,k_3}}{n}\sim (1-\bar{\lambda})\left[\dfrac{(1-\bar{k})}{1-\bar{\lambda}}\log(1-\bar{\lambda})+\sum_{i=1}^{3}\dfrac{\bar{k_i}}{1-\bar{\lambda}}\log a_i +H\left(\dfrac{k_i}{1-\bar{\lambda}},\dfrac{1-\bar{k}}{1-\bar{\lambda}}\right)\right]\\
-(3-2\bar{\lambda}-\bar{k})H\left(\dfrac{\Delta\bar{\alpha}_i-\bar{k_i}}{3-2\bar{\lambda}-\bar{k}}\right)+(3-2\bar{\lambda})H\left(\dfrac{\Delta\bar{\alpha}_i}{3-2\bar{\lambda}}\right)\\
-(1-\bar{\lambda}-\bar{k})\log(1-\bar{\lambda}-\bar{k})+(3-2\bar{\lambda}-\bar{k})\log(3-2\bar{\lambda}-\bar{k})-(3-2\bar{\lambda})\log(3-2\bar{\lambda}),
\end{multline}
where $\bar{x}=x/n$ and we have used the notation $\alpha_1=\alpha,\alpha_2=\beta,\alpha_3=\gamma$ and $a_1=a,a_2=b,a_3=c$. The term $\Delta\alpha_i=1-2\alpha_i$ and in all cases $i=1,2,3$. The first term in square brackets can be interpreted as a relative entropy between the normalized variables $\bar{k_i}/(1-\lambda)$ and $a_i$ which asymptotically will tend to be minimum when $\bar{k_i}/(1-\lambda)\approx a_i$. The second term is a Shannon entropy which is minimum when one of the variables attain its maximum and the other two are zero; this condition can be expressed as
\begin{eqnarray}
\Delta \alpha_i-\bar{k}_i=3-2\bar{\lambda}-\bar{k}\rightarrow \sum_{j\neq i}\bar{k}_{j}=\sum_{j\neq i}(1-2\bar{\alpha}_j),\\
\Delta \alpha_{j\neq i}-\bar{k}_{j\neq i}=0\rightarrow \bar{k}_{j\neq i}=1-2\alpha_{j\neq i},
\end{eqnarray} 
which implies $\alpha_{i}=0$ and the remaining $\alpha_{j}$ are equal. Thus, this term is only maximum when we are in the edges of the polytope that corresponds to the biseparable states. The last terms in the third line of \eqref{eqbeastasymp} are maximum when $\bar{k}=0$, i.e. $\bar{k}_i=0$ for all $\bar{\lambda}$. As we can see, it is not clear which term will dominate asymptotically for arbitrary $\alpha,\beta,\gamma$ and $a,b,c$. In the light of this intractability, we tun to numerical simulations to help us understand how \eqref{eqbeast} depends on the parameters $a,b,c$ and the triplets $\alpha,\beta,\gamma$, some cases are shown in Fig. \ref{fignumerasymp}. 
The behaviour we observed during this numerical simulations is that the maximum value attained in the permitted region for the $\bar{k_1}$ is found in the facets or edges of the region with exceptional cases in which we have exchangeable symmetry among all qubits $a=b=c=d$ and $\alpha=\beta=\gamma$. This particular case can be seen in Fig. \ref{fignumerasymp} (d). What it was easy to see is that the term corresponding to the relative entropy in \eqref{eqbeastasymp} is mostly dominant as we see in Fig \ref{fignumerasymp} (a). Note also that the maximum will also tend to a certain $\bar{k}_i$ if the associated $\alpha_i$ is larger than the others; however this effect is less appreciable, probably for the reasons we discuss previously about the second term in \eqref{eqbeastasymp} not reaching its maximum.
\\
\\
In spite of the insight of the numerical simulations, we note that the maximum follows a non-trivial behaviour and thus we are compelled to write the rate for general triplets and parameters as a maximization problem. That is
\begin{multline}\label{eqsup}
\lim\limits_{n\to\infty}-\log\dfrac{\braket{\Phi_{\psi_W}}{\Phi_{\psi_W}}}{n}=-\bar{\alpha}\log a-\bar{\beta}\log b-\bar{\gamma}\log c -(1-\bar{\alpha}-\bar{\beta}-\bar{\gamma})\log d\\
-\sup_{k_1,k_2,k_3}\bar{s}_{k_1,k_2,k_3}(\bar{\alpha},\bar{\beta},\bar{\gamma};a,b,c),
\end{multline}
and in a similar fashion for 
\begin{equation}\label{eqsupW}
\lim\limits_{n\to\infty}-\log\dfrac{\braket{\Phi_W}{\Phi_W}}{n}=\log 3 -\sup_{k_1,k_2,k_3}\bar{s}^{W}_{k_1,k_2,k_3}(\bar{\alpha},\bar{\beta},\bar{\gamma}),
\end{equation}
where 
\begin{multline}\label{eqbeastasympW}
\bar{s}^{W}_{k_1,k_2,k_3}(\bar{\alpha},\bar{\beta},\bar{\gamma})\sim (1-\bar{\lambda})H\left(\dfrac{k_i}{1-\bar{\lambda}}\right)
-(2-\bar{\lambda})H\left(\dfrac{\Delta\bar{\alpha}_i-\bar{k_i}}{2-\bar{\lambda}}\right)+(3-2\bar{\lambda})H\left(\dfrac{\Delta\bar{\alpha}_i}{3-2\bar{\lambda}}\right)\\
+(2-\bar{\lambda})\log(2-\bar{\lambda})-(3-2\bar{\lambda})\log(3-2\bar{\lambda}),
\end{multline}
with the restriction $\bar{k}=1-\bar{\lambda}$.
\\
\\
We can analyse a particular case for $\alpha,\beta,\gamma$ given by the Keyl-Werner theorem. In this case the local spectrum are $\{a,1-a\}$ for the first qubit and $\{b,1-b\},\{c,1-c\}$ for the other two qubits respectively. Thus if we choose $a_i$ such that $a_i<1-a_i$ we will have asymptotically
 \begin{multline}\label{eqbeastasympkeylw}
 \lim\limits_{n\to\infty}\dfrac{s_{k_1,k_2,k_3}}{n}\sim d\left[\dfrac{(1-\bar{k})}{d}\log d+\sum_{i=1}^{3}\dfrac{\bar{k_i}}{d}\log a_i +H\left(\dfrac{k_i}{d},\dfrac{1-\bar{k}}{d}\right)\right]\\
 -(1-2d-\bar{k})H\left(\dfrac{\Delta\bar{a}_i-\bar{k_i}}{1-2d-\bar{k}}\right)+(1-2d)H\left(\dfrac{\Delta\bar{a}_i}{1-2d}\right)\\
 -(d-\bar{k})\log(d-\bar{k})+(1-2d-\bar{k})\log(1-2d-\bar{k})-(1-2d)\log(1-2d),
 \end{multline}
which is simplified to
 \begin{multline}\label{eqbeastasympkeylw2}
 \lim\limits_{n\to\infty}\dfrac{s_{k_1,k_2,k_3}}{n}\sim dD\left(\dfrac{\bar{k}_i}{d},1-\dfrac{\bar{k}}{d}||a_i,d\right)
 -(1-2d-\bar{k})H\left(\dfrac{\Delta\bar{a}_i-\bar{k_i}}{1-2d-\bar{k}}\right)+(1-2d)H\left(\dfrac{\Delta\bar{a}_i}{1-2d}\right)\\
 -(d-\bar{k})\log(d-\bar{k})+(1-2d-\bar{k})\log(1-2d-\bar{k})-(1-2d)\log(1-2d),
 \end{multline}
where now we see explicitly the relative entropy $D(p_i||q_i)=\sum_{i}p_i\log (p_i/q_i)$. It is still however non trivial to describe its supremum although with numerical simulations it is shown that the first term is dominant to determine the triplet $(k_1^*,k_2^*,k_3^*)$ where the maximum can be found.
\\
\\

\subsection{GHZ class}
We will apply the same treatment for states in the GHZ entanglement class for which the effective Kronecker coefficient is one. Because for the general state ($g_{\alpha\beta\gamma}\geq1$) in the GHZ class this approach is intractable we will make calculations for such states that lie in the facet of the compatibility polytope which have the property $g_{\alpha\beta\gamma}=1$. 
We will begin by calculating the covariant $\Phi_{GHZ}$ for the particular state $\ket{GHZ}$, explicitly
\begin{equation}
\Phi_{GHZ}=A_{111}^{n_1}B_{200}^{n_2}B_{020}^{n_3}B_{002}^{n_4}C_{111}^{n_5}D_{000}^{n_6},
\end{equation}
which takes the form
\begin{multline}
\Phi_{GHZ}=\left(\dfrac{x_0 y_0 z_0+x_1 y_1 z_1}{\sqrt{2}}\right)^{n_1}\left(\frac{x_0 x_1}{2}\right)^{n_2}\left(\dfrac{y_0y_1}{2}\right)^{n_3}\\
\times\left(\dfrac{z_0z_1}{2}\right)^{n_4}\left(\frac{x_0 y_0 z_0-x_1 y_1 z_1}{4 \sqrt{2}}\right)^{n_5}\left(\dfrac{1}{4}\right)^{n_6},
\end{multline}
where we will take $n_5=0$ as we have seen it does not have an effect in the total state, then
\begin{equation}
\Phi_{GHZ}=\dfrac{1}{2^{2n_6+n_1/2+n_2+n_3+n_4}}\sum_{k=0}^{n_1}\binom{n_1}{k}x_0^{k+n_2}x_1^{n_1+n_2-k}y_0^{k+n_3}y_1^{n_1+n_3-k}z_0^{k+n_4}z_1^{n_1+n_4-k},
\end{equation}
as we have seen in section \ref{secKronecker} for this entanglement class we have the possibility of $g_{\alpha\beta\gamma}>1$, thus the elements on $\vec{n}$ will be given by equations \eqref{ecghz1}-\eqref{ecghz2} for a certain value of $k$ (which we will call $i$ for the remaining of the section)
\begin{equation}
\Phi_{GHZ}=\dfrac{1}{2^{n/2}}\sum_{k=0}^{n_1+2i}\binom{n_1+2i}{k}x_0^{k+n_2-i}x_1^{n_1+n_2-k+i}y_0^{k+n_3-i}y_1^{n_1+n_3-k+i}z_0^{k+n_4-i}z_1^{n_1+n_4-k+i},
\end{equation}
Translating this to a state we obtain
\begin{equation}
\ket{\Phi_{GHZ}}=\dfrac{1}{2^{n/2}}\sum_{k=0}^{n_1+2i}\binom{n_1+2i}{k}\dfrac{\ket{k+n_2-i}\ket{k+n_3-i}\ket{k+n_4-i}}{\sqrt{\binom{n_1+2n_2}{k+n_2-i}\binom{n_1+2n_3}{k+n_3-i}\binom{n_1+2n_4}{k+n_4-i}}},
\end{equation}
which in terms of the triplet $(\alpha,\beta,\gamma)$ with $\lambda=\alpha+\beta+\gamma$ is
\begin{multline}
\ket{\Phi_{GHZ}}=\dfrac{1}{2^{n/2}}\sum_{k=0}^{n-\lambda+2i}\binom{n-\lambda+2i}{k}\\
\times\dfrac{\ket{k+(\beta+\gamma-\alpha)/2-i}\ket{k+(\alpha+\gamma-\beta)/2-i}\ket{k+(\alpha+\beta-\gamma)/2-i}}{\sqrt{\binom{n-2\alpha}{k+(\beta+\gamma-\alpha)/2-i}\binom{n-2\beta}{k+(\alpha+\gamma-\beta)/2-i}\binom{n-2\gamma}{k+(\alpha+\beta-\gamma)/2-i}}}.
\end{multline}
The inner product will have the not so trivial form
\begin{multline}
\braket{\Phi_{GHZ}}{\Phi_{GHZ}}=\dfrac{1}{2^{n}}\sum_{k=0}^{n-\lambda+2i}\sum_{k'=0}^{n-\lambda+2i'}\binom{n-\lambda+2i'}{k'}\binom{n-\lambda+2i}{k}\\
\times\dfrac{\delta_{k-i,k'-i'}}{\sqrt{\binom{n-2\alpha}{k'+(\beta+\gamma-\alpha)/2-i'}\binom{n-2\alpha}{k+(\beta+\gamma-\alpha)/2-i}\binom{n-2\beta}{k'+(\alpha+\gamma-\beta)/2-i'}}}\\
\times\dfrac{1}{\sqrt{\binom{n-2\beta}{k+(\alpha+\gamma-\beta)/2-i}\binom{n-2\gamma}{k'+(\alpha+\beta-\gamma)/2-i'}\binom{n-2\gamma}{k+(\alpha+\beta-\gamma)/2-i}}},
\end{multline}
where the triplet $(\alpha,\beta,\gamma)$ is such that $g_{\alpha\beta\gamma}=1$. One of the cases where this happens is when  one of the diagrams has only boxes in the first row which we will denote by $\bullet$ ($\alpha=0$). In this case $g_{\bullet\beta\gamma}=1$ and $i=0$,
then 
\begin{multline}
\braket{\Phi_{GHZ}}{\Phi_{GHZ}}=\dfrac{1}{2^{n}}\sum_{k=0}^{n-\beta-\gamma}\binom{n-\beta-\gamma}{k}^2\dfrac{1}{\binom{n}{k+(\beta+\gamma)/2}\binom{n-2\beta}{k+(\gamma-\beta)/2}\binom{n-2\gamma}{k+(\beta-\gamma)/2}},
\end{multline}
and expanding the binomial terms
\begin{multline}
\braket{\Phi_{GHZ}}{\Phi_{GHZ}}=\dfrac{1}{2^{n}}\sum_{k=0}^{n-\beta-\gamma}\dfrac{(n-\beta-\gamma)!^2}{k!^2(n-\beta-\gamma-k)!^2}\dfrac{(k+(\beta+\gamma)/2)!(n-k-(\beta+\gamma)/2)!}{n!}\\
\times\dfrac{(k+(\gamma-\beta)/2)!(n-2\beta-(k+(\gamma-\beta)/2))!(k+(\beta-\gamma)/2)!(n-2\gamma-(k+(\beta-\gamma)/2))!}{(n-2\beta)!(n-2\gamma)!}.
\end{multline}
The terms $(k+(\gamma-\beta)/2)!$ and $(k+(\beta-\gamma)/2)!$ when $k=0$ must be positive to have a non-diverging amplitude. Then we must demand $\beta=\gamma$ which further simplifies the expression
\begin{equation}
\braket{\Phi_{GHZ}}{\Phi_{GHZ}}=\dfrac{1}{2^{n}}\sum_{k=0}^{n-2\beta}\dfrac{(k+\beta)!(n-k-\beta)!}{n!}=\dfrac{\, _2F_1(1,\beta +1;\beta -n;-1)}{\binom{n}{\beta }}.
\end{equation}
whose asymptotics are (using the Laplace method),
\begin{equation}
\braket{\Phi_{GHZ}}{\Phi_{GHZ}}=\dfrac{1}{2^{n}}\sum_{k=0}^{n-2\beta}\dfrac{(k+\beta)!(n-k-\beta)!}{n!}\sim 2^{-2n};
\end{equation}
hence, the probability will be
\begin{equation}
p(\bullet,\beta,\beta|{GHZ})\sim f^{\beta}2^{-2n},
\end{equation}
and rate function
\begin{equation}
\phi(\bullet,\bar{\beta},\bar{\beta}|{GHZ})=\log 4 -H(\bar{\beta}),
\end{equation}
which lies in the lower part of the entanglement polytope.
\\
\\
The other case where the effective Kronecker coefficient $g_{\alpha\beta\gamma}=1$ is when one of the diagrams has the same length for the first and second row which we will denote as $\boxplus$ ($\alpha=n/2$). An assertion we will prove using theorem \ref{teo2}. If we put $\alpha=n/2$ in theorem \ref{teo2} we obtain that for the Kronecker coefficient to be non-zero $\beta+\gamma\geq n/2+2i$ and $\beta+\gamma\leq n/2+2i$ for some $i\geq0$. Since $\beta$ and $\gamma$ are fixed, this condition is only fulfilled once, hence $g_{\boxplus\beta\gamma}=1$. Furthermore, the term we defined here as $i$ will take the value $i=\max((\beta+\gamma-n/2)/2,0)$,
\\
\\
Then the inner product simplifies to
\begin{multline}
\braket{\Phi_{GHZ}}{\Phi_{GHZ}}=\dfrac{1}{2^{n}}\sum_{k=0}^{n/2-\beta-\gamma+2i}\binom{n/2-\beta-\gamma+2i}{k}^2\\
\times\dfrac{1}{\binom{0}{k+(\beta+\gamma-n/2)/2-i}\binom{n-2\beta}{k+(n/2+\gamma-\beta)/2-i}\binom{n-2\gamma}{k+(n/2+\beta-\gamma)/2-i}},
\end{multline}
we will have two cases, $i> 0$ and $ i=0$. In the former case we see that $k=0$, then
\begin{equation}
\braket{\Phi_{GHZ}}{\Phi_{GHZ}}|_{i> 0}=
\dfrac{(n/2-\beta)!^2(n/2-\gamma)!^2}{2^{n}(n-2\beta)!(n-2\gamma)!},
\end{equation}
and in the latter case we obtain
\begin{equation}
\braket{\Phi_{GHZ}}{\Phi_{GHZ}}|_{i= 0}=\dfrac{\gamma!(n-2\beta-\gamma!)\beta!(n-2\gamma-\beta)!}{2{n}(n-2\beta)!(n-2\gamma)!},
\end{equation}
which in the asymptotic limit, the rate goes as
\begin{multline}
-\lim\limits_{n\to\infty}\log\dfrac{\braket{\Phi_{GHZ}}{\Phi_{GHZ}}|_{i> 0}}{n}= {\log 2}+(1-2\bar{\beta})H\left(\dfrac{1}{2},\dfrac{1}{2}\right)
+(1-2\bar{\gamma})H\left(\dfrac{1}{2},\dfrac{1}{2}\right),
\end{multline}
\begin{multline}
-\lim\limits_{n\to\infty}\log\dfrac{\braket{\Phi_{GHZ}}{\Phi_{GHZ}}|_{i=0}}{n}= {\log 2}+(1-2\bar{\beta})H\left(\dfrac{\bar{\gamma}}{1-2\bar{\beta}},\dfrac{1-\bar{\gamma}-2\bar{\beta}}{1-2\bar{\beta}}\right)\\
+(1-2\bar{\gamma})H\left(\dfrac{\bar{\beta}}{1-2\bar{\gamma}},\dfrac{1-\bar{\beta}-2\bar{\gamma}}{1-2\bar{\gamma}}\right).
\end{multline}
Now we will turn to a general state in the GHZ entanglement class, as we have seen it depends on five angular parameters. The covariants read 
\begin{multline}
A_{111}=\sqrt{K} \left(x_0 y_0 z_0 \cos (\delta )+e^{i \phi } \sin (\delta ) \left(x_1 \sin ({\epsilon})+x_0 \cos ({\epsilon})\right) \left(y_1 \sin (\theta )+y_0 \cos (\theta )\right)\right.\\
\times \left.\left(z_1 \sin (\varphi )+z_0 \cos (\varphi )\right)\right),
\end{multline}
\begin{equation}
B_{200}=K x_0 e^{i \phi } \sin (\delta ) \cos (\delta ) \sin (\theta ) \sin (\varphi ) \left(x_1 \sin ({\epsilon})+x_0 \cos ({\epsilon})\right),
\end{equation}
\begin{equation}
B_{020}=K y_0 e^{i \phi } \sin (\delta ) \cos (\delta ) \sin ({\epsilon}) \sin (\varphi ) \left(y_1 \sin (\theta )+y_0 \cos (\theta )\right),
\end{equation}
\begin{equation}
B_{002}=K z_0 e^{i \phi } \sin (\delta ) \cos (\delta ) \sin ({\epsilon}) \sin (\theta ) \left(z_1 \sin (\varphi )+z_0 \cos (\varphi )\right),
\end{equation}
\begin{multline}
C_{111}=\frac{1}{2} K^{3/2} e^{i \phi } \sin (\delta ) \cos (\delta ) \sin ({\epsilon}) \sin (\theta ) \sin (\varphi )\times\\
\left(x_0 y_0 z_0 \cos (\delta )-e^{i \phi } \sin (\delta ) \left(x_1 \sin ({\epsilon})+x_0 \cos ({\epsilon})\right) \left(y_1 \sin (\theta )+y_0 \cos (\theta )\right) \left(z_1 \sin (\varphi )+z_0 \cos (\varphi )\right)\right),
\end{multline}
\begin{equation}
D_{000}=K^2 \left(-e^{2 i \phi }\right) \sin ^2(2 \delta ) \sin ^2({\epsilon}) \sin ^2(\theta ) \sin ^2(\varphi ),
\end{equation}
and the general covariant will be ($n_5=0$, that is, $n$ is even)
\begin{multline}
\Phi_{\psi_{GHZ}}=\left(x_0 y_0 z_0 \cos (\delta )+e^{i \phi } \sin (\delta ) \left(x_1 \sin ({\epsilon})+x_0 \cos ({\epsilon})\right) \left(y_1 \sin (\theta )+y_0 \cos (\theta )\right) \right.\\
\left.\left(z_1 \sin (\varphi )+z_0 \cos (\varphi )\right)\right)^{n_1}\\
\times \left( x_0 e^{i \phi } \sin (\delta ) \cos (\delta ) \sin (\theta ) \sin (\varphi ) \left(x_1 \sin ({\epsilon})+x_0 \cos ({\epsilon})\right)\right)^{n_2}\\
\left(y_0 e^{i \phi } \sin (\delta ) \cos (\delta ) \sin ({\epsilon}) \sin (\varphi ) \left(y_1 \sin (\theta )+y_0 \cos (\theta )\right)\right)^{n_3}\\
\left(z_0 e^{i \phi } \sin (\delta ) \cos (\delta ) \sin ({\epsilon}) \sin (\theta ) \left(z_1 \sin (\varphi )+z_0 \cos (\varphi )\right)\right)^{n_4}\\
K^{n_1/2+n_2+n_3+n_4+2n_6}\left(\left(-e^{2 i \phi }\right) \sin ^2(2 \delta ) \sin ^2({\epsilon}) \sin ^2(\theta ) \sin ^2(\varphi )\right)^{n_6},
\end{multline}
which after expanding and regrouping terms we have
\begin{multline}
\Phi_{\psi_{GHZ}}=4^{n_6}K^{n/2}\sum_{j=0}^{n_1}\binom{n_1}{j}x_0^{n_1+n_2-j} y_0^{n_1+n_3-j} z_0^{n_1+n_4-j} (\cos (\delta )\sin(\delta))^{n_1+n_2+n_3+n_4+2n_6}\\
e^{i \phi(n_2+n_3+n_4-2n_6+ j)}\tan(\delta)^{j}\\
\times  \left(x_1 \sin ({\epsilon})+x_0 \cos ({\epsilon})\right)^{n_2+j} \left(y_1 \sin (\theta )+y_0 \cos (\theta )\right)^{n_3+j} \left(z_1 \sin (\varphi )+z_0 \cos (\varphi )\right)^{n_4+j}\\
\times \sin(\epsilon)^{n_2+n_3+2n_6}\sin(\theta)^{n_2+n_4+2n_6}\sin(\varphi)^{n_3+n_4+2n_6},
\end{multline}
and expanding further we obtain
\begin{multline}
\Phi_{\psi_{GHZ}}=4^{n_6}K^{n/2}\sum_{j=0}^{n_1}\binom{n_1}{j}x_0^{n_1+n_2-j} y_0^{n_1+n_3-j} z_0^{n_1+n_4-j} (\cos (\delta )\sin(\delta))^{n_1+n_2+n_3+n_4+2n_6}\\
e^{i \phi(n_2+n_3+n_4-2n_6+ j)}\tan(\delta)^{j}\\
\sum_{k_1=0}^{n_2+j}\binom{n_2+j}{k_1}x_0^{n_2+j-k_1} \cos ({\epsilon})^{n_2+j-k_1}x_1^{k_1} \sin ({\epsilon})^{k_1} \sum_{k_2=0}^{n_3+j}\binom{n_3+j}{k_2}y_0^{n_3+j-k_2} \cos (\theta )^{n_3+j-k_2}y_1^{k_2} \sin (\theta )^{k_2}\\
\sum_{k_3=0}^{n_4+j}\binom{n_4+j}{k_3}z_0^{n_4+j-k_3} \cos (\varphi )^{n_4+j-k_3}z_1^{k_3} \sin (\varphi )^{k_3}\\
\times \sin(\epsilon)^{n_2+n_3+2n_6}\sin(\theta)^{n_2+n_4+2n_6}\sin(\varphi)^{n_3+n_4+2n_6},
\end{multline}
which translates to the state
\begin{multline}
\ket{\Phi_{\psi_{GHZ}}}=\eta_{\alpha\beta\gamma}4^{n_6}K^{n/2}\sum_{j=0}^{n_1}\binom{n_1}{j} (\cos (\delta )\sin(\delta))^{n/2+n_1/2}e^{i \phi(n_2+n_3+n_4-2n_6+ j)}\tan(\delta)^{j}\\
\sum_{k_1=0}^{n_2+j}\binom{n_2+j}{k_1} \cos ({\epsilon})^{n_2+j-k_1} \sin ({\epsilon})^{k_1} \sum_{k_2=0}^{n_3+j}\binom{n_3+j}{k_2} \cos (\theta )^{n_3+j-k_2} \sin (\theta )^{k_2}\\
\sum_{k_3=0}^{n_4+j}\binom{n_4+j}{k_3} \cos (\varphi )^{n_4+j-k_3}\sin (\varphi )^{k_3}
\\\times \sin(\epsilon)^{n_2+n_3+2n_6}\sin(\theta)^{n_2+n_4+2n_6}\sin(\varphi)^{n_3+n_4+2n_6}
\times\dfrac{\ket{j_1,m_1}\ket{j_2,m_2}\ket{j_3,m_3}}{\sqrt{\binom{n_1+2n_2}{k_1}\binom{n_1+2n_3}{k_2}\binom{n_1+2n_4}{k_3}}},
\end{multline}
or in terms of the triplet $(\alpha,\beta,\gamma)$ (taking into account the $i$ introduced in the GHZ case) for $g_{\alpha,\beta,\gamma}=1$
\begin{multline}
\ket{\Phi_{\psi_{GHZ}}}=\eta_{\alpha\beta\gamma}4^{i}K^{n/2}\sum_{j=0}^{n-\alpha-\beta-\gamma+2i}\binom{n-\alpha-\beta-\gamma+2i}{j} (\cos (\delta )\sin(\delta))^{n-\alpha/2-\beta/2-\gamma/2+i}\\
e^{i \phi(\alpha+\beta+\gamma-5i+ j)}\tan(\delta)^{j}\\
\sum_{k_1=0}^{(\beta+\gamma-\alpha)/2+j-i}\binom{(\beta+\gamma-\alpha)/2+j-i}{k_1} \cos ({\epsilon})^{(\beta+\gamma-\alpha)/2+j-i-k_1} \sin ({\epsilon})^{k_1} \\
\sum_{k_2=0}^{(\alpha+\gamma-\beta)/2+j-i}\binom{(\alpha+\gamma-\beta)/2+j-i}{k_2} \cos (\theta )^{(\alpha+\gamma-\beta)/2+j-i-k_2} \sin (\theta )^{k_2}\\
\sum_{k_3=0}^{(\alpha+\beta-\gamma)/2+j-i}\binom{(\alpha+\beta-\gamma)/2+j-i}{k_3} \cos (\varphi )^{(\alpha+\beta-\gamma)/2+j-i-k_3}\sin (\varphi )^{k_3}
\\\times \sin(\epsilon)^{\gamma}\sin(\theta)^{\beta}\sin(\varphi)^{\alpha}
\times\dfrac{\ket{j_1,m_1}\ket{j_2,m_2}\ket{j_3,m_3}}{\sqrt{\binom{n-2\alpha}{k_1}\binom{n-2\beta}{k_2}\binom{n-2\gamma}{k_3}}}.
\end{multline}
This monstrous expression will be simplified in a few particular cases of interest. The first case will the triplet $(\bullet,\beta,\beta)$ where we have $i=0$ and hence
\begin{multline}
\ket{\Phi_{\psi_{GHZ}}}=\eta_{\bullet\beta\beta}K^{n/2}\sum_{j=0}^{n-2\beta}\binom{n-2\beta}{j} (\cos (\delta )\sin(\delta))^{n-\beta}e^{i \phi(2\beta+j)}\tan(\delta)^{j}\\
\sum_{k_1=0}^{\beta+j}\binom{\beta+j}{k_1} \cos ({\epsilon})^{\beta+j-k_1} \sin ({\epsilon})^{k_1} \\
\sum_{k_2=0}^{j}\binom{j}{k_2} \cos (\theta )^{j-k_2} \sin (\theta )^{k_2}\\
\sum_{k_3=0}^{j}\binom{j}{k_3} \cos (\varphi )^{j-k_3}\sin (\varphi )^{k_3}
\\\times \sin(\epsilon)^{\beta}\sin(\theta)^{\beta}
\times\dfrac{\ket{j_1,m_1}\ket{j_2,m_2}\ket{j_3,m_3}}{\sqrt{\binom{n}{k_1}\binom{n-2\beta}{k_2}\binom{n-2\beta}{k_3}}}.
\end{multline}
The inner product in the origin of the polytope takes the form
\begin{equation}
\braket{\Phi_{\psi_{GHZ}}}{\Phi_{\psi_{GHZ}}}|_{\bullet,\bullet,\bullet}=\braket{A_{111}^{n}}{A_{111}^n},
\end{equation}
where te explicit expression for the inner product is rather complicated. However in the limit we expect that the geometric entanglement is obtained.
\\
The really interesting case will be when one of the diagrams is a box $\alpha=\boxplus$ in which case we have (see that $i=\max((\beta+\gamma-n/2)/2,0)$ from theorem \ref{teo2})
\begin{multline}
\ket{\Phi_{\psi_{GHZ}}}=\eta_{\boxplus\beta\gamma}4^{i}K^{n/2}\sum_{j=0}^{n/2-\beta-\gamma+2i}\binom{n/2-\beta-\gamma+2i}{j} (\cos (\delta )\sin(\delta))^{n-n/4-\beta/2-\gamma/2+i}\\
e^{i \phi(n/2+\beta+\gamma-5i+ j)}\tan(\delta)^{j}\\
\sum_{k_1=0}^{(\beta+\gamma-n/2)/2+j-i}\binom{(\beta+\gamma-n/2)/2+j-i}{k_1} \cos ({\epsilon})^{(\beta+\gamma-n/2)/2+j-i-k_1} \sin ({\epsilon})^{k_1} \\
\sum_{k_2=0}^{(n/2+\gamma-\beta)/2+j-i}\binom{(n/2+\gamma-\beta)/2+j-i}{k_2} \cos (\theta )^{(n/2+\gamma-\beta)/2+j-i-k_2} \sin (\theta )^{k_2}\\
\sum_{k_3=0}^{(n/2+\beta-\gamma)/2+j-i}\binom{(n/2+\beta-\gamma)/2+j-i}{k_3} \cos (\varphi )^{(n/2+\beta-\gamma)/2+j-i-k_3}\sin (\varphi )^{k_3}
\\\times \sin(\epsilon)^{\gamma}\sin(\theta)^{\beta}\sin(\varphi)^{n/2}
\times\dfrac{\ket{j_1,m_1}\ket{j_2,m_2}\ket{j_3,m_3}}{\sqrt{\binom{0}{k_1}\binom{n-2\beta}{k_2}\binom{n-2\gamma}{k_3}}},
\end{multline}
let us first analyse the case $i=(\beta+\gamma-n/2)/2$ (the upper region of the compatibility polytope) where we obtain the simplified expression
\begin{multline}
\ket{\Phi_{\psi_{GHZ}}}|_{i>0}=\eta_{\boxplus\beta\gamma}4^{i}K^{n/2}(\cos (\delta )\sin(\delta))^{n/2}e^{i \phi(n/2+\beta+\gamma-5i)}\\
\sum_{k_2=0}^{n/2-\beta}\binom{n/2-\beta}{k_2} \cos (\theta )^{n/2-\beta-k_2} \sin (\theta )^{k_2}\\
\sum_{k_3=0}^{n/2-\gamma}\binom{n/2-\gamma}{k_3} \cos (\varphi )^{n/2-\gamma-k_3}\sin (\varphi )^{k_3}
\\\times \sin(\epsilon)^{\gamma}\sin(\theta)^{\beta}\sin(\varphi)^{n/2}
\times\dfrac{\ket{j_1,m_1}\ket{j_2,m_2}\ket{j_3,m_3}}{\sqrt{\binom{n-2\beta}{k_2}\binom{n-2\gamma}{k_3}}},
\end{multline}
and for the inner product
\begin{multline}
\braket{\Phi_{\psi_{GHZ}}}{\Phi_{\psi_{GHZ}}}|_{i>0}=(\eta_{\boxplus\beta\gamma})^2 4^{\beta+\gamma-n/2}K^{n}(\cos (\delta )\sin(\delta))^{n}\\
\sum_{k_2=0}^{n/2-\beta}\binom{n/2-\beta}{k_2}^2 |\cos^2 (\theta )|^{n/2-\beta-k_2} |\sin^2 (\theta )|^{k_2}\\
\sum_{k_3=0}^{n/2-\gamma}\binom{n/2-\gamma}{k_3}^2 |\cos^2 (\varphi )|^{n/2-\gamma-k_3}|\sin^2 (\varphi )|^{k_3}
\\\times \sin(\epsilon)^{2\gamma}\sin(\theta)^{2\beta}\sin(\varphi)^{n}
\times\dfrac{1}{\binom{n-2\beta}{k_2}\binom{n-2\gamma}{k_3}}.
\end{multline}
Now we will analyse the rate of this inner product in the vertex of the facet of the compatibility polytope defined by $\bar{\alpha}=1/2$. For the vertex $(\boxplus,\boxplus,\boxplus)$ we obtain
\begin{equation}
\braket{\Phi_{\psi_{GHZ}}}{\Phi_{\psi_{GHZ}}}|_{i>0}=(\eta_{\boxplus\boxplus\boxplus})^2 2^{n}K^{n}(\cos (\delta )\sin(\delta))^{n} \sin(\epsilon)^{n}\sin(\theta)^{n}\sin(\varphi)^{n},
\end{equation}
a result proportional to the tangle and to the LU invariant 
\begin{equation}
\braket{D_{000}}{D_{000}}=K^4 \sin ^4( \delta )\cos^4(\delta) \sin ^4(\epsilon) \sin ^4(\theta ) \sin ^4(\varphi ),
\end{equation}
thus
\begin{equation}
-\log\braket{\Phi_{\psi_{GHZ}}}{\Phi_{\psi_{GHZ}}}|_{i>0}=-n\log 2 \braket{D_{000}}{D_{000}}^{1/4}-2\log(\eta_{\boxplus\boxplus\boxplus}),
\end{equation}
Now for the vertex $(\boxplus,\bullet,\boxplus)$ we obtain
\begin{multline}
\braket{\Phi_{\psi_{GHZ}}}{\Phi_{\psi_{GHZ}}}|_{i>0}=(\eta_{\boxplus\bullet\boxplus})^2 K^{n}(\cos (\delta )\sin(\delta))^{n}
\sum_{k_2=0}^{n/2}\binom{n/2}{k_2}^2 |\cos^2 (\theta )|^{n/2-k_2} |\sin^2 (\theta )|^{k_2}
\\\times \sin(\epsilon)^{n}\sin(\varphi)^{n}
\times\dfrac{1}{\binom{n}{k_2}},
\end{multline}
where we calculate the asymptotic limit of the sum using the Laplace method 
\begin{equation}
\sum_{k_2=0}^{n/2}\binom{n/2}{k_2}^2 |\cos^2 (\theta )|^{n/2-k_2} |\sin^2 (\theta )|^{k_2}\dfrac{1}{\binom{n}{k_2}}\sim\left(\dfrac{1+\cos\theta}{2}\right)^n.
\end{equation}

Therefore,
\begin{equation}
\braket{\Phi_{\psi_{GHZ}}}{\Phi_{\psi_{GHZ}}}|_{i>0}\approx(\eta_{\boxplus\bullet\boxplus})^2 K^{n}(\cos (\delta )\sin(\delta))^{n}
\left(\dfrac{1+\cos\theta}{2}\right)^n \sin(\epsilon)^{n}\sin(\varphi)^{n},
\end{equation}
where we compare it to the LU invariant 
\begin{equation}
\braket{B_{020}}{B_{020}}=K^2 \sin ^2(\delta ) \cos ^2(\delta )  \dfrac{(1+\cos^2 ( \theta ))}{2}\sin ^2(\epsilon) \sin ^2(\varphi ).
\end{equation}
Analogously we have that for the triplet $(\boxplus,\boxplus,\bullet)$
\begin{equation}
\braket{\Phi_{\psi_{GHZ}}}{\Phi_{\psi_{GHZ}}}|_{i>0}\approx(\eta_{\boxplus\boxplus\bullet})^2 K^{n}(\cos (\delta )\sin(\delta))^{n}
\left(\dfrac{1+\cos\varphi}{2}\right)^n \sin(\epsilon)^{n}\sin(\theta)^{n},
\end{equation}
where we compare it to the LU invariant 
\begin{equation}
\braket{B_{002}}{B_{002}}=K^2 \sin ^2(\delta ) \cos ^2(\delta )  \dfrac{(1+\cos^2 ( \varphi ))}{2}\sin ^2(\epsilon) \sin ^2(\theta ).
\end{equation}
\\
Putting everything together we have that the relative rate 
\begin{equation}\label{eqGHzyeah}
\delta \phi(\boxplus,\bar{\beta},\bar{\gamma}|\psi_{GHZ})=-\log\braket{\Phi_{\psi_{GHZ}}}{\Phi_{\psi_{GHZ}}}|_{i>0}+\log\braket{\Phi_{{GHZ}}}{\Phi_{{GHZ}}}|_{i>0},
\end{equation}
Now we will evaluate the rate at the vertex of the facet on the upper region of the polytope, that is $(\boxplus,\bullet,\boxplus)$, $(\boxplus,\boxplus,\bullet)$ and $(\boxplus,\boxplus,\boxplus)$. We then have
\begin{equation}
\delta\phi(\boxplus,\bullet,\boxplus|\psi_{GHZ})=-\log \left(K\cos (\delta )\sin(\delta)
\left(\dfrac{1+\cos\theta}{2}\right) \sin(\epsilon)\sin(\varphi)\right)-2\log 2,
\end{equation}
\begin{equation}
\delta\phi(\boxplus,\boxplus,\bullet|\psi_{GHZ})=-\log \left(K\cos (\delta )\sin(\delta)
\left(\dfrac{1+\cos\varphi}{2}\right) \sin(\epsilon)\sin(\theta)\right)-2\log 2,
\end{equation}
\begin{equation}
\delta\phi(\boxplus,\boxplus,\boxplus|\psi_{GHZ})=-\log 2K\cos (\delta )\sin(\delta) \sin(\epsilon)\sin(\theta)\sin(\varphi)-\log 2
\end{equation}
After some algebra, it can be shown that we can rewrite equation \eqref{eqGHzyeah} as 
\begin{multline}
\delta\phi(\boxplus,\bar{\beta},\bar{\gamma}|\psi_{GHZ})=(1-\Delta\bar{\beta}-\Delta\bar{\gamma})\delta\phi(\boxplus,\boxplus,\boxplus|\psi_{GHZ})+\Delta\bar{\beta}\delta\phi(\boxplus,\bullet,\boxplus|\psi_{GHZ})\\
+\Delta\bar{\gamma}\phi(\boxplus,\boxplus,\bullet|\psi_{GHZ}),
\end{multline}
where the convex structure is again elucidated.
In the next sections we will focus on an approach that will allows us to obtain the  rates $\phi(\bar{\alpha},\bar{\beta},\bar{\gamma},\psi)$ instead of the relative rate. In order to do that we will compute the probabilities explicitly by means of the Louck polynomials described in section \ref{secLouckstate}. We will calculate the asymptotic rates of states in the GHZ class and verify our calculations for the W entanglement class in the $\alpha+\beta+\gamma=n$ plane. What is remarkable is that the rates at the vertex of the polytope are given in terms of the very close expressions to LU invariants (a different invariant for each vertex). It is also noteworthy that in the facet of the polytope, the (relative) rates are expressed as a convex combination of the rates at the vertex aforementioned as we have seen in this section.
\section{Asymptotic Rates based on Louck polynomials approach}\label{secasymplouck}
In this section we will calculate the asymptotic rate (as defined in section \ref{secAsymptotictheory})  for the probabilty $p(\alpha,\beta,\gamma|\psi)$ for different types of states $\psi$ using a polynomial approach. As this will prove to be a laborious task for a general $\psi$ and general $(\alpha,\beta,\gamma)$, we will only calculate explicitly the rates at the vertex and two facets of the entanglement polytope. We will find that in most cases the asymptotic rate will be a convex combination of the asymptotic rates at the vertex of the polytope as we evidenced in the previous section. This approach will be useful to verify the results obtained with the \emph{covariant-to-state} approach. We begin with the highly symmetric GHZ state.
\subsection{The rate for the GHZ state}\label{secloukghz}
In this section we will calculate the rate for the GHZ entanglement class. We know a general state in this class can be written in the form
\begin{equation}
\ket{\psi_{GHZ}}=A\otimes B\otimes C\ket{GHZ},
\end{equation}
where
\begin{equation}
A\otimes B\otimes C=\sqrt{2K}\left(\begin{array}{cc}
c_\delta & s_\delta c_\epsilon e^{i\phi}\\
0 & s_\delta s_\epsilon e^{i\phi}
\end{array}\right)\otimes \left(\begin{array}{cc}
1 & c_\theta\\
0 & s_\theta
\end{array}\right) \otimes \left(\begin{array}{cc}
1 & c_\varphi\\
0 & s_\varphi
\end{array}\right).
\end{equation}
Thus the $n$-th tensor power is
\begin{equation}
\ket{\psi_{GHZ}}^{\otimes n}=A^{\otimes n}\otimes B^{\otimes n}\otimes C^{\otimes n}\ket{GHZ}^{\otimes n}.
\end{equation}
We can write the expression for $\ket{GHZ}^{\otimes n}$ in terms of sequences as
\begin{equation}
\ket{GHZ}^{\otimes n}=\left(\dfrac{\ket{000}+\ket{111}}{\sqrt{2}}\right)^{\otimes n}=\dfrac{1}{2^{n/2}}\sum_{s}\ket{s}\ket{s}\ket{s},
\end{equation}
where the sum is over all the sequences $s$ of length $n$. Now we apply the Schur transform to each of the sequences
\begin{equation}
\ket{GHZ}^{\otimes}=2^{-n/2}\sum_{\alpha,\beta,\gamma}\sum_{m_1,m_2,m_3}\sum_{\mu_1,\mu_2,\mu_3}\sum_{s}\braket{\alpha, m_1, \mu_1}{s}\braket{\beta, m_2, \mu_2}{s}\braket{\gamma, m_3, \mu_3}{s}\ket{\vec{\lambda},\vec{m},\vec{\mu}},
\end{equation}
where $\ket{\vec{\lambda},\vec{m},\vec{\mu}}=\ket{\alpha,m_1,\mu_1}\ket{\beta,m_2,\mu_2}\ket{\gamma,m_3,\mu_3}$. This product can be simplified by the fact that the only non-zero terms will be the ones in which $s\sim m_1$, $s\sim m_2$ and $s\sim m_3$, thus $m_1=m_2=m_3=m$
\begin{equation}
\ket{GHZ}^{\otimes}=2^{-n/2}\sum_{\alpha,\beta,\gamma}\sum_{m}\sum_{\mu_1,\mu_2,\mu_3}\sum_{s\sim m}\braket{\alpha, m, \mu_1}{s}\braket{\beta, m, \mu_2}{s}\braket{\gamma, m, \mu_3}{s}\ket{\vec{\lambda},\vec{m},\vec{\mu}}.
\end{equation}
Then the probability to be in a representation $(\alpha,\beta,\gamma)$ will be given by
\begin{equation}
p(\alpha,\beta,\gamma|\psi_{GHZ})=2^{-n}\sum_{m,m'}D^{\alpha}(A^{\dagger}A)_{m'm}D^{\beta}(B^{\dagger}B)_{m'm}D^{\gamma}(C^{\dagger}C)_{m'm}R^{\alpha\beta\gamma}_{m,m'}.
\end{equation}
where
\begin{equation}
R^{\alpha\beta\gamma}_{m,m'}=\sum_{s\sim m}\sum_{s'\sim m'}\sum_{\mu_1,\mu_2,\mu_3}\braket{\alpha, m, \mu_1}{s}\braket{\beta, m, \mu_2}{s}\braket{\gamma, m, \mu_3}{s}\braket{s'}{\alpha, m', \mu_1}\braket{s'}{\beta, m', \mu_2}\braket{s'}{\gamma, m', \mu_3},
\end{equation}
which can be compactly written using \eqref{eqLouckident1} as
\begin{equation}
R^{\alpha\beta\gamma}_{m,m'}=f^{\alpha}f^{\beta}f^{\gamma}\sum_{s\sim m}\sum_{s'\sim m'}C^{\alpha}(\Omega)_{m',m}C^{\beta}(\Omega)_{m',m}C^{\gamma}(\Omega)_{m',m},
\end{equation}
where $\Omega=W(s'\circ s)$ the weight matrix of the bisequence $s'\circ s$. Alternatively we can replace the sum over the sequences for the sum over the matrix $\Omega$, thus
\begin{equation}
R^{\alpha\beta\gamma}_{m,m'}=f^{\alpha}f^{\beta}f^{\gamma}\sum_{\Omega}\binom{n}{\Omega}C^{\alpha}(\Omega)_{m',m}C^{\beta}(\Omega)_{m',m}C^{\gamma}(\Omega)_{m',m},
\end{equation}
where the multinomial factor was added to count the number of bisequences that produce the same $\Omega$. The quantity $R^{\alpha\beta\gamma}_{m,m'}$ is in general unmanageable in the asymptotic regime. However, there are some cases in which it can be calculated with some effort, for example, when one of the Young frames is $(n,0)$ which we will denote as $\bullet$. In this case,
\begin{equation}
C^{\bullet}(\Omega)_{m,m'}=\dfrac{1}{\sqrt{\binom{n}{m}\binom{n}{m'}}},
\end{equation}
so
\begin{equation}
R^{\bullet\beta\gamma}_{m,m'}=\dfrac{f^{\beta}f^{\gamma}}{\sqrt{\binom{n}{m}\binom{n}{m'}}}\sum_{\Omega}\binom{n}{\Omega}C^{\beta}(\Omega)_{m',m}C^{\gamma}(\Omega)_{m',m}=\delta_{\beta,\gamma}\dfrac{f^{\beta}}{\sqrt{\binom{n}{m}\binom{n}{m'}}}.
\end{equation}
The other case in which $R^{\alpha\beta\gamma}_{m,m'}$ simplifies is when one of the Young frames is $(n/2,n/2)$, which we will denote as $\boxplus$. Note that in this case we require $j=0$ then $m'=m=0$. It is shown in Appendix \ref{AppendixA} (after many pages of algebra and polynomial identities) that asymptotically we can write
\begin{equation}
\lim\limits_{n\to\infty}R^{\boxplus\beta\gamma}_{0,0}\sim\exp\left[\dfrac{n}{2}H\left(\bar{\gamma}_1\bar{\gamma}_2+\bar{\beta}_1^2-\frac{1}{4},\bar{\gamma}_1\bar{\gamma}_2+\bar{\beta}_2^2-\frac{1}{4},\bar{\beta}_1\bar{\beta}_2+\bar{\gamma}_1^2-\frac{1}{4} , \bar{\beta}_1\bar{\beta}_2+\bar{\gamma}_2^2-\frac{1}{4}\right) \right],
\end{equation} 
where $H$ is the Shannon entropy defined as
\begin{equation}
H({p_i})=-\sum_i p_i \log p_i,
\end{equation}
and $\bar{\lambda}=\lambda/n$ are the normalized partitions. In order for the rate of the probability to be complete, we must also calculate the rate for the representation matrices $D^{\boxplus}(A^{\dagger}A)_{0,0}$, $D^{\beta}(B^{\dagger}B)_{0,0}$ and $D^{\gamma}(C^{\dagger}C)_{0,0}$. To see the detailed calculations refer to Appendix \ref{ApRep}. After some algebra we obtain the rate of a state in the GHZ entanglement class
\begin{multline}
\phi(\boxplus,\bar{\beta},\bar{\gamma}|\psi_{GHZ})=\log 2-\log|\det (A^{\dagger}A)|-2\bar{\beta}_2\log|\det(B^{\dagger}B)|-2\bar{\gamma}_2\log|\det(C^{\dagger}C)|\\
-\Delta\bar{\beta}\log(1+c_{\theta})-\Delta\bar{\gamma}\log(1+c_{\varphi})\\
-\dfrac{1}{2}H\left(\bar{\gamma}_1\bar{\gamma}_2+\bar{\beta}_1^2-\frac{1}{4},\bar{\gamma}_1\bar{\gamma}_2+\bar{\beta}_2^2-\frac{1}{4},\bar{\beta}_1\bar{\beta}_2+\bar{\gamma}_1^2-\frac{1}{4} , \bar{\beta}_1\bar{\beta}_2+\bar{\gamma}_2^2-\frac{1}{4}\right),
\end{multline}
where $\Delta \bar{\lambda}=\bar{\lambda}_1-\bar{\lambda}_2$. Note that this result is slightly different from the one obtain in previous sections.
With this formula for the asymptotic rate we can calculate the rates at the vertices of the polytope (except for the origin).
\\
For $\beta=\gamma=\boxplus$ we obtain
\begin{multline}
\phi(\boxplus,\boxplus,\boxplus|\psi_{GHZ})=\log 2-\log|\det (A^{\dagger}A)|-\log|\det(B^{\dagger}B)|-\log|\det(C^{\dagger}C)|
-\dfrac{1}{2}H\left(\frac{1}{4},\frac{1}{4},\frac{1}{4} , \frac{1}{4}\right),
\end{multline}
which simplifies to
\begin{eqnarray}
\phi(\boxplus,\boxplus,\boxplus|\psi_{GHZ})&=&-\log\left(|\det (A^{\dagger}A)||\det(B^{\dagger}B)||\det(C^{\dagger}C)|\right),\\
&=&-2\log(2Ks_{\epsilon}s_{\theta}s_{\varphi}c_\delta s_\delta),\\
&=&-\log(\tau_3)=-\log(4|D_{000}|),
\end{eqnarray}
where we have obtained the 3-tangle. This is a remarkable result since it is manifestly SLOCC invariant.
\\
\\
Now we will consider the case $\beta=\boxplus$ and $\gamma=\bullet$ where we have 
\begin{equation}
\phi(\boxplus,\boxplus,\bullet|\psi_{GHZ})=\log 2-\log\left(|\det (A^{\dagger}A)||\det(B^{\dagger}B)|\right)-\log(1+c_\varphi)-\dfrac{1}{2}H\left(0,0,1,0\right),
\end{equation}
which simplifies to
\begin{equation}
\phi(\boxplus,\boxplus,\bullet|\psi_{GHZ})=-\log\left(2^2K^2 s_\delta^2c_\delta^2s_\epsilon^2s_\theta^2\dfrac{(1+c_\varphi)}{2}\right),
\end{equation}
which is to be compared with the invariant $\braket{B_{002}}{B_{002}}=K^2 s_\delta^2c_\delta^2s_\epsilon^2s_\theta^2\frac{(1+c_\varphi^2)}{2}$, where the only difference is in the $\cos(\theta)^2$. This is again a surprising result since it is almost an LU invariant. Note that in order to obtain this result we had to perform a limit (see Appendix \ref{AppendixA}, equation \eqref{limitLU}) and a Laplace approximation where we may have lost the LU invariance. Analogously
\begin{equation}
\phi(\boxplus,\bullet,\boxplus|\psi_{GHZ})=-\log\left(2^2K^2 s_\delta^2c_\delta^2s_\epsilon^2s_\varphi^2\dfrac{(1+c_\theta)}{2}\right),
\end{equation}
which is to be compared with $\braket{B_{020}}{B_{020}}=2^2K^2 s_\delta^2c_\delta^2s_\epsilon^2s_\varphi^2\frac{(1+c_\theta^2)}{2}$. Furthermore we can calculate the rates for a particular case of interest, namely the state $\ket{GHZ}$ we obtain
\begin{multline}
\phi(\boxplus,\bar{\beta},\bar{\gamma}|GHZ)=\log 2\\
-\dfrac{1}{2}H\left(\bar{\gamma}_1\bar{\gamma}_2+\bar{\beta}_1^2-\frac{1}{4},\bar{\gamma}_1\bar{\gamma}_2+\bar{\beta}_2^2-\frac{1}{4},\bar{\beta}_1\bar{\beta}_2+\bar{\gamma}_1^2-\frac{1}{4} , \bar{\beta}_1\bar{\beta}_2+\bar{\gamma}_2^2-\frac{1}{4}\right),\\
\end{multline}
\begin{eqnarray}
\phi(\boxplus,\boxplus,\boxplus|GHZ)=0,\\
\phi(\boxplus,\bullet,\boxplus|GHZ)=\log 2,\\
\phi(\boxplus,\boxplus,\bullet|GHZ)=\log 2.
\end{eqnarray}
This allows us to write the rate $\phi(\boxplus,\bar{\beta},\bar{\gamma}|\psi_{GHZ})$ as a convex combination
\begin{multline}
\phi(\boxplus,\bar{\beta},\bar{\gamma}|\psi_{GHZ})=\phi(\boxplus,\bar{\beta},\bar{\gamma}|GHZ)+(1-\Delta\bar{\beta}-\Delta\bar{\gamma})\phi(\boxplus,\boxplus,\boxplus|\psi_{GHZ})\\
+\Delta\bar{\beta}(\phi(\boxplus,\bullet,\boxplus|\psi_{GHZ})-\log 2)+\Delta\bar{\gamma}(\phi(\boxplus,\boxplus,\bullet|\psi_{GHZ})-\log 2),
\end{multline}
or in a more succinct form
 \begin{multline}
 \delta\phi(\boxplus,\bar{\beta},\bar{\gamma}|\psi_{GHZ})=(1-\Delta\bar{\beta}-\Delta\bar{\gamma})\delta\phi(\boxplus,\boxplus,\boxplus|\psi_{GHZ})\\
 +\Delta\bar{\beta}\delta\phi(\boxplus,\bullet,\boxplus|\psi_{GHZ})+\Delta\bar{\gamma}\delta\phi(\boxplus,\boxplus,\bullet|\psi_{GHZ}),
 \end{multline}
where $\delta\phi(\boxplus,\bar{\beta},\bar{\gamma}|\psi_{GHZ})=\phi(\boxplus,\bar{\beta},\bar{\gamma}|\psi_{GHZ})-\phi(\boxplus,\bar{\beta},\bar{\gamma}|GHZ)$. This outstanding feature permits us to express the rate of a state in the facet of the polytope with vertex $\{(1/2,1/2,0),(1/2,0,1/2),(1/2,1/2,1/2)\}$ as a convex combination of the rates in the vertex. This is also valid for the other two facets which have two common points with the one we have just analysed. Now we will explore the case of the W entanglement class, where we lack the symmetry of the GHZ state and in consequence the calculations are formidable.
\section{The rate for the W entanglement class}
Following the same line of though of the previous section we would like to calculate the asymptotic rate for states in the W entanglement class
\begin{equation}
\ket{\psi_{W}}=A\otimes B\otimes \mathbb{I}\ket{W},
\end{equation}
with 
\begin{equation}
A\otimes B\otimes \mathbb{I}=\left(\begin{array}{cc}
\sqrt{c} & \sqrt{d}\\
0 & \sqrt{a}
\end{array}\right)\otimes \left(\begin{array}{cc}
\sqrt{3} & 0\\
0 &\sqrt{\dfrac{3b}{c}} 
\end{array}\right) \otimes \left(\begin{array}{cc}
1 & 0\\
0 & 1
\end{array}\right).
\end{equation}
For the $n$ copy expansion we have that
\begin{equation}
\ket{W}^{\otimes n}=\left(\dfrac{\ket{001}+\ket{010}+\ket{001}}{\sqrt{3}}\right)^{\otimes n},
\end{equation}
which we can write as in section \ref{secLouckstate} as
\begin{equation}
\ket{W}^{\otimes n}=\sum_{\Omega}\sqrt{\binom{n}{\Omega}}[W]^{\Omega}\ket{\Omega},
\end{equation}
the $2\times2\times2$ tensor $[W]$ has non-zero components $W_{001}=W_{010}=W_{001}=1/\sqrt{3}$. In this case, we have that
\begin{equation}
\ket{\Omega}=\dfrac{1}{\sqrt{\binom{n}{\Omega}}}\sum_{s_1,s_2,s_3:W(s_1\circ s_2 \circ s_3)=\Omega}\ket{s_1}\ket{s_2}\ket{s_3}.
\end{equation}
Note that the tensor $\Omega$ is the weight of the \emph{tri-sequence} $s_1\circ s_2\circ s_3$. It is also easy to see that the only non-vanishing components of $\Omega$ are $\Omega_{001},\Omega_{010},\Omega_{100}$. As a consequence the third sequence $s_3$ is completely determined by the first two sequences $s_1$ and $s_2$. For example, it is easy to see that if $\ket{s_1}=\ket{00001110}$ and $\ket{s_2}=\ket{11110000}$ then $\ket{s_3}=\ket{00000001}$. The compatibility conditions over $\Omega$  are
\begin{equation}\label{eqcondW}
\Omega_{010}+\Omega_{001}=w^{1}_0\quad
\Omega_{100}+\Omega_{001}=w^{2}_0\quad
\Omega_{010}+\Omega_{100}=w^{3}_0\quad
\Omega_{010}+\Omega_{001}+\Omega_{100}=n,
\end{equation}
where $w^{i}_0$ is the number of zeroes in the $i$th sequence. To relate this quantities to the ones from the angular momentum used in the Schur basis we have the relation $m_{i}=w^{i}_1-n/2$. From conditions \eqref{eqcondW} it follows that
\begin{equation}
\Omega_{100}=m_1+\dfrac{n}{2}\quad \Omega_{010}=m_2+\dfrac{n}{2}\quad\Omega_{001}=m_3+\dfrac{n}{2}\quad m_1+m_2+m_3=-\dfrac{n}{2},
\end{equation}
Applying the Schur transform to $\ket{W}$ we obtain
\begin{equation}
\ket{W}^{\otimes}=\dfrac{1}{3^{n/2}}\sum_{\vec{\lambda}}\sum_{\vec{m}}\sum_{\vec{\mu}}\sum_{s_1,s_2,s_3:W(s_1\circ s_2 \circ s_3)=\Omega}\braket{\alpha,m_1,\mu_1}{s_1}\braket{\beta,m_2,\mu_2}{s_2}\braket{\gamma,m_3,\mu_3}{s_3}\ket{\vec{j},\vec{m},\vec{\mu}}.
\end{equation}
Thus the probability to be in a triplet $(\alpha,\beta,\gamma)$ will be in general
\begin{equation}
p(\alpha,\beta,\gamma|\psi_{W})=\dfrac{1}{3^{n}}\sum_{\vec{m},\vec{m}'}D^{\alpha}(A^{\dagger}A)_{m_1,m_1'}D^{\beta}(B^{\dagger}B)_{m_2,m_2'}\mathbb{I}_{m_3,m_3'}R^{\alpha\beta\gamma}_{\vec{m},\vec{m}'}
\end{equation}
which because $B^{\dagger}B$ is diagonal and $m_1+m_2+m_3=m_1'+m_2'+m_3'=-n/2$ is fixed it reduces to
\begin{equation}
p(\alpha,\beta,\gamma|\psi_{W})=\dfrac{1}{3^{n}}\sum_{m_1,m_2,m_3}D^{\alpha}(A^{\dagger}A)_{m_1,m_1}D^{\beta}(B^{\dagger}B)_{m_2,m_2}R^{\alpha\beta\gamma}_{m_1,m_2,m_3},
\end{equation}
with
\begin{multline}
R^{\alpha\beta\gamma}_{m_1,m_2,m_3}=\sum_{\vec{\mu}}\sum_{s_1,s_2,s_3\sim\Omega}\sum_{s_1',s_2',s_3'\sim\Omega'}\braket{\alpha,m_1,\mu_1}{s_1}\braket{s_1'}{\alpha,m_1,\mu_1}\braket{\beta,m_2,\mu_2}{s_2}\\
\times\braket{s_2'}{\beta,m_2,\mu_2}\braket{\gamma,m_3,\mu_3}{s_3}\braket{s_3'}{\gamma,m_3,\mu_3},
\end{multline}
where we have used the notation $s_1,s_2,s_3\sim\Omega$ to imply $s_1,s_2,s_3:W(s_1\circ s_2\circ s_3)=\Omega$.
In terms of the Louck polynomials it can be written as 
\begin{equation}
R^{\alpha\beta\gamma}_{m_1,m_2,m_3}=f^{\alpha}f^{\beta}f^{\gamma}\sum_{s_1,s_2,s_3}\sum_{s_1',s_2',s_3'}C^{\alpha}(\Omega_1)_{m_1,m_1}C^{\beta}(\Omega_2)_{m_2,m_2}C^{\gamma}(\Omega_3)_{m_3,m_3},
\end{equation}
where $\Omega_i=W(s_i\circ s_i')$ for $i=1,2,3$. Because the sum over sequences can be replaced over the sum of matrices $\Omega$ if we include the appropriate multiplicity factor we have
\begin{equation}
R^{\alpha\beta\gamma}_{m_1,m_2,m_3}=f^{\alpha}f^{\beta}f^{\gamma}\sum_{\Omega_1,\Omega_2,\Omega_3}Z(\Omega_1,\Omega_2,\Omega_3)C^{\alpha}(\Omega_1)_{m_1,m_1}C^{\beta}(\Omega_2)_{m_2,m_2}C^{\gamma}(\Omega_3)_{m_3,m_3},
\end{equation}
where $Z$ counts the number of sequences sets ${s_1,s_2,s_3,s_1',s_2',s_3'}$ compatible with the constrains over $\Omega$ and $\Omega'$ and over $\Omega_1,\Omega_2,\Omega_3$. Each matrix $\Omega_i$ is only dependent on one free parameter which we called $x_i$, explicitly
\[
\Omega_i=\left(\begin{array}{cc}
n-w_i-x_i &w_i-w_i'+ x_i\\
x_i & w_i'-x_i
\end{array}\right),
\]
where $w_i$ is the weight of the sequence (the number of ones in the $i$-th sequence) or in terms of the $m_i$ we have
\[
\Omega_i=\left(\begin{array}{cc}
n/2-m_i-x_i & m_i-m_i'+x_i\\
x_i & m_i'+n/2-x_i
\end{array}\right).
\]
To calculate the factor $Z$ we introduce the tensor tensor $Q$ with six indices indicating the \emph{tri-sequences} $s_1\circ s_2\circ s_3$ and $s_1'\circ s_2'\circ s_3'$.  We will focus on entries of $Q$ which are non-zero, that is the ones of the form $i,j$ for $i,j=r_1,r_2,r_3$ with $r_1:=100, r_2:=010$ and $r_3:=001$. To make $Q$ compatible with ${\Omega_i}$ we must ask
\begin{equation}
{\Omega_1}_{(i,i')}=\sum_{j,j'}\sum_{k,k'}Q_{ijk,i'j'k'},\quad
{\Omega_2}_{(j,j')}=\sum_{i,i'}\sum_{k,k'}Q_{ijk,i'j'k'},\quad
{\Omega_3}_{(k,k')}=\sum_{i,i'}\sum_{j,j'}Q_{ijk,i'j'k'}.
\end{equation}
These constraints result in
\[
{\Omega_1}_{(1,1)}=Q_{100,100}=w_1'-x_1,\quad {\Omega_2}_{(1,1)}=Q_{010,010}=w_2'-x_2,\quad {\Omega_3}_{(1,1)}=Q_{001,001}=w_3'-x_3.
\]
\[
{\Omega_1}_{(1,0)}=Q_{100,001}+Q_{100,010}=x_1,\quad {\Omega_2}_{(1,0)}=Q_{010,100}+Q_{010,001}=x_2,\quad
\]
\[ {\Omega_3}_{(1,0)}=Q_{001,100}+Q_{001,010}=x_3,
\]
\[
{\Omega_1}_{(0,0)}=Q_{010,001}+Q_{010,010}+Q_{001,001}+Q_{001,010}=n-w_1-x_1,
\]
\[
{\Omega_2}_{(0,0)}=Q_{001,001}+Q_{100,100}+Q_{001,100}+Q_{100,001}=n-w_2-x_2,
\]
\[
{\Omega_3}_{(0,0)}=Q_{010,010}+Q_{100,100}+Q_{010,100}+Q_{100,010}=n-w_3-x_3,
\]
\[
{\Omega_1}_{(0,1)}=Q_{001,100}+Q_{010,100}=w_1-w_1'+x_1,
\]
\[
{\Omega_2}_{(0,1)}=Q_{100,010}+Q_{001,010}=w_2-w_2'+x_2,
\]
\[
{\Omega_3}_{(0,1)}=Q_{100,001}+Q_{010,001}=w_3,-w_3'+x_3,
\]
and the additional constraints
\[
w_1+w_2+w_3=n,\quad w_1'+w_2'+w_3'=n.
\]
This set of equations can be solved in terms of a single free parameter $k$, explicitly (with $w_i=w_i'$ because $m_i=m_i'$)
\begin{eqnarray*}
	Q_{100,010}=k,\\
	Q_{100,100}=w_1-x_1,\\
	Q_{100,001}=x_1-k,\\
	Q_{010,100}=x_1+x_2-x_3-k,\\
	Q_{010,010}=w_2-x_2,\\
	Q_{010,001}=x_3-x_1+k,\\
	Q_{001,100}=x_3-x_2+k,\\
	Q_{001,010}=x_2-k,\\
	Q_{001,001}=w_3-x_3.
\end{eqnarray*}
Therefore we can calculate $Z(\Omega_1,\Omega_2,\Omega_3)$ as a multinomial term
\begin{equation}
Z=\left(\prod_{i=1}^{3}\dfrac{1}{(w_i-x_i)!}\right)\sum_{k}\dfrac{n!}{k!(x_1-k)!(x_2-k)!(x_1+x_2-x_3-k)!(x_3-x_1+k)!(x_3-x_2+k)!}.
\end{equation}
As we can see, the factor $Z$ makes the approach we made in the GHZ case not feasible. However, we can consider the vertex of the polytope or a particular facet and see if we can gain some information on the term $R^{\alpha\beta\gamma}_{m_1,m_2,m_3}$. This is why we will consider the facet in the polytope $\alpha_2+\beta_2+\gamma_2=n$ which is the one that separates the $W$ region of the polytope with the upper region only accessible to states in the GHZ class.
\subsection{The $\alpha_2+\beta_2+\gamma_2=n$ plane }
\begin{figure}[h]
	\centering
	\includegraphics[scale=0.3]{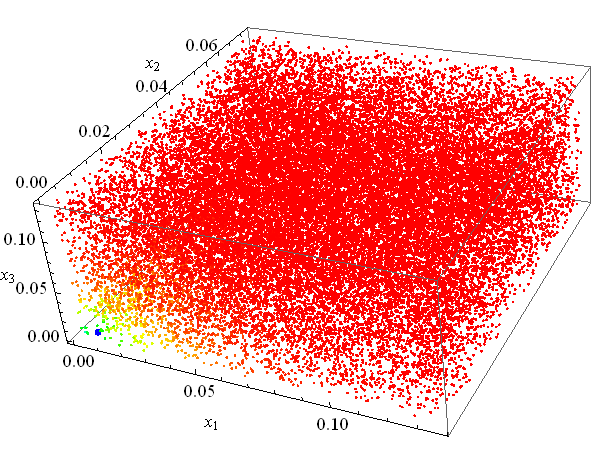}
	\caption{Simulation for the sum in equation \eqref{eqsimula}. Colder colors are higher values in the space $(x_1,x_2,x_3)$. For all our tests on the values of $\alpha,\beta,\gamma$ the maximum was located at the origin.}
	\label{finNume}
\end{figure}

When we restrict our analysis to the plane  $\alpha_2+\beta_2+\gamma_2=n$, together with the condition $w_1+w_2+w_3=n$, or equivalently $m_1+m_2+m_3=-n/2$, and $w_1\geq \alpha_2$, $w_2\geq \beta_2$ , $w_3\geq \gamma_2$ the only possible value for the Young frames is $\alpha_2=w_1$, $w_2= \beta_2$ and $w_3= \gamma_2$. In terms of the angular momentum variables $\vec{j}$ and $\vec{m}$ this implies
$m_1=-j_1$, $m_2=-j_2$ and $m_3=-j_3$.
\\
\\
Thus, the Louck polynomials take the form( see Appendix B for more details)
\begin{equation}
C^{\alpha}(\Omega_1)_{-j,-j}=\dfrac{(-1)^{x_1}}{\binom{n}{\alpha_2}\binom{n-\alpha_2}{x_1}},
\end{equation}
then we have that
\begin{equation}\label{eqsimula}
R^{\alpha\beta\gamma}_{m_1,m_2,m_3}=\dfrac{f^{\alpha}f^{\beta}f^{\gamma}}{\binom{n}{\alpha_2}\binom{n}{\beta_2}\binom{n}{\gamma_2}}\sum_{x_1,x_2,x_3}\dfrac{(-1)^{x_1+x_2+x_3}Z(x_1,x_2,x_3)}{\binom{n-\alpha_2}{x_1}\binom{n-\beta_2}{x_2}\binom{n-\gamma_2}{x_3}},
\end{equation}
where we have changed the sum over $\Omega$ for a sum over $x$. To describe the asymptotics of this function we perform a numerical analysis that showed us that the maximum of the sum over $x_1,x_2,x_3$ is attained at $x_1=x_2=x_3=0$ as can be seen in Fig. \ref{finNume}.
Then, we have that, asymptotically 
\begin{equation}
R^{\alpha\beta\gamma}_{m_1,m_2,m_3}\approx\dfrac{f^{\alpha}f^{\beta}f^{\gamma}}{\binom{n}{\alpha_2}\binom{n}{\beta_2}\binom{n}{\gamma_2}}\dfrac{Z(0,0,0)}{\binom{n-\alpha_2}{0}\binom{n-\beta_2}{0}\binom{n-\gamma_2}{0}}=\dfrac{f^{\alpha}f^{\beta}f^{\gamma}}{\binom{n}{\alpha_2}\binom{n}{\beta_2}\binom{n}{\gamma_2}}\dfrac{n!}{\alpha_2!\beta_2!\gamma_2!}\sim \exp[nH(\bar{\alpha}_2,\bar{\beta}_2,\bar{\gamma}_2)].
\end{equation}
Therefore, we have that
\begin{equation}
p(\alpha,\beta,\gamma|\psi_{W})\approx D^{\alpha}(A^{\dagger}A)_{m_1,m_1}D^{\beta}(B^{\dagger}B)_{m_2,m_2}R^{\alpha\beta\gamma}_{m_1,m_2,m_3},
\end{equation}
with $m_1=\alpha_2-n/2$, $m_2=\beta_2-n/2$ and $m_3=\gamma_2-n/2$. Then from the asymptotics of the representation matrices we have by Appendix \ref{ApRep} (in the angular momentum notation)
\begin{equation}\label{ecRepW}
D^{\alpha}(X)_{-j_1,-j_1}=\det(X)^{\alpha_2}X_{11}^{\alpha_1-\alpha_2}.
\end{equation}
Thus the asymptotic rate at the plane $\alpha_2+\beta_2+\gamma_2=n$ is
\begin{multline}\label{eqrateWyeah}
\phi(\alpha,\beta,\gamma|\psi_{W})=\log 3-H(\bar{\alpha}_2,\bar{\beta}_2,\bar{\gamma}_2)-\bar{\alpha}_2\log|\det(A^{\dagger}A)|-\Delta\bar{\alpha}\log (A^{\dagger}A)_{11}\\
-\bar{\beta}_2\log|\det(B^{\dagger}B)|-\Delta\bar{\beta}\log(B^{\dagger}B)_{11},
\end{multline}
which can be simplified using explicitly $A$ and $B$
\begin{equation}
\phi(\alpha,\beta,\gamma|\psi_{W})=\log 3-H(\bar{\alpha}_2,\bar{\beta}_2,\bar{\gamma}_2)-\bar{\alpha}_2\log(a c)-\Delta\bar{\alpha}\log(c)-\bar{\beta}_2\log\left(\dfrac{9b}{c}\right)-\Delta\bar{\beta}\log\left(3\right),
\end{equation}
and into a more symmetric expression (using $\bar{\gamma}_2=1-\bar{\alpha}_2-\bar{\beta}_2$)
\begin{equation}\label{eqrateWgeneral}
\phi(\bar{\alpha},\bar{\beta},\bar{\gamma}|\psi_{W})=-H(\bar{\alpha}_2,\bar{\beta}_2,\bar{\gamma}_2)-\bar{\alpha}_2\log(a )-\bar{\beta}_2\log\left(b\right)-\bar{\gamma}_2\log(c),
\end{equation}
which is to be compared to equation \eqref{ecWabc} to obtain
\begin{equation}
\phi(\bar{\alpha},\bar{\beta},\bar{\gamma}|{W})=-H(\bar{\alpha}_2,\bar{\beta}_2,\bar{\gamma}_2)+\log3.
\end{equation}
\section{Summary of Results}
In this chapter we focused on calculating the constituents of the expression
\begin{equation}
p(\alpha,\beta,\gamma|\psi_1)=p(\alpha,\beta,\gamma|\psi_2)\dfrac{\braket{\Phi(\psi_1)}{\Phi(\psi_1)}}{\braket{\Phi(\psi_1)}{\Phi(\psi_1)}},
\end{equation}
and its asymptotics for different entanglement classes using two techniques that as far as we know are new to attack this problem. We obtained the asymptotic rates for the probability at some of the facets (where $g_{\alpha\beta\gamma}=1$) of the entanglement polytope for the W and GHZ class. We evidenced that the rate at one facet can be given as a convex combination of the vertex of the compatibility polytope coplanar with the facet. We also observed that the rates at the vertex are given in terms of LU invariants constructed from the covariant associated to each vertex. Although we saw that the correspondence is not perfect in some cases,e.g. 
\begin{equation}
\braket{B_{200}}{B_{200}}\neq\lim\limits_{n\to\infty}\dfrac{\braket{B_{200}^{n/2}}{B_{200}^{n/2}}^2}{n},
\end{equation}
it is due to the asymptotic limit taken where we dealt with polynomials of infinite degree, thus losing part of the LU invariance. The method used to calculate the desired quantities involves a map from SLOCC covariants to a state in the Shur-Weyl decomposition, which proves efficient in this three qubit state because we have a \emph{small} set of covariants. The method also deals with the Louck polynomials to calculate the probability directly; however it will prove to be a laborious task to find the asymptotics for the cases where the expression can be simplified and intractable for a general case. A more elaborate and succinct analysis will be given in the next chapter. 

%% file: Chapter4.tex

\chapter{Conclusions and Outreach} 

\label{Chapter4} 
In this chapter we want to remark the important features of this work, as well as give some insights on what may become future work and research lines. 
\lhead{Chapter 4. \emph{Conclusions and Outreach}} 

\section{Outreach: four qubits}
Throughout this work we have calculated the asymptotic rates for the distribution $p(\alpha,\beta,\gamma|\psi)$ for different entanglement classes. To that end we have used two approaches; the computation making the Schur transform and using the Louck polynomials, which is rather complicated for the general case, and the \emph{covariant-to-state} approach which gives us an explicit formula for the ratio $p(\alpha,\beta,\gamma|\psi_1)/p(\alpha,\beta,\gamma|\psi_2)$ by calculating the ratio $\braket{\Phi_{\psi_1}}{\Phi_{\psi_1}}/\braket{\Phi_{\psi_2}}{\Phi_{\psi_2}}$. The asymptotics are not easily calculated and involves non-trivial maximization problems. 

These two approaches can in principle be extended to four or more qubit systems. The cons are that as we may expect the complexity increases vastly if we add another qubit; for example, the entanglement classes are now 9 \emph{families} in which it is not clear how to organize them in some hierarchy (separable and entangled states ca belong to the same family) \cite{Verstraete_2002}. Additionally, the generators of the algebra of covariants, which  are given in \cite{Briand4}, are 170 covariants. They can be calculated by means of a computer search using \emph{transvectans} and guided by the corresponding Hilbert multivariate series (with $u_1=u_2=u_3=u_4=u$ for simplicity)
\begin{equation}
	\dfrac{P(u,t)}{(1-tu^2)(1-tu^4)(1-t^2)(1-t^2u^2)^2(1-t^2u^4)^3(1-t^4)(1-t^4u^2)(1-t^4u^4)(1-t^6)},
\end{equation}
where $P(t,u)$ is a polynomial of degree 20 in $t$ and $u$ which we will not write here but can be found at \cite{Briand4}. Nevertheless as we have mentioned before, not all the covariants are necessary for an entanglement classification. Advances in this classification can be seen in \cite{Zimmerman} where the authors give a smaller subset of \emph{covariant vectors} and attempt to classify what they called \emph{nilpotent orbits}, which is nothing more than orbits in which all the invariants vanish. See also \cite{Chen2,Luque} and a review \cite{Borsten_2012} with interesting relations to black holes entropy.
\\
\\
Regarding proposition \ref{prop1} and theorem \ref{teo2} for four qubits, we have verified them numerically so far as the first 47 of the 170 covariants in \cite{Briand4} taking into account some additional restrictions given by the syzygys. The complete set is ongoing work. This has lead us to think that proposition \ref{prop1} may be general for any $n$ qubit system. However, to make this assertion into a formal proof seems rather impractical as well as the \emph{covariant-to-state} approach for this case. For \emph{qudit} systems it is not clear yet how the relation between \emph{fundamental tuples} of Young diagrams can be constructed from the covariants.
\section{Conclusions}
In his work we have calculated the asymptotic rates $\phi(\bar{\alpha},\bar{\beta},\bar{\gamma}|\psi)$ of the probability $p(\alpha,\beta,\gamma|\psi)$ in different regions of the compatibility polytope. To calculate the probability $p(\alpha,\beta,\gamma|\psi)$ we employed two techniques. The first technique involves taking the $n$-th tensor product of the state and write it explicitly to then apply the Schur transformation. This approach involves dense algebra dealing with Louck and Han-Eberlein polynomials which we adapt to fit our problem. The second technique is what we called the \emph{covariant-to-state} approach and involves $\vec{n}$-th powers of the covariants to describe a state $\ket{\Phi_{\psi}}\in V_\lambda^{d}$ in the Wedderburn decomposition of $(\mathbb{C}^{2}\otimes\mathbb{C}^{2}\otimes\mathbb{C}^{2})^{\otimes n}$, a technique we consider is a novelty and that can bring a more geometrical/algebraic insight into entanglement. With this approach we were able to calculate explicitly the internal product ratio $\braket{\Phi_{\psi_1}}{\Phi_{\psi_1}}/\braket{\Phi_{\psi_2}}{\Phi_{\psi_2}}$ for the GHZ and W entanglement class in the facets of the polytope. We calculated for the W class its asymptotics in the $\alpha+\beta+\gamma=n$ plane as well as an expression for general $\alpha,\beta,\gamma$, and in the GHZ class the rates at the facet defined by $\bar{\alpha}=1/2$.

In the facets of the polytope where $g_{\alpha\beta\gamma}=1$, we show that the rate (relative rate) can be written as a convex combination of the rates at the vertex of the polytope that lie in the facet. Furthermore, the rates at the vertex are given in terms of LU invariants relative to each vertex for the W class. In the GHZ class we obtained that the rates at the vertex are obtained as the asymptotic expression of LU invariants, which may not be LU invariant themselves, for example
\begin{equation}
\braket{B_{200}}{B_{200}}\neq\lim\limits_{n\to\infty}\dfrac{\braket{B_{200}^{n/2}}{B_{200}^{n/2}}^2}{n}.
\end{equation}
However, this result coincides with the asymptotic rate calculated by making the Schur transform directly and expressing the probability $p(\alpha,\beta,\gamma|\psi)$ as a non-trivial sum over Louck polynomials, whose linearisation coefficients are so-far unknown, making the general problem practically intractable.

We have also found a relation in the case of three qubits (and probably extendible to $n$ qubit systems) between the Kronecker coefficients and a set of fundamental Young frames constructed from the covariants. This result is summarized in theorem \ref{teo2} and basically consists on counting the number of solutions to a given set of equations (see proposition \ref{prop1}). This relation also showed us, using the Keyl-Werner theorem, a correspondence in the asymptotic limit between the convex regions in the entanglement polytope  and the regions where $g_{\alpha,\beta,\gamma}\neq0$ using the restrictions from theorem \ref{teo2}. We successfully verified our results by comparing to what is known about Kronecker coefficients in the literature. 
\\
\\
For future work we would like to see if this results hold in the four or more qubits case. Although we have mentioned the dimensions of the problem escalate exponentially with increasing number of qubits, we will still be able to obtain some insight about the entanglement distribution in these systems. We would also like to see how to extend our results to values of the Kronecker coefficient larger than one . That is, we would like to investigate the bulk of the polytope, specially the GHZ entanglement class in order to generalize our results. We hope that by doing this, the asymptotic rate will be written as a convex combination of the rates at the vertex of the polytope. Another interesting variation to the problem will be to study identical particle systems, where now the Hilbert space changes and thus the covariants must change. Finally, one could also study systems of qudits, for example two qubits and a qutrit and see how this asymmetry in the dimensions of the Hilbert space affects entanglement classification.


%% file: AppendixA.tex

\chapter{Detailed Calculations} 

\label{AppendixA} 

\lhead{Appendix A. \emph{Detailed Calculations}} 

\section{Representations of GL in the Schur basis}\label{ApRep}
So far, we have studied the Schur transform for a state $\ket{\psi}^{\otimes n}$. We will now turn to the representations of $g\in GL(2,\mathbb{C})^{\times n}$. Remember that each element of the group acts on one copy of the state $\ket{\psi}$. In the Schur basis we have vectors of the form $\ket{j,m}$ where in terms of Young Diagrams can be seen as a diagram with $\lambda_1-\lambda_2=2j$ \emph{free boxes} and $m$ denotes the number of zeros in the corresponding free boxes. We call free boxes the ones that remain in the first row without boxes below, e.g. the nest tableaux has
3 free boxes 
\[
\yng(7,4).
\]
Remember we can always write the state in the Schur base as a linear combination of sequence vectors $\ket{s}$
\[
\ket{j,m}=\dfrac{1}{\sqrt{\binom{n'}{n_0}}}\sum_{s\sim(n_0,n_1)}\ket{s},
\]
where $s\sim(n_0,n_1)$ indicates that the sequence (in the $n'$ free boxes) has $n_0$ zeros and $n_1$ ones. In terms of $j$ and $m$ we have that 
\[
\ket{j,m}=\dfrac{1}{\sqrt{\binom{2j}{j+m}}}\sum_{s\sim(j+m,j-m)}\ket{s}.
\]
What we want to find is the representation of the element $g^{\otimes n}$ in the Schur basis, i.e. we want
\begin{equation}
D^{\lambda}_{m',m}(g)\equiv\bra{j,m'}g^{\otimes n}\ket{j,m}=\det(g)^{\lambda_2}\dfrac{\sqrt{(j+m)!(j-m)!(j+m')!(j-m')!}}{(2j)!}\sum_{s\sim m,s\sim m'}\bra{s'}g^{\otimes}\ket{s},
\end{equation}
with $2j=\lambda_1-\lambda_2$.
We know that $g^{\otimes}$ acts individually on each part of the sequence, thus we will obtain a form
\begin{multline}
D^{\lambda}_{m',m}(g)=\det(g)^{\lambda_2}\dfrac{\sqrt{(j+m)!(j-m)!(j+m')!(j-m')!}}{(2j)!}\\
\times\sum_{\vec{n}}W(n,n_{00},n_{01},n_{10},n_{11})g_{00}^{n_{00}}g_{01}^{n_{01}}g_{10}^{n_{10}}g_{11}^{n_{11}},
\end{multline}
where the sum is performed over all the values of $\vec{n}=(n_{00},n_{01},n_{10},n_{11})$ compatible with the sequences $s,s'$. The combinatorial factor $W$ indicates how many of the sequences are compatible with $\vec{n}$, i.e.,
\[
W(n,n_{00},n_{01},n_{10},n_{11})=\dfrac{n!}{n_{00}!n_{01}!n_{10}!n_{11}!}.
\]
The numbers in $\vec{n}$ are subject to the restrictions impose by the values of $m$ and $m'$ or equivalently $n_0,n_1,n_0'$ and $n_1'$. This restrictions can be easily understood if we think of a matrix $N$
\[
N=\left(\begin{array}{cc}
n_{00} & n_{01}\\
n_{10} & n_{11}
\end{array}\right).
\]
We have that if we sum over rows 1 and 2 we must obtain $n_0'$ and $n_1'$ respectively; if we sum over the columns 1 and two we must obtain $n_0$ and $n_1$ respectively. Thus 
\begin{eqnarray*}
	n_{00}+n_{01}=n_0',\\
	n_{10}+n_{10}=n_1',\\
	n_{00}+n_{10}=n_0,\\
	n_{01}+n_{11}=n_1.
\end{eqnarray*}
In consequence the matrix $N$ can be written in term of an independent parameter $x$ as
\[
\Omega=\left(\begin{array}{cc}
n_0'-x & x\\
n_0-n_0'+x & n_1-x
\end{array}\right),
\]
or in terms of $j$ and $m$ we have
\[
\Omega=\left(\begin{array}{cc}
j+m'-x & x\\
m-m'+x & j-m-x
\end{array}\right).
\]
Then we have a compact expression for the representation of $g$

\begin{multline}\label{ecRepD}
D^{\lambda}_{m',m}(g)=\det(g)^{\lambda_2}\dfrac{\sqrt{(j+m)!(j-m)!(j+m')!(j-m')!}}{(2j)!}\\
\sum_{x=\max(m'-m,0)}^{\min(j-m,j+m')}\dfrac{(2j)!g_{01}^{x}g_{10}^{m-m'+x}g_{11}^{j-m-x}g_{00}^{j+m'-x}}{(j+m'-x)!(x)!(m-m'+x)!(j-m-x)!},
\end{multline}
or in a notation introduced in Chapter \ref{Chapter2}
\begin{equation}
D^{\lambda}_{m',m}(g)=\det(g)^{\lambda_2}\sum_{\Omega}\binom{2j}{\Omega}g^{\Omega}.
\end{equation}
Now we are interested in calculating its asymptotic rate in the particular cases considered throughout Chapter \ref{Chapter3}.
First we consider the case $g=A^{\dagger}A$ and $j=0$, it read explicitly from \eqref{ecRepD} that
\begin{equation}
D^{\boxplus}_{0,0}(A^{\dagger}A)=\det(A^{\dagger}A)^{n/2}.
\end{equation}
Now for the matrices of the form 
\[
B^{\dagger}B=\left(\begin{array}{cc}
1 & c_\theta\\
c_{\theta} & 1
\end{array}\right),
\]
and an arbitrary $\lambda$ we have
\begin{equation}
D^{\lambda}_{0,0}(g)=\det(g)^{\lambda_2}\dfrac{j!^2}{(2j)!}\sum_{\Omega}\binom{2j}{\Omega}g^{\Omega},
\end{equation}
where in the asymptotic limit we obtain that the sum goes as a relative entropy if the matrix entries are normalized to one, making the normalization
\begin{equation}
D^{\lambda}_{0,0}(g)=\det(g)^{\lambda_2}\dfrac{j!^2}{(2j)!}(2+2c_\theta)^{2j}\sum_{\Omega}\binom{2j}{\Omega}\left(\dfrac{g}{2+2c_\theta}\right)^{\Omega}.
\end{equation}
In the asymptotic limit, the sum can be approximated by the Laplace method where the dominating term will be
\begin{equation}
\sum_{\Omega}\binom{2j}{\Omega}\left(\dfrac{g}{2+2c_\theta}\right)^{\Omega}\sim \exp\left[-2j D(\tilde{\Omega},\tilde{g})\right]\sim 1,
\end{equation}
where $\tilde{\Omega}=\Omega/2j$ and $\tilde{g}=g/(2+2c_\theta)$. Thus we have for the whole expression
\begin{equation}\label{limitLU}
\lim\limits_{n\to\infty}\dfrac{ \log D^{\lambda}_{0,0}(g)}{n}=\bar{\lambda_{2}}\det(g)+(\bar{\lambda}_1-\bar{\lambda}_2)\log(1+c_\theta).
\end{equation}
Now we turn to expression \eqref{ecRepW} where we have that $m=-j$
\begin{equation}
D^{\lambda}_{-j,-j}(g)=\det(g)^{\lambda_2}g_{11}^{2j}=\det(g)^{\lambda_2}g_{11}^{\lambda_1-\lambda_2},
\end{equation}
which completes the expression we used in Chapter \ref{Chapter3}.
\section{Hahn-Eberlein polynomials and the asymptotics of $R^{\boxplus\beta\gamma}_{0,0}$}\label{ApHEpol}
\begin{figure}[h]
	\centering
	\includegraphics[scale=0.6]{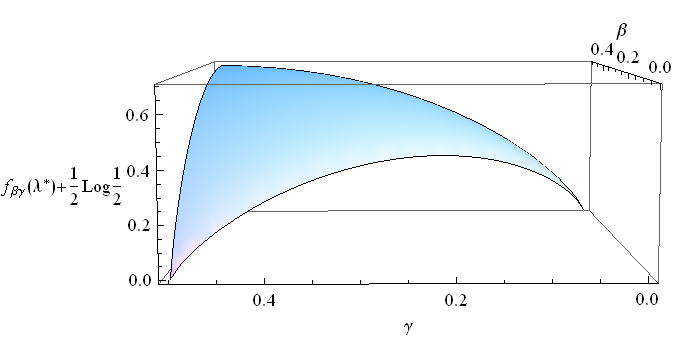}
	\caption{Function $f_{\bar{\beta}\bar{\gamma}}(\bar{\lambda}^{*})+\frac{1}{2}\log\frac{1}{2}$ plotted in the $\beta-\gamma$ plane.}
	\label{fbg}
\end{figure}

In this section we will make with details the calculation of the asymptotics of the term $R^{\boxplus\beta\gamma}_{0,0}$. First, let us recall the definition of $R^{\alpha\beta\gamma}$
\begin{equation}
R^{\alpha\beta\gamma}_{m,m'}=f^{\alpha}f^{\beta}f^{\gamma}\sum_{\Omega}\binom{n}{\Omega}C^{\alpha}(\Omega)_{m',m}C^{\beta}(\Omega)_{m',m}C^{\gamma}(\Omega)_{m',m},
\end{equation}
thus we have to calculate
\begin{equation}
R^{\boxplus\beta\gamma}_{0,0}=f^{\boxplus}f^{\beta}f^{\gamma}\sum_{\Omega}\binom{n}{\Omega}C^{\boxplus}(\Omega)_{0,0}C^{\beta}(\Omega)_{0,0}C^{\gamma}(\Omega)_{0,0}.
\end{equation}
In order to be able to perform such calculation, we must introduce the Hahn-Eberlein polynomials $E_{\lambda_1}^{w,w'}(x)$ which are related to the Locuk polynomials via the relation (in terms of the weights $w,w'$)
\begin{equation}
C^{\lambda}_{w',w}(\Omega)=\dfrac{(w_<)!(n-w_{>})!}{n!}\sqrt{\dfrac{(\lambda_1-w_<)!(w_>-\lambda_2)!}{(\lambda_1-w_>)!(w_<-\lambda_2)!}}E_{\lambda_2}^{(w_<,w_>)}(x),
\end{equation}
where 
\begin{equation}
\Omega=\left(\begin{array}{cc}
n-w-x &w_i-w'+ x\\
x & w'-x
\end{array}\right),
\end{equation}
and 
\begin{equation}
E_{\lambda_2}^{w',w}(x)={}_3F_2\left(\begin{array}{ccc}
-\lambda_2,&-x,&\lambda_2-n-1\\
 & -w',& w-n
\end{array}; 1\right).
\end{equation}
In our particular case, $m=m'=0$ which implies $w=w'=n/2$. In this case the relation between the Louck and the Hahn-Eberlein polynomials simplifies to
\begin{equation}
C^{\lambda}_{n/2,n/2}(\Omega)=\dfrac{1}{\binom{n}{n/2}}E_{\lambda_2}^{(n/2,n/2)}(x),
\end{equation}
and the binomial term becomes
\begin{equation}
\binom{n}{\Omega}=\dfrac{n!}{(x!)^2(n/2-x)!^2}=\binom{n}{n/2}\binom{n/2}{x}^2.
\end{equation}
Therefore we have that
\begin{equation}
R^{\boxplus\beta\gamma}_{0,0}=\dfrac{f^{\boxplus}f^{\beta}f^{\gamma}}{\binom{n}{n/2}^2}\sum_{x}\binom{n/2}{x}^2E_{n/2}^{(n/2,n/2)}(x)E_{\beta}^{(n/2,n/2)}E_{\gamma}^{(n/2,n/2)}.
\end{equation}
We can calculate explicitly the first Hahn-Eberlein polynomial using the formula form \cite{McEliece_1977} that reads in our notation
\begin{equation}
E_{\lambda}^{(w',w)}(x)=\dfrac{1}{\binom{w'}{x}\binom{n-w}{x}}\text{coef}_{(yz)^x}\left[(1-yz)^{\lambda}(1+y)^{w'-\lambda}(1+z)^{n-w-\lambda}\right],
\end{equation}
where the function $\text{coef}_{(yz)^x}[f]$ extracts the coefficient of the monomial $(yz)^x$ in the expansion of $f$. In the case $w=w'=\lambda=n/2$ we obtain
\begin{equation}
E_{n/2}^{(n/2,n/2)}=\dfrac{1}{\binom{n/2}{x}^2}\text{coef}_{(yz)^x}\left[(1-yz)^{n/2}\right]=\dfrac{1}{\binom{n/2}{x}^2}\text{coef}_{(yz)^x}\left[\sum_{k=0}^{n/2}\binom{n/2}{k}(-yz)^{k}\right]=\dfrac{(-1)^x}{\binom{n/2}{x}}.
\end{equation}
Therefore we have that
\begin{equation}\label{eqHEsum}
R^{\boxplus\beta\gamma}_{0,0}=\dfrac{f^{\boxplus}f^\beta f^\gamma}{\binom{n}{n/2}^2}\sum_{x}(-1)^x\binom{n/2}{x}E_{\beta}^{(n/2,n/2)}(x)E_{\gamma}^{(n/2,n/2)}(x).
\end{equation}
In order to continue with the calculations, we must introduce the so-called Krawtchouk polynomials (see the Appendix in \cite{McEliece_1977} and references within). This set of discrete polynomials obeys the orthogonality relation with respect to the weight/measure $\binom{n}{x}$
\begin{equation}\label{eqorthoK}
\sum_{x=0}^{n}\binom{n}{x}K_{\lambda}^{(n)}(x)K_{\mu}^{(n)}(x)=\binom{n}{\lambda}\delta_{\lambda,\mu},
\end{equation} 
and generating function
\begin{equation}
(1-y)^x(1+y)^{n-x}=\sum_{\lambda=0}^{n}K_{\lambda}^{(n)}(x)y^{\lambda}.
\end{equation}
As \eqref{eqorthoK} is similar to the sum in \eqref{eqHEsum} but only differing from an alternating sign $(-1)^x$ which we may include on one of the Krawtchouk polynomials. The relation reads explicitly $(-1)^xK_{\lambda}^{(n)}(x)=2^{n}K_{n-\lambda}^{(n)}(x)$ as we will show in the next lines.
\\
Consider the expansion 
\begin{eqnarray}
(2(y+z))^{n/2}&=&\sum_{x=0}^{n/2}(-1)^{x}\binom{n/2}{x}(1-y)^{x}(1+y)^{n/2-x}(1-z)^{x}(1+z)^{n/2-x},\\
&=&\sum_{\lambda,\mu=0}^{n/2}\left[\sum_{x=0}^{n/2}(-1)^{x}\binom{n/2}{x}K_{\lambda}^{(n/2)}(x)K_{\mu}^{(n/2)}(x)\right]y^{\lambda}z^{\mu},\\
&=&2^{n/2}\sum_{\lambda=0}^{n/2}\binom{n/2}{\lambda}y^{\lambda}z^{n/2-\lambda},
\end{eqnarray}
then it is easy to see that
\begin{equation}
\sum_{x=0}^{n/2}(-1)^{x}\binom{n/2}{x}K_{\lambda}^{(n/2)}(x)K_{\mu}^{(n/2)}(x)=2^{n/2}\binom{n/2}{\lambda}\delta_{\mu,n/2-\lambda}.
\end{equation}
Thus if we can find the connection coefficients between the Hahn-Eberlein and the Krawtchouk polynomials we will be able to make the sum in \eqref{eqHEsum} and find its asymptotics. The connection coefficients $c(\lambda|\mu)$ are defined as
\begin{equation}
E_{\lambda}^{(n/2,n/2)}(x)=\sum_{\mu}c(\lambda|\mu)K_{\mu}^{(n/2)}(x),
\end{equation}
and the sum  in \eqref{eqHEsum} translates to
\begin{equation}
\sum_{x}(-1)^x\binom{n/2}{x}E_{\beta}^{(n/2,n/2)}(x)E_{\gamma}^{(n/2,n/2)}(x)=2^{n/2}\sum_{\lambda}\binom{n/2}{\lambda}c(\beta|\lambda)c(\gamma|n/2-\lambda).
\end{equation}
Now our goal is to calculate the connection coefficients between the Hahn-Eberlein and the Krawtchouk polynomials. Fortunately in \cite{McEliece_1977} we have the relations
\begin{equation}
E_{\lambda}^{w',w}(x)=2^{-\lambda}\sum_{\mu=0}^{\lambda}\dfrac{\binom{\lambda}{k}}{\binom{w'}{\lambda-\mu}\binom{n-w}{\mu}}K_{\lambda-\mu}^{(w')}(x)K_{\mu}^{(n-w)}(x),
\end{equation}
and
\begin{equation}
K_{\lambda}^{(n/2)}(x)K_{\mu}^{(n/2)}(x)=\sum_{\nu=0}^{n/2}\binom{n/2-\nu}{(\lambda+\mu-\nu)/2}\binom{\nu}{(\lambda+\nu-\mu)/2}K_{\nu}^{(n/2)}(x).
\end{equation}
Using these two relations we obtain
\begin{equation}
E_{\lambda}^{n/2,n/2}=2^{-\lambda}\sum_{\nu=0}^{n/2}\binom{n/2-\nu}{(\lambda-\nu)/2}\left[\sum_{\mu=0}^{\lambda}\dfrac{\binom{\lambda}{\mu}\binom{\nu}{(\lambda-2\mu+\nu)/2}}{\binom{n/2}{\lambda-\mu}\binom{n/2}{\mu}}\right]K_{\nu}^{(n/2)}(x).
\end{equation}
From this expression the connection coefficients can be read explicitly, however we will first simplify the term in brackets
\begin{equation}
\sum_{\mu=0}^{\lambda}\dfrac{\binom{\lambda}{\mu}\binom{\nu}{(\lambda-2\mu+\nu)/2}}{\binom{n/2}{\lambda-\mu}\binom{n/2}{\mu}}=\dfrac{\lambda!\nu!}{(n/2)!^2}\sum_{\mu=0}^{\lambda}\dfrac{(n/2-\lambda+\mu)!(n/2-\mu)!}{(\nu-(\lambda-2\mu+\nu)/2)!((\lambda-2\mu+\nu)/2)!},
\end{equation}
and making the change of variable $\zeta=(\lambda-2\mu+\nu)/2$ we obtain
\begin{equation}
\sum_{\mu=0}^{\lambda}\dfrac{\binom{\lambda}{\mu}\binom{\nu}{(\lambda-2\mu+\nu)/2}}{\binom{n/2}{\lambda-\mu}\binom{n/2}{\mu}}=\dfrac{\lambda!\nu!}{(n/2)!^2}\sum_{\zeta=(\lambda+\nu)/2}^{(\nu-\lambda)/2}\dfrac{((n+\nu-\lambda)/2-\zeta)!((n-\lambda-\nu)/2+\zeta)!}{(\nu-\zeta)!\zeta!}.
\end{equation}
However, there is not much we can do to simplify this expression. Then 
\begin{multline}
2^{n/2}\sum_{\lambda}\binom{n/2}{\lambda}c(\beta|\lambda)c(\gamma|n/2-\lambda)=2^{n/2-\beta-\gamma}\sum_{\lambda}\binom{n/2}{\lambda}\binom{n/2-\lambda}{(\beta-\lambda)/2}\binom{\lambda}{(\gamma-n/2+\lambda)/2}\\
\dfrac{\beta!\lambda!\gamma!(n/2-\lambda)!}{(n/2)!^4}\sum_{\mu=0}^{\beta}\dfrac{(n/2-\beta+\mu)!(n/2-\mu)!}{((\lambda-\beta)/2+\mu)!((\beta+\lambda)/2-\mu)!}\\
\sum_{\mu'=0}^{\gamma}\dfrac{(n/2-\gamma+\mu')!(n/2-\mu')!}{((n/2-\lambda-\gamma)/2+\mu')!((\gamma+n/2-\lambda)/2-\mu')!}.
\end{multline}
We will analyse its asymptotic limit, to proceed we will first use the Laplace method on the sum over $\mu$ and $\mu'$. We have that
\begin{equation}
\sum_{\mu=0}^{\beta}\dfrac{(n/2-\beta+\mu)!(n/2-\mu)!}{((\lambda-\beta)/2+\mu)!((\beta+\lambda)/2-\mu)!}\sim \exp\left[(n-\beta)\log\frac{n-\beta}{2}-\lambda\log\frac{\lambda}{2}-n+\beta+\lambda\right],
\end{equation}
\begin{multline}
\sum_{\mu'=0}^{\gamma}\dfrac{(n/2-\gamma+\mu')!(n/2-\mu')!}{((n/2-\lambda-\gamma)/2+\mu')!((\gamma+n/2-\lambda)/2-\mu')!}\sim\\
 \exp\left[(n-\gamma)\log\frac{n-\gamma}{2}-(n/2-\lambda)\log\frac{n/2-\lambda}{2}-\frac{n}{2}+\gamma-\lambda\right].
\end{multline}
Now from the asymptotics of binomial terms we have that the sum over $\lambda$ is maximized when we take the supreme of the next quantity
\begin{multline}
f_{\beta\gamma}^{n}(\lambda)=\dfrac{n}{2}\log\dfrac{n}{2}-\left(\dfrac{n}{2}-\lambda\right)\log\left(\dfrac{n}{2}-\lambda\right)-\lambda\log\lambda+\left(\dfrac{n}{2}-\lambda\right)\log\left(\dfrac{n}{2}-\lambda\right)-\left(\dfrac{\beta-\lambda}{2}\right)\log\left(\dfrac{\beta-\lambda}{2}\right)\\
-\left(\dfrac{n-\beta-\lambda}{2}\right)\log\left(\dfrac{n-\beta-\lambda}{2}\right)+\lambda\log\lambda-\left(\dfrac{\gamma-n/2+\lambda}{2}\right)\log\left(\dfrac{\gamma-n/2+\lambda}{2}\right)\\
-\left(\dfrac{\lambda-\gamma+n/2}{2}\right)\log\left(\dfrac{\lambda-\gamma+n/2}{2}\right)+\lambda\log\lambda+\left(\dfrac{n}{2}-\lambda\right)\log\left(\dfrac{n}{2}-\lambda\right)\\
-\lambda\log\dfrac{\lambda}{2}-\left(\dfrac{n}{2}-\lambda\right)\log\dfrac{n/2-\lambda}{2},
\end{multline}
where we have only included the terms containing $\lambda$ in the exponential rate of the elements of the sum. The expression above can be simplified further to
\begin{multline}
f_{\beta\gamma}^{n}(\lambda)=-\left(\dfrac{\beta-\lambda}{2}\right)\log\left(\dfrac{\beta-\lambda}{2}\right)-\left(\dfrac{n-\beta-\lambda}{2}\right)\log\left(\dfrac{n-\beta-\lambda}{2}\right)\\
-\left(\dfrac{\gamma-n/2+\lambda}{2}\right)\log\left(\dfrac{\gamma-n/2+\lambda}{2}\right)-\left(\dfrac{\lambda-\gamma+n/2}{2}\right)\log\left(\dfrac{\lambda-\gamma+n/2}{2}\right)+\dfrac{n}{2}\log n.
\end{multline} 
Differentiating with respect to $\lambda$ and solving for the maximum we obtain
\begin{equation}
\lambda^{*}=\beta -\gamma +\frac{\gamma ^2-\beta ^2}{n}+\frac{n}{4},
\end{equation}
and replacing on the asymptotic expression we obtain
\begin{multline}
2^{n/2}\sum_{\lambda}\binom{n/2}{\lambda}c(\beta|\lambda)c(\gamma|n/2-\lambda)\sim 2^{n/2-\beta-\gamma}\exp\left[f_{\beta\gamma}^{n}(\lambda^{*})-{2n}\log\dfrac{n}{2}+\beta\log\beta+\gamma\log\gamma\right.\\
\left.+(n-\beta)\log\dfrac{n-\beta}{2}+(n-\gamma)\log\dfrac{n-\gamma}{2}\right].
\end{multline}
Putting everything together we have that
\begin{multline}
R_{0,0}^{\boxplus\beta\gamma}\sim \exp\left[nH(\bar{\beta})+nH(\bar{\gamma})-\left(\dfrac{n}{2}+\beta+\gamma\right)\log 2+f^{n}_{\beta\gamma}(\lambda^{*})-2n\log\dfrac{n}{2}+\beta\log\beta+\gamma\log\gamma\right.\\
\left.+(n-\beta)\log\dfrac{n-\beta}{2}+(n-\gamma)\log\dfrac{n-\gamma}{2} \right].
\end{multline}
For the asymptotic rate we finally have
\begin{equation}
\lim\limits_{n\to\infty} \dfrac{\log R^{\boxplus\beta\gamma}_{0,0}}{n}=\dfrac{1}{2}\log\dfrac{1}{2}+f_{\bar{\beta}\bar{\gamma}}(\bar{\lambda}^*),
\end{equation}
where
\begin{multline}
f_{\bar{\beta}\bar{\gamma}}(\bar{\lambda})=-\left(\dfrac{\bar{\beta}-\bar{\lambda}}{2}\right)\log\left(\dfrac{\bar{\beta}-\bar{\lambda}}{2}\right)-\left(\dfrac{1-\bar{\beta}-\bar{\lambda}}{2}\right)\log\left(\dfrac{1-\bar{\beta}-\bar{\lambda}}{2}\right)\\
-\left(\dfrac{\bar{\gamma}-1/2+\bar{\lambda}}{2}\right)\log\left(\dfrac{\bar{\gamma}-1/2+\bar{\lambda}}{2}\right)-\left(\dfrac{\bar{\lambda}-\bar{\gamma}+1/2}{2}\right)\log\left(\dfrac{\bar{\lambda}-\bar{\gamma}+1/2}{2}\right),
\end{multline}
and
\begin{equation}
\bar{\lambda}^*=\bar{\beta} -\bar{\gamma} +\bar{\gamma} ^2-\bar{\beta} ^2+\frac{1}{4}.
\end{equation}
Let us now see how does this function behaves for $\bar{\beta}$ and $\bar{\gamma}$ in Figure \ref{fbg}, notice that it is only defined for $\beta+\gamma\geq n/2$ as we expected since we are calculating the rate in the upper region of the compatibility polytope.
Note that the values at the extreme points are
\begin{equation}
\lim\limits_{n\to\infty} \dfrac{\log R^{\boxplus\boxplus\bullet}_{0,0}}{n}=0,
\end{equation}
\begin{equation}
\lim\limits_{n\to\infty} \dfrac{\log R^{\boxplus\bullet\boxplus}_{0,0}}{n}=0,
\end{equation}
and
\begin{equation}
\lim\limits_{n\to\infty} \dfrac{\log R^{\boxplus\boxplus\boxplus}_{0,0}}{n}=\log 2.
\end{equation}
We can write the asymptotic expression for general $\bar{\beta}$ and $\bar{\gamma}$ (after some algebra) in terms of Shannon entropy as
\begin{equation}
\dfrac{1}{2}\log\dfrac{1}{2}+f_{\bar{\beta}\bar{\gamma}}(\bar{\lambda}^*)=H\left(\bar{\gamma}_1\bar{\gamma}_2+\bar{\beta}_1^2-\frac{1}{4},\bar{\gamma}_1\bar{\gamma}_2+\bar{\beta}_2^2-\frac{1}{4},\bar{\beta}_1\bar{\beta}_2+\bar{\gamma}_1^2-\frac{1}{4} , \bar{\beta}_1\bar{\beta}_2+\bar{\gamma}_2^2-\frac{1}{4}\right),
\end{equation} 
where $H$ is the Shannon entropy defined as
\begin{equation}
H({p_i})=-\sum_i p_i \log p_i.
\end{equation}